# Bespoke Nanoparticle Synthesis and Chemical Knowledge Discovery Via Autonomous Experimentations

*Hyuk Jun Yoo,[1,2†] Nayeon Kim,[1,3†] Heeseung Lee,[1,4] Daeho Kim,[1,2] Leslie Tiong Ching Ow,[1] Hyobin Nam,[5] Chansoo Kim,[1,6] Seung Yong Lee,[5] Kwan-Young Lee,[2*] Donghun Kim,[1*] and Sang Soo Han[1*]*

[1]Computational Science Research Center, Korea Institute of Science and Technology, Seoul 02792, Republic of Korea

[2]Department of Chemical and Biological Engineering, Korea University, Seoul 02841, Republic of Korea

[3]Department of Chemistry, Korea University, Seoul 02841, Republic of Korea

[4]Department of Materials Science and Engineering, Korea University, Seoul 02841, Republic of Korea

[5]Materials Architecturing Research Center, Korea Institute of Science and Technology, Seoul 02792, Republic of Korea

[6]AI-Robot Department, University of Science and Technology (UST), Seoul 02792, Republic of Korea

†These authors contributed equally.

*Correspondence to: sangsoo@kist.re.kr (S.S.H.); donghun@kist.re.kr (D.K.); kylee@korea.ac.kr (K.-Y.L.).

# Abstract

The optimization of nanomaterial synthesis using numerous synthetic variables is considered to be extremely laborious task because the conventional combinatorial explorations are prohibitively expensive. In this work, we report an autonomous experimentation platform developed for the *bespoke* design of nanoparticles (NPs) with targeted optical properties. This platform operates in a closed-loop manner between a batch synthesis module of NPs and a UV-Vis spectroscopy module, based on the feedback of the AI optimization modeling. With silver (Ag) NPs as a representative example, we demonstrate that the Bayesian optimizer implemented with the early stopping criterion can efficiently produce Ag NPs precisely possessing the desired absorption spectra within only 200 iterations (when optimizing among five synthetic reagents). In addition to the outstanding material developmental efficiency, the analysis of synthetic variables further reveals a novel chemistry involving the effects of citrate in Ag NP synthesis. The amount of citrate is a key to controlling the competitions between spherical and plate-shaped NPs and, as a result, affects the shapes of the absorption spectra as well. Our study highlights both capabilities of the platform to enhance search efficiencies and to provide a novel chemical knowledge by analyzing datasets accumulated from the autonomous experimentations.

# Main

Nanoparticles (NPs) have extensively been utilized in various applications including solar cells[1–4], catalysis[5–7] and chemical sensors[8–10] due to the wide tunability of electronic and optical properties through the manipulation of their size, shape and surface states[11-14]. In particular, the wet chemical synthesis of colloidal NPs has attracted much attention mainly because of its low-cost and solution-based processing. The properties of colloidal NPs are known to be highly affected by numerous synthetic variables, including solution volume[15,16], concentration[17–19], injection rate[20,21], synthetic sequences[22] and aging time[23,24]. Currently, combinatorial experiments (*aka* Edisonian approach) are often used to design NPs that exhibit desired properties, but these trial-and-error-based methods are extremely laborious and time-consuming, which inspires the need for an approach of more intelligent explorations over large chemical spaces.

Autonomous material developments based on artificial intelligence (AI) have recently emerged as a promising direction to maximize the search efficiency over various classes of materials, including organic molecule[25,26,27], perovskite[28–30], colloidal quantum dots[31], and nanoparticles[32-35]. For example, Abolhasani and coworkers[31] adopted AI in their quantum dot developmental systems, where only 920 trial experiments were sufficient to reach the target properties. Similarly, Cooper and coworkers[36] reported an autonomous developmental system for finding photocatalytic materials where ten environmental synthetic conditions were successfully optimized with less than 700 experiments in a week, which is a surprisingly small number given the vast parameter spaces over ten dimensions. Such dramatic enhancements in the search efficiency were achieved mainly because AI models effectively learn the correlations between synthetic conditions and properties and, as a result, suggest improved synthetic recipes as next trial conditions.

Certainly, recent reports regarding autonomous NP synthesis have demonstrated the accelerated experiments aided by robotic data collections[25-36] and AI optimization modeling[37,38]. Although the significantly enhanced search efficiency is regarded as the utmost value of an autonomous laboratory, we believe that a novel chemical knowledge discovery could naturally be created since the AI robotic systems are destined to explore not only the broad range of chemical spaces but also the relationship between the reaction conditions and the corresponding material properties. In typical laboratory environments (e.g., without robotics and AI), researchers often intuitively exclude the majority of the available chemical input spaces to intentionally avoid negative results and expedite experimentation, which severely hinders the new chemistry discovery. On the other hand, since no prior knowledge is provided, the AI robotic platforms in autonomous laboratories are capable of conducting broader exploration toward identifying both positive and negative results, which should be well-suited for novel chemistry discovery by revealing the effects and interdependency of certain parameters of synthetic ingredients. Despite these possible benefits, previous reports focused mainly on improving the material developmental efficiency but lacked efforts to find chemical insights from the operations of AI robotic platforms.

In the process of developing new materials, a primary goal is often to discover materials that exhibit the most superior properties. For instance, when developing catalysts, the aim is to discover materials that maximize the catalytic activity, selectivity, and stability. In addition, in certain cases, one may need to develop a material that satisfies multiple specified properties simultaneously. For example, researchers may seek metallic NPs with sizes of 5 nm and spherical shapes, or they may desire plate-like NPs with sizes of 10 nm. Moreover, there are situations where materials with known properties are needed but the synthesis condition remain undisclosed. Such scenarios call for a *bespoke* approach to tailor materials according to specific user requirements. To accelerate materials discovery, computer simulations such as density functional theory have become prevalent[39]. While these methods efficiently identify materials with the maximized properties based on the structure-property relationship, achieving *bespoke* design with the current computer simulation techniques remains quite challenging. In contrast, an autonomous laboratory offers a more comprehensive approach to materials development. By utilizing an autonomous laboratory, it can be feasible to determine the synthesis conditions for materials possessing desired properties on the basis of the relationship between the synthesis parameters and the material properties (i.e., a process-property relationship).

Herein, we report an autonomous laboratory for the *bespoke* NP developments with target optical properties. The platform operates based on robotic executions of the NP synthesis module and ultraviolet-visible (UV-Vis) spectroscopy module as well as AI-driven exploration of synthetic recipes. Using Ag NPs as representative materials, we demonstrate that our Bayesian optimizer with the early stopping criterion successfully generates Ag NPs with target absorption spectra within only 200 iterations when optimizing across five synthetic reagents without any prior knowledge given. Quantitative estimations revealed that the number of required experiments increases approximately in a linear manner with the increasing number of synthesis variables, in contrast to the grid-based search scheme in an exponential manner. In addition to the search optimization efficiency, our analysis of synthetic variables further unraveled a novel theory regarding the effects of citrate in Ag NP synthesis: The amount of citrate is key to controlling the competition between spherical and plate-shaped Ag NPs, which affects the peak intensity and peak sharpness in the absorption spectra. This work highlights that AI robotic platforms can significantly enhance the material developmental efficiency and can unveil previously unknown chemistry based on datasets collected using the autonomous experimentations.

# Results

**Closed-loop NP design in the autonomous laboratory.** Our closed-loop operations for the NP development are schematized in Fig. 1a. Our autonomous laboratory was designed to output Ag NPs possessing a given target absorption spectrum based on the robotic automation of NP batch synthesis, UV-Vis spectroscopic characterization, and AI modeling. The system functions as follows: (1) the target property is given as an input: for example, Ag NPs with a maximum-intensity-wavelength ($\lambda_{max}$) of 573 nm, (2) robotic platforms operate to obtain synthesis and characterization datasets, (3) the AI model learns those datasets to suggest a better synthetic recipe or conditions, and finally, steps (2) and (3) iterate until the target property is achieved.

**Hardware setting and automation.** Fig. 1b shows the bird's eye view of our hardware, where each NP synthesis module (left part of Fig. 1b) and UV-Vis spectroscopy module (right part of Fig. 1b) is separately shown. A video depicting hardware operations is available in Supplementary Video 1 for easier understanding of our system components and their interactions. The batch synthesis module is composed of vial storage, a stirring machine, and a solution dispensing system. The vial storage system is a self-designed two-story vial holder system where unused (empty) and used (solution-filled) vials are spatially separated in the bottom and top floors, respectively. Both floors have some gradients along *z*-axis and are equipped with a serve motor with Arduino Uno so that the vials continuously and flawlessly move down by gravity. The detailed design principles of our vial storage system are provided in Supplementary Figure S2a, which we believe would be useful when applying to other types of chemical vessels. The stirring machine supports 16 vial holders which enables the simultaneous execution of multiple synthetic experiments simultaneously. The solution dispensing system consists of a stock solution, syringe pump, dispenser, and XYZ linear actuator, as shown in Fig. 1b and Supplementary Figure S2. Each stock solution was stored in an ice bath to prevent the degradation of chemicals. Each syringe pump and the XYZ linear actuator precisely controls the volume/rate and target positions, respectively, for the task of solution movements from the stock into vials on the stirring machine. Moreover, the UV-Vis spectroscopy module (right part of Fig. 1b) is composed of a chemical vessel container, pipetting system, and UV-Vis spectrometer system. Note that a cuvette is a transparent and very small scale (2 mL) vessel that is prototypically used for optical measurements, and thus the pipetting system is necessary to sample the colloidal NP solutions into cuvettes based on OpenLH[40]. The movements of chemical vessels between the synthesis and characterization modules are controlled by the robotic arm with OnRobot gripper. More detailed information about our hardware settings are provided in Supplementary Figures S1-S3.

It is worth noting the vision system of our autonomous laboratory (although not main focus of this study), which is designed to detect any machine failure cases based on object detection techniques. Our systems operate without human interventions, which may lead to dangerous situation, such as imperfect vial movement and positioning. To address these safety problems and to improve the fidelity of our system, vision systems were introduced to continuously monitor operational errors at several sites including stirring machines (for vial positioning detection), cuvette holders (for cuvette positioning detection), and pipetting system (for pipette tipping detection). Our object detection model implemented in the laboratory (named as DenseSSD[41]) was previously reported and was demonstrated to be much more efficient than traditional computer vision models for detecting transparent objects such as chemical vessels. At least a few hundred real scenes were collected at each monitoring site for the machine learning training and validation of the DenseSSD model. When the model detects failure cases, it is programmed such that the whole operation is immediately halted to prevent any possible dangers. Despite the recent increasing attention to autonomous developmental systems in materials science community, we importantly note that there is a lack of effort to address safety issues in fully-robotic (researcher-free) environments. We propose that our vision-based safety alert system can become an essential component for the democratization of the autonomous material synthesis laboratory without human surveillance. More detailed information about the vision system is provided in Supplementary Figure S3c as well as our previous work.[41]

Measuring instrumental errors is critical to understand the reliability of the hardware systems. The precisions of the hardware components in both the NP synthesis and characterization modules were tested as follows. First, for the solution dispensing systems, the injection volume was precisely controlled as evidenced by the 0.999 $R^2$ value, which substantially outperforms conventional pipettes as shown in Supplementary Figure S4a and S4b. Second, a precision test was performed for the UV-Vis spectrometer, where the deviation in the wavelength peak position was measured to be very small at approximately only 4.35 nm, as shown in Supplementary Figure S4c. Using these reliable instruments, we performed the NP synthesis and UV-Vis spectroscopic characterizations 100 times with identical recipes, and the deviations of three optical properties including $\lambda_{max}$, full width at half maximum (FWHM), and peak intensity were measured to be very small as shown in Supplementary Figure S4, indicating that our experimental systems are highly reliable.

**Synthesis recipe optimization modeling.** Our autonomous laboratory platform sought to synthesize Ag NPs with target optical properties that are tailored to predefined requirements. Ag NPs with specific optical properties are well known to be functional in diverse applications such as biomedical application[42], sensors[43,44], light emission diodes[45], and sensitizers of solar cells[1-4]. In particular, when mixed Ag NPs, consisting of red (absorption spectrum with $\lambda_{max}$ = 513 nm), purple ($\lambda_{max}$ = 573 nm), and blue ($\lambda_{max}$ = 667 nm) NPs, have been used as light-trapping materials for broad light-harvesting in organic solar cells, and achieved remarkable enhancement in efficiency.[3] However, since all of the three Ag NPs are commercialized, their synthesis recipes are unknown. Herein, we aim to reproduce commercialized Ag NPs via our autonomous laboratory in which the target absorption spectra are obtained from the literature[3]. We used five reagents, $AgNO_3$, $H_2O_2$, $NaBH_4$, citrate and $H_2O$, as ingredients for Ag NP synthesis, and their volumes were parameterized in the optimization process. More details of the absorption spectra and synthesis process are shown in Supplementary Figures S5 and S6. In addition, we introduced a fitness function, ranging between -1 and zero, based on the matches of three features of $\lambda_{max}$, FWHM, and peak intensity to measure the similarity of the produced absorption spectrum and the targeted one. The mathematical equation for the fitness function is provided in the Methods and Supplementary Figure S7. Out of three factors that construct the fitness function, the heaviest weights (90%) were given to the match of $\lambda_{max}$ while much smaller weights (<10%) were assigned to the other two factors. We note that in general applications, these weight assignments would depend on the relevant experimental targets.

We used a Bayesian optimization model, and Fig. 2a shows the evolution of the fitness using the Bayesian optimizer for the example of the 573 nm ($\lambda_{max}$) target property. Since the acquisition function of the upper confidence bound (UCB) was applied[28,36], exploitation and exploration were simultaneously considered, and as a result, some fluctuations in fitness were observed in Fig. 2a, rather than a monotonic increase in the fitness values. To maximize the search efficiency without much sacrificing the accuracy, we implemented a stopping rule named the early stopping criterion, which was conceived from the early stopping regularization method[46,47] to prevent overfitting issues in machine learning. In our early stopping scheme, when the best fitness value (between -0.1 and zero) is not updated even after five consecutive iterations (i.e., the patience of 5), then the optimization search stops. More details about the early stopping criterion are described in the Methods, Supplementary Table S1 and Figure S8.

Figs. 2b and 2c show the evolutions of the absorption spectra and five reagent volumes, respectively, as the experimental iterations proceed. For this exemplary case of the 573 nm ($\lambda_{max}$) target property, it took less than 200 iterations for the AI model to reach the optimal result. Until approximately 100 iterations, the produced absorption spectra are found to be much different from the target in terms of both $\lambda_{max}$ and FWHM. When the fitness first became larger than -0.1 at approximately 120 iterations, we observed that $\lambda_{max}$ was well matched at 573 nm after the volume control of $AgNO_3$, $H_2O_2$, and $NaBH_4$, but the match of FWHM was still not satisfactory. Then, further optimizations led to improved fitness at approximately 175 iterations where both $\lambda_{max}$ and FWHM were excellently fit after citrate volume control from 4,000 μL to 100 μL. This tendency is generally observed that the AI model first optimizes $\lambda_{max}$ with the controls of $AgNO_3$, $H_2O_2$, and $NaBH_4$ reagents and then later tunes FWHM with citrate controls.

Similar results for the other two cases of 513 nm and 667 nm target wavelengths are provided in Supplementary Figures S9 and S10. For these cases, it also took less than 200 iterations for optimization, which is a surprisingly small number given that the theoretical count for a grid-based exploration for five parameters of reagent volumes is at least on the order of $10^9$. The comparison of efficiency between autonomous search and grid-based search will be further discussed in the last result section. Due to the dominant weight of $\lambda_{max}$ in the fitness function, the early iterations fit $\lambda_{max}$ and latter iterations more closely fit other factors of FWHM mainly based on the control of citrate volumes.

**Analysis of absorption spectrum and NP morphology.** The absorption spectra obtained by our Bayesian optimizer with the early stopping criterion are presented in Fig. 3a, and are compared to the targeted spectra obtained from the literature for three cases of 513 nm, 573 nm and 667 nm. Overall, they match excellently, although some differences were observed at the shoulder peak in the lower wavelength range of 300-450 nm. These differences are attributed to the fitness function design, where the match of the main peak position ($\lambda_{max}$) and width (FWHM) are the most important parameters, while matching of other subpeaks is not considered.

Figs. 3b and 3c highlight the morphologies of Ag NPs for each target $\lambda_{max}$ case based on transmission electron microscopy (TEM) results. The average size of Ag NPs increases from 20.1 nm to 32.9 nm for absorption spectra with $\lambda_{max}$ of 513 nm to $\lambda_{max}$ of 667 nm, respectively, and this correlating trend between $\lambda_{max}$ and NP size is consistent with previous reports[1,48,49]. The double peaks (main and shoulder peaks) in the absorption spectra indicate the existence of two different morphologies of NPs, i.e., spherical shapes and plate shapes. It is well known that the different shapes of Ag NPs cause absorptions at different wavelengths: spherical shapes lead to absorptions at 300-450 nm[50-52] while plate shapes should be responsible for absorptions in the larger wavelength range of 450-700 nm due to the in-plane (lateral direction of the plate) dipolar resonance[49,53,54], as agreed with the deconvolution results of UV-Vis absorption spectra in Supplementary Figure S11. The mixing of two different shapes (spheres and plates) was confirmed from the TEM results in Fig. 3c and Supplementary Figure S12 where both triangular nanoplates with {111} surfaces and spherical NPs were observed at different ratios. We note that for the case of 513 nm, both shapes were comparably found, whereas nanoplates were dominantly found for the other two cases of 573 nm and 667 nm. This finding well supports that the FWHM, which is closely related to the NP size uniformity, was found to be large at 214.2 nm for the 513 nm target wavelength case, and relatively much smaller (<200 nm) for the other two cases[52-54] because the convolution of two peaks could result in a single broader peak.

**SHAP-based interpretation of synthesis variables.** Using the datasets accumulated from the autonomous experiments, we performed SHapley Additive exPlanations (SHAP)[55] analysis to measure the influence of synthesis variables of five input reagents, and the results for each target wavelength ($\lambda_{max}$) are presented in Fig. 4. Independent of the target wavelengths, the volumes of $AgNO_3$ and $H_2O_2$ are strongly influential parameters whereas those of citrate and $H_2O$ are weakly influential. The significant effects of $AgNO_3$ (metal source), $H_2O_2$ (oxidant), and $NaBH_4$ (reductant) reagents on NP morphologies and properties are well documented in many previous reports[50,53,56–58], and agrees well with our SHAP analysis.

Based on the SHAP analysis, the trajectories of synthesis variables during optimizations are investigated, as shown in Fig. 4 where the strongly influential variables ($AgNO_3$, $H_2O_2$) and the weakly influential variables (citrate, $H_2O$) are separately visualized in two-dimensional fitness maps. The data clearly show that AI broadly explores the full volume range for $AgNO_3$ and $H_2O_2$, and that, on the contrary, the explorations are relatively limited for citrate and $H_2O$. These conclusions are confirmed by dense search points only in the corners in the corresponding fitness maps of Fig. 4. Such different observations likely occur because the volume changes of citrate or $H_2O$ have relatively little effect on $\lambda_{max}$, and thus there are weaker driving forces for the AI model to explore the full volume range of these reagents broadly. For all target wavelength cases, it was consistently found that the fitness becomes terrible when the $H_2O_2$ volume is too large likely because NPs are too oxidized and, as a result, decompose into Ag ions ($Ag^+$)[43,58]. It was also found that the volume ratio of $H_2O_2$ to $AgNO_3$ ($H_2O_2/AgNO_3$) is strongly correlated with $\lambda_{max}$ as shown in Supplementary Figure S13 and other literature[56-58] and thus would serve as a crucial descriptor for controlling peak positions in the absorption spectra.

**Discovery of novel chemical knowledge from autonomous experimentations.** Although the SHAP analysis in Fig. 4 revealed that, out of five synthetic ingredients, the volume of citrate is the least influential parameter on $\lambda_{max}$ of the absorption spectra, its effects on FWHM and absorption peak intensity were found clearly noticeable as shown in Figs. 2b and 2c. In Fig. 5a where the absorption spectrum results of two extreme citrate volumes (100 μL and 4,000 μL) were compared with all other conditions being identical, the lower volume of citrate causes stronger and sharper absorption peaks, i.e., higher peak intensity and smaller FWHM in Fig 5a. This trend is attributed to different NP morphologies confirmed from TEM images where NPs tend to be spherical for the 100 μL case and were mainly formed as plate shapes for 4,000 μL. We importantly note that the citrate effects have not been clearly reported thus far in the literature,[50,53,56–58] and the comparative results lead us to propose novel chemical theory regarding the citrate effects on NP formation and resultant UV-Vis absorption spectra.

To systematically explore the effects of citrate in a broader range, we parametrized the citrate concentrations from 0 to 160 mM in Fig. 5b instead of controlling the citrate volume since the volume control range is limited up to the vessel capacity. From the deconvolution analyses of the absorption spectra, we found that a peak around the wavelength of 400 nm dominates over the other peak at approximately 600 nm for higher concentration cases. The peak approximately 400 nm is contributed by the spherical NPs while that around 600 nm is attributed to the nanoplates, and these two peaks formed with different ratios depending on the citrate concentrations, as shown in Fig. 5b and Supplementary Figure S14. For the 160 mM case, spherical NPs were almost exclusively found while nanoplates were dominantly found for the other extreme case of 0.5 mM. Note that NPs were not produced without citrate (0 mM case), which was also well reported in previous studies[50,53]. This analysis adequately explains the experimental observation that the FWHM is the largest at approximately 300 nm at intermediate concentration levels (10-20 mM) due to the convolution of comparably strong peaks. A more detailed analysis of citrate concentration controls and their effects is shown in Supplementary Figure S15.

The effects of citrate concentrations on NP shapes can be readily understood from the growth kinetics, as schematized in Fig. 5c. Citrate tends to preferably adsorb on Ag {111} surfaces due to the stronger binding energy[59-62], which results in anisotropic NP growth, i.e., plate shapes. However, if sufficient amounts of citrates were given during the NP synthesis, they adsorb on all available binding sites on both {111} and other surfaces such as {100}, which likely leads to isotropic NP growth, i.e., spherical shapes. We emphasize that, unlike many other recent studies of autonomous material synthesis[25-36] which mainly focused on enhancing the search efficiency based on the combined approach of robotics and AI modeling, our study highlights the two capabilities for improved search efficiency and uncovering a novel chemical knowledge (in this case, the effect of citrate on NP formation) by analyzing the datasets collected autonomous experiments.

**Quantitative estimations of search efficiency.** To quantitatively understand the search efficiency of our Bayesian optimizer, we measured the number of experiments required to complete the optimization with an increasing number of synthetic reagents, as shown in Fig. 6. These synthesis variables were progressively added from two to five in the order of their previously determined SHAP impact (Fig. 4), i.e., $AgNO_3$ → $H_2O_2$ → $NaBH_4$ → citrate → $H_2O$. The efficiency of the AI model was estimated by comparing it to theoretically computed numbers of the grid-based search scheme which is also known as a full factorial design. Here, the grid-based search indicates that, if each parameter range is divided into $N$ grids for $M$ synthetic reagents, then the exploration number over the whole chemical space would scale exponentially as $N^M$. Note that $N$ is set as 79 in our study given that the large volume control ranges between 100 μL and 4,000 μL. As the number of synthesis variables linearly increases grid-based explorations would become exponentially expensive, which severely limits practical uses of such grid-based schemes. On the other hand, our AI-based search is significantly more efficient, completing the optimizations within 200 iterations for all three cases of different target wavelengths (513 nm, 573 nm, and 667 nm), as shown in Fig. 6 and Supplementary Table S2. We emphasize that this search efficiency depends on the number of variables. Unlike the grid-based search, the number of required experiments increases approximately in a linear manner with the increasing number of synthesis variables, and this feature should be highly beneficial when exploring complex chemical spaces involving multiple parameters.

We also note particular features of some traditional approaches in the design of experiments (DOE), including Latin hypercube sampling (LHS)[63] and the Taguchi method[64]. LHS[63] is basically a type of random sampling approaches that uniformly distributes the sampling points over the whole parameter space. Although LHS ensures wide explorations, its search efficiency is similar to that of grid-based search. The Taguchi method[64] is also one of the widely employed approaches in DOE. This method nicely reflects the variations of parameters but is limited when exploring interdependent synthesis variables[65], which is common case in most material synthesis experiments (for an example, the interdependency between the reagent volume parameter and concentration parameter). Unlike traditional DOE approaches, AI modeling, such as Bayesian optimizations, exhibits evident benefits, including a balanced search mechanism between exploitation and exploration, and significantly enhanced optimization efficiency particularly in high-dimensional parameter spaces.

# Discussion

Although the demonstrations presented in this work were performed on synthesizing Ag NPs and characterizing optical properties, our presented platforms and closed-loop experimental workflows can be readily expanded to other materials such as multicomponent NPs and other properties that may be desired in respective applications such as catalysis, solar cells, and sensors. The study of autonomous material development is only its infancy, and thus, several limitations need to be resolved in the near future to enable broad applicability. First, from a hardware point of view, some experimental tools and protocols need to be designed to be more robot-friendly rather than human-in-the-loop. In the autonomous laboratory, human interventions will be only minimal and robotic executions will be dominant. However, most of the experimental hardware equipment that is commercially available today requires manual intervention, which sometimes makes the robotic operations extremely challenged to execute simple human-enabled tasks. For example, rotating the cap of vials (a type of chemical vessel) is an easy task for humans, but requires difficult and time-consuming processes for robots. Second, from a software point of view, the AI models need to be substantially advanced. Efforts to find the global minimum with the best search efficiency, even in a complex chemical space, need to be continued. Additionally, the next-generation AI models should solve for categorical variables (synthesis sequences, selections/sequences of operations and chemicals, etc.), in addition to continuous variables (such as solution volumes, concentrations, and injection rate).

In summary, we developed and reported an autonomous laboratory platform for the *bespoke* design of NPs with target optical properties. This platform operates in a closed-loop manner between the NP synthesis module and UV-Vis spectroscopy module, in which the experimental planning and decision-making were performed by a Bayesian optimizer implemented with the early stopping criterion. Our autonomous laboratory excellently optimized Ag NPs to yield desired absorption spectra within 200 iterations when considering five synthetic reagents. We also find that the number of required experiments increases approximately in a linear manner with the increasing number of synthesis variables, as opposed to the grid-based search scheme in an exponential manner. In addition to the outstanding search efficiency, further analysis of synthesis variables revealed previously unknown chemical knowledge regarding the effect of citrate on Ag NP synthesis. The amount of citrate is key to controlling the competition between spherical and plate-shaped NPs. As a result, it also determines the resulting Ag NPs absorption spectra. Our work emphasizes the value of autonomous experimentation platform, which offer twofold benefits of enhancing material developmental efficiency and elucidating novel chemical knowledge by analyzing the datasets accumulated from the operations of AI robotic platforms.

# Methods

**Chemical reagents.** Silver nitrate ($AgNO_3$), sodium citrate dihydrate (citrate), 30 wt% hydrogen peroxide solution ($H_2O_2$) and sodium borohydride ($NaBH_4$) were purchased from Sigma–Aldrich. All reagents were used as received. All stock solutions were based on distilled water. As shown in Supplementary Figure S2c, solutions of $NaBH_4$ and $H_2O_2$ were stored in ice bucket and changed every 12 hours to delay the potential of the reductant and oxidant.

**Batch synthesis.** $AgNO_3$ solution (1.25 mM), citrate solution (20 mM), $H_2O_2$ solution (0.375 wt%), $NaBH_4$ (10 mM) and DI water were stored in brown bottles to protect them from light illumination. The chemical reactants were chosen from the literature[50,53]. The reason for controlling the discrete variable is that our solution dispensing system has an error of approximately 1~2 µL, so that it is given discretely as 50 µL to make the AI model robust if several dispensing errors occur in the hardware. A robotic arm picks vials up and puts inside in a stirrer to prepare nanoparticle synthesis. In the synthesis process sequence, $NaBH_4$, DI water, citrate, and $H_2O_2$ were sequentially injected with a solution dispensing system and stirred for five minutes to mix evenly on a stirrer. After that, $AgNO_3$ was added and stirred for 40 minutes at 800 rpm at room temperature. After the nanoparticle synthesis is finished, the reaction is terminated, and vials with synthesized nanoparticles are moved in the UV-Vis spectroscopy module. All processes of the batch synthesis module are shown in Supplementary

Video 1.

**Optical spectroscopic characterizations.** After NP synthesis is completed, the robotic arm places the solution-filled vials into a vial holder. Then, the arm locates the cuvette from storage to the holder and places the cuvette into the UV-Vis spectroscopic testing holder to obtain optical properties. The absorption spectra data were extracted by a spectrometer through an optical fiber, and a light source was used to obtain reference and absorbance data. The cuvette was filled to 2 mL with DI water, and the colloidal nanoparticle solution was injected to 0.8 mL, and mixed three times via a robotic arm and linear actuator pump. The pipetting system was followed by OpenLH[40]. The Scipy library, including the *find_peaks, peak_prominences,* and *peak_widths* functions, was used for the extraction of optical properties. All processes of the UV-Vis spectroscopy module are also shown in Supplementary Video 1.

**Fitness function.** Fitness is an evaluation function to measure the degree of match between two absorption spectra. This can be computed as follows:

$$Fitness\ function = -0.9 * \frac{(\lambda_{max} - \lambda_{max,target})}{A_{\lambda_{max}}} - 0.07 * (1 - intenisty) - 0.03 * \frac{FWHM}{A_{FWHM}} \quad (1)$$

More detailed information on the fitness function is described in Supplementary Figure S6.

**Bayesian optimization with the early stopping criterion.** We implemented Bayesian optimization with a Gaussian process, which interpolates each data point continuously. Bayesian optimization can use several types of acquisition functions, such as expected improvement and entropy search. Among them, the upper confidence bound (UCB) was implemented to consider exploitation and exploration simultaneously in our autonomous laboratory[36]. Let a UCB acquisition function be defined as the following equation:

$$UCB = \mu(X) + \kappa\sigma(X), (\kappa = 10) \quad (2)$$

where $X = \{x_1, \ldots, x_n\}$ is the vector of variables, $\mu(X)$ is the mean function, $\sigma(X)$ is the deviation and $\kappa$ is the weight of exploration. To generate initial data points, grid sampling or random sampling was used for conventional implementation. However, grid sampling fixes some parameters and can mislead initial data without including variations in synthesis conditions. On the other hand, random sampling can complement grid sampling, but the spread of data points is localized depending on random seeds with respect to exploration. Therefore, in this study, Latin hypercube sampling[63] was implemented to consider some parts of the data distribution evenly for each parameter, devoid of conventional sampling artifacts. It can reflect the variation in synthesis conditions with a small amount of data. We used the Matern kernel function, constant scaling and noise, which can allow interpolation smoothness and experimental errors. The code of Bayesian optimization was followed by past research[36]. The early stopping function was modified to add a filter exploration process. The value of patience was set to 5, and the filter value was set to -0.1 in Supplementary Table S1 and Figure S7. Details of our Bayesian optimization with early stopping code implementation can be found in the Data availability.

**TEM analysis.** TEM equipment (FEI Tecnai, FEI Titan) was used to obtain the morphology data of the Ag NPs. ImageJ was used to measure the size distribution in TEM images[66].

**SHAP analysis.** SHapely Addictive exPlanations[55] (SHAP) analysis is a game theoretic approach to calculate the influence order of variables when the model predicts some outputs. A higher SHAP value implies a large impact of the variable on the model. *KernelExplainer* with *LogitLink* utilizes weighted linear regression to compute SHAP values from models for each epoch cumulatively through experimental data. The SHAP values were implemented to draw the beeswarm plot colored by the value of the synthesis variable via the *summary_plot* function. We also visualized the fitness surface based on the accumulated experimental data points, as shown in Fig. 4. Here, each experimental datum is composed of a synthesis recipe and fitness value. The fitness surface was made based on the linear interpolation libraries implemented in MATLAB.

# Data Availability

Several examples of our result data are provided in the following GitHub repository (https://github.com/KIST-CSRC/MasterPlatform/tree/main/DB). Each datum consists of synthesis recipe, raw spectra data, and optical properties.

# Code Availability

The codes and relevant explanations that are required for the autonomous laboratory functioning are available in the GitHub repository. They were prepared in three platforms of master platform (https://github.com/KIST-CSRC/MasterPlatform), batch synthesis platform (https://github.com/KIST-CSRC/BatchSynthesisPlatform) and UV-Vis spectroscopy platform (https://github.com/KIST-CSRC/UVPlatform). All codes are written over Python 3.7 and all environments could be created through requirements.txt file.

# References

1. Putnin, T. *et al.* Enhanced organic solar cell performance: Multiple surface plasmon resonance and incorporation of silver nanodisks into a grating-structure electrode. *Opto-Electronic Adv.* **2**, 1–11 (2019).

2. Mallah, A. R. *et al.* Experimental study on the effects of multi-resonance plasmonic nanoparticles for improving the solar collector efficiency. *Renew. Energy* **187**, 1204–1223 (2022).

3. Phengdaam, A. *et al.* Improvement of organic solar cell performance by multiple plasmonic excitations using mixed-silver nanoprisms. *J. Sci. Adv. Mater. Devices* **6**, 264–270 (2021).

4. Wang, D. H. *et al.* Enhanced light harvesting in bulk heterojunction photovoltaic devices with shape-controlled Ag nanomaterials: Ag nanoparticles versus Ag nanoplates. *RSC Adv.* **2**, 7268 (2012).

5. Hvolbæk, B. *et al.* Catalytic activity of Au nanoparticles. *Nano Today* **2**, 14–18 (2007).

6. Reier, T., Oezaslan, M. & Strasser, P. Electrocatalytic oxygen evolution reaction (OER) on Ru, Ir, and pt catalysts: A comparative study of nanoparticles and bulk materials. *ACS Catal.* **2**, 1765–1772 (2012).

7. Lee, H. W. *et al.* Solid-solution alloying of immiscible Pt and Au boosts catalytic performance for $H_2O_2$ direct synthesis. *Acta Mater.* **205**, 116563 (2021).

8. Howes, P. D., Chandrawati, R. & Stevens, M. M. Colloidal nanoparticles as advanced biological sensors. *Science (80-. ).* **346**, (2014).

9. Khaniani, Y. *et al.* A gold nanoparticle-protein G electrochemical affinity biosensor for the detection of SARS-CoV-2 antibodies: a surface modification approach. *Sci. Rep.* **12**, (2022).

10. Neal, C. J. *et al.* Picomolar detection of hydrogen peroxide using enzyme-free inorganic nanoparticle-based sensor. *Sci. Rep.* **7**, (2017).

11. Xia, Y., Xiong, Y., Lim, B. & Skrabalak, S. E. Shape-controlled synthesis of metal nanocrystals: Simple chemistry meets complex physics? *Angew. Chemie Int. Ed.* **48**, 60–103 (2009).

12. Roduner, E. Size matters: Why nanomaterials are different. *Chem. Soc. Rev.* **35**, 583–592 (2006).

13. Galati, E. et al. Shape- specific patterning of polymer-functionalized nanoparticles. *ACS Nano* **11**, 4995–5002 (2017).

14. Abolhasani, M., Oskooei, A., Klinkova, A., Kumacheva, E. & Günther, A. Shaken, and stirred: Oscillatory segmented flow for controlled size evolution of colloidal nanomaterials. *Lab on a Chip* **14**, 2309–2318 (2014).

15. Xu, Z., Gao, H. & Guoxin, H. Solution-based synthesis and characterization of a silver nanoparticle–graphene hybrid film. *Carbon N. Y.* **49**, 4731–4738 (2011).

16. Kim, D.-Y. *et al.* Green synthesis of silver nanoparticles using Laminaria japonica extract: Characterization and seedling growth assessment. *J. Clean. Prod.* **172**, 2910–2918 (2018).

17. Chen, D., Qiao, X., Qiu, X. & Chen, J. Synthesis and electrical properties of uniform

silver nanoparticles for electronic applications. *J. Mater. Sci.* **44**, 1076–1081 (2009).

18. Agnihotri, S., Mukherji, S. & Mukherji, S. Size-controlled silver nanoparticles synthesized over the range 5–100 nm using the same protocol and their antibacterial efficacy. *RSC Adv.* **4**, 3974–3983 (2014).

19. Bastús, N. G., Merkoçi, F., Piella, J. & Puntes, V. Synthesis of highly monodisperse citrate-stabilized silver nanoparticles of up to 200nm: kinetic control and catalytic properties. *Chem. Mater.* **26**, 2836–2846 (2014).

20. Kwak, C. H. *et al.* Customized microfluidic reactor based on droplet formation for the synthesis of monodispersed silver nanoparticles. *J. Ind. Eng. Chem.* **63**, 405–410 (2018).

21. Kim, D., Jeong, S. & Moon, J. Synthesis of silver nanoparticles using the polyol process and the influence of precursor injection. *Nanotechnology* **17**, 4019–4024 (2006).

22. Ojea-Jiménez, I., Bastús, N. G. & Puntes, V. Influence of the sequence of the reagents addition in the citrate-mediated synthesis of gold nanoparticles. *J. Phys. Chem. C* **115**, 15752–15757 (2011).

23. Izak-Nau, E. *et al.* Impact of storage conditions and storage time on silver nanoparticles physicochemical properties and implications for their biological effects. *RSC Adv.* **5**, 84172–84185 (2015).

24. Jin, W. *et al.* The influence of CTAB-capped seeds and their aging Time on the morphologies of silver nanoparticles. *Nanoscale Res. Lett.* **14**, 81 (2019).

25. Coley, C. W. *et al.* A robotic platform for flow synthesis of organic compounds informed by AI planning. *Science (80-. ).* **365**, (2019).

26. Steiner, S. *et al.* Organic synthesis in a modular robotic system driven by a chemical programming language. *Science (80-. ).* **363**, (2019).

27. Langner, S. *et al.* Beyond Ternary OPV: High-throughput experimentation and self-driving laboratories optimize multicomponent Systems. *Adv. Mater.* **32**, 1907801 (2020).

28. Epps, R. W. *et al.* Artificial chemist: an autonomous quantum dot synthesis bot. *Adv. Mater.* **32**, 2001626 (2020).

29. Li, J. *et al.* Autonomous discovery of optically active chiral inorganic perovskite

nanocrystals through an intelligent cloud lab. *Nat. Commun.* **11**, 2046 (2020).

30. Higgins, K., Valleti, S. M., Ziatdinov, M., Kalinin, S. V. & Ahmadi, M. Chemical robotics enabled exploration of stability in multicomponent lead halide perovskites via machine learning. *ACS Energy Lett.* **5**, 3426–3436 (2020).

31. Volk, A. A. *et al.* AlphaFlow: autonomous discovery and opti-mization of multi-step chemistry using a self-driven fluidic lab guided by reinforcement learning. *Nat. Commun.* **14**, 1403 (2023).

32. Salley, D. *et al.* A nanomaterials discovery robot for the Darwinian evolution of shape programmable gold nanoparticles. *Nat. Commun.* **11**, 2771 (2020).

33. Tao, H. *et al.* Self-driving platform for metal nanoparticle synthesis: combining microfluidics and machine learning. *Adv. Funct. Mater.* **31**, 2106725 (2021).

34. Mekki-Berrada, F. *et al.* Two-step machine learning enables optimized nanoparticle synthesis. *npj Comput. Mater.* **7**, 55 (2021).

35. Jiang, Y. *et al.* An artificial intelligence enabled chemical synthesis robot for exploration and optimization of nanomaterials. *Sci. Adv.* **8**, 1–12 (2022).

36. Burger, B. *et al.* A mobile robotic chemist. *Nature* **583**, 237–241 (2020).

37. Häse, F., Roch, L. M., Kreisbeck, C. & Aspuru-Guzik, A. Phoenics: A Bayesian optimizer for chemistry. *ACS Cent. Sci.* **4**, 1134–1145 (2018).

38. Häse, F., Aldeghi, M., Hickman, R. J., Roch, L. M. & Aspuru-Guzik, A. Gryffin: An algorithm for Bayesian optimization of categorical variables informed by expert knowledge. *Appl. Phys. Rev.* **8**, 031406 (2021)

39. Yeo, B.C., Nam, H., Nam, H. *et al.* High-throughput computational-experimental screening protocol for the discovery of bimetallic catalysts. *npj Comput Mater* **7**, 137 (2021).

40. Gome, G., Waksberg, J., Grishko, A., Wald, I. Y. & Zuckerman, O. OpenLH: Open liquid-handling system for creative experimentation with biology. in *TEI 2019 - Proceedings of the 13th International Conference on Tangible, Embedded, and Embodied Interaction* 55–64 (2019).

41. Tiong, L. C. O. *et al.* Machine vision for vial positioning detection toward the safe automation of material synthesis. Preprint at https://arxiv.org/abs/2206.07272 (2022)

42. Farooq, S., Dias Nunes, F. & de Araujo, R. E. Optical properties of silver nanoplates and perspectives for biomedical applications. *Photonics Nanostructures - Fundam. Appl.* **31**, 160–167 (2018).

43. Zannotti, M. *et al.* Tuning of hydrogen peroxide etching during the synthesis of silver nanoparticles. An application of triangular nanoplates as plasmon sensors for $Hg^{2+}$ in aqueous solution. *J. Mol. Liq.* **309**, (2020).

44. Mayer, K. M. & Hafner, J. H. Localized surface plasmon resonance sensors. *Chem. Rev.* **111**, 3828–3857 (2011).

45. Saade, J. & De Araújo, C. B. Synthesis of silver nanoprisms: A photochemical approach using light emission diodes. *Mater. Chem. Phys.* **148**, 1184–1193 (2014).

46. Dai, Z., Yu, H., Low, K. H. & Jaillet, P. Bayesian optimization meets Bayesian optimal stopping. in *International Conference on Machine Learning* (2019).

47. Makarova, A. *et al.* Automatic Termination for Hyperparameter Optimization. *Proceedings of the First International Conference on Automated Machine Learning*, PMLR 188:7/1-21 (2021).

48. Aherne, D., Ledwith, D. M., Gara, M. & Kelly, J. M. Optical properties and growth aspects of silver nanoprisms produced by a highly reproducible and rapid synthesis at room temperature. *Adv. Funct. Mater.* **18**, 2005–2016 (2008).

49. Pastoriza-Santos, I. & Liz-Marzán, L. M. Colloidal silver nanoplates. State of the art and future challenges. *J. Mater. Chem.* **18**, 1724–1737 (2008).

50. Zhang, Q., Li, N., Goebl, J., Lu, Z. & Yin, Y. A systematic study of the synthesis of silver nanoplates: Is citrate a “magic” reagent? *J. Am. Chem. Soc.* **133**, 18931–18939 (2011).

51. Linic, S., Christopher, P. & Ingram, D. B. Plasmonic-metal nanostructures for efficient conversion of solar to chemical energy. *Nat. Mater.* **10**, 911–921 (2011).

52. Kelly, K. L., Coronado, E., Zhao, L. L. & Schatz, G. C. The optical properties of metal

nanoparticles: The influence of size, shape, and dielectric environment. *J. Phys. Chem. B* **107**, 668–677 (2003).

53. Tsuji, M. *et al.* Rapid transformation from spherical nanoparticles, nanorods, cubes, or bipyramids to triangular prisms of silver with PVP, citrate, and $H_2O_2$. *Langmuir* **28**, 8845–8861 (2012).

54. Brioude, A. & Pileni, M. P. Silver nanodisks: Optical properties study using the discrete dipole approximation method. *J. Phys. Chem. B* **109**, 23371–23377 (2005).

55. Martin, T. *Interpretable machine learning*. (2019).

56. Li, N. *et al.* $H_2O_2$-Aided Seed-Mediated Synthesis of Silver Nanoplates with Improved Yield and Efficiency. *ChemPhysChem* **13**, 2526–2530 (2012).

57. Parnklang, T. *et al.* $H_2O_2$-triggered shape transformation of silver nanospheres to nanoprisms with controllable longitudinal LSPR wavelengths. *RSC Adv.* **3**, 12886 (2013).

58. Wongravee, K. *et al.* Chemometric analysis of spectroscopic data on shape evolution of silver nanoparticles induced by hydrogen peroxide. *Phys. Chem. Chem. Phys.* **15**, 4183–4189 (2013).

59. Xu, H. & Wiley, B. J. The roles of citrate and defects in the anisotropic growth of Ag nanostructures. *Chem. Mater.* **33**, 8301–8311 (2021).

60. Jiang, X. C., Chen, C. Y., Chen, W. M. & Yu, A. B. Role of citric acid in the formation of silver nanoplates through a synergistic reduction approach. *Langmuir* **26**, 4400–4408 (2010).

61. Zeng, J. *et al.* A mechanistic study on the formation of silver nanoplates in the presence of silver seeds and citric acid or citrate ions. *Chem. - An Asian J.* **6**, 376–379 (2011).

62. Kilin, D. S., Prezhdo, O. V. & Xia, Y. Shape-controlled synthesis of silver nanoparticles: Ab initio study of preferential surface coordination with citric acid. *Chem. Phys. Lett.* **458**, 113–116 (2008).

63. Iman, R. L. Latin Hypercube sampling. In *Encyclopedia of Quantitative Risk Analysis and Assessment* (John Wiley & Sons, Ltd, 2008).

64. Saeed, M. Guttekin, O. Victoria, J. Chen-Hsiu, H. Strengths and limitations of Taguchi's contributions to quality, manufacturing, and process engineering. *Journal of Manufacturing Systems* **23**, 73-126 (2004).

65. Tao, H., Wu, T., Aldeghi, M. et al. Nanoparticle synthesis assisted by machine learning. *Nat Rev Mater* **6**, 701–716 (2021).

66. Rueden. C, Dietz. C, Horn. M, Schindelin. J, Northan. B, B. M. & E. K. ImageJ Ops. (2016).

# Acknowledgements

This work was supported by the National Research Foundation of Korea funded by the Ministry of Science and ICT [NRF-2022M3H4A7046278] and the IITP grant [No. 2021-0-02076] funded by the Korean government (MSIT).

# Author Contributions

S.S.H, Donghun Kim, and K.-Y.Lee conceived the idea and supervised the project. H.J.Y., N.K. performed hardware settings for the automations of NP synthesis and characterizations. H.J.Y. developed Bayesian optimizer with the early stopping criterion. N.K implemented an XAI approach to enable interpretable data insights. H.J.Y., N.K., H.L, Daeho Kim performed experiments for data collections. H.J.Y., N.K., H.L, Daeho Kim, H.N., C.K. and S.Y.L verified experimental data and TEM analysis. L.C.O.T developed vision systems based on object detection models. All authors contributed to results discussion and manuscript writing.

# Competing Interests

The authors declare no competing financial or non-financial interests.

# Figures

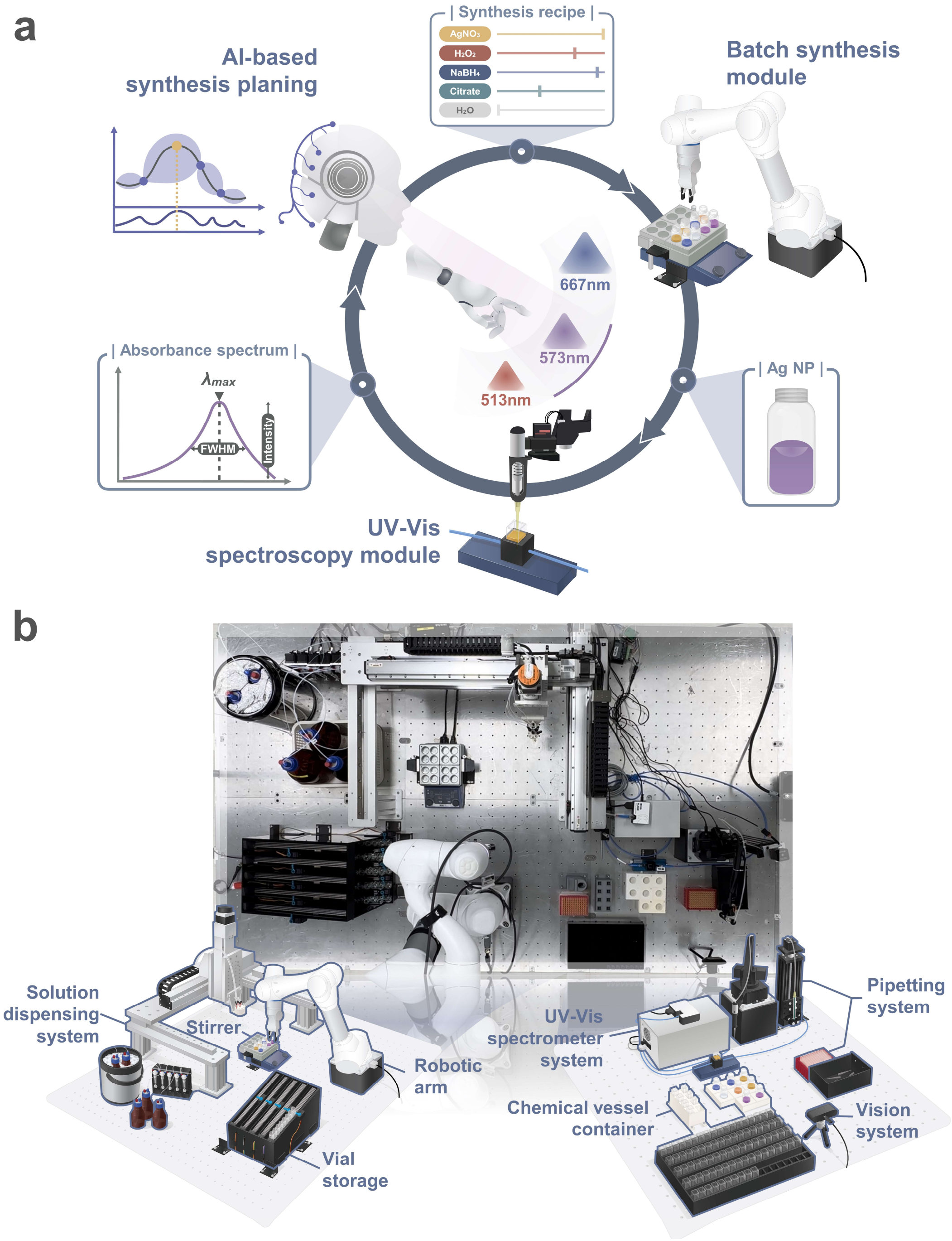

**Fig. 1 The autonomous laboratory platform for *bespoke* NP design with target optical properties.** (a) Scheme of the closed-loop operations for the development of Ag NPs with desired absorption spectra, as exemplified by $\lambda_{max}$ of 513 nm, 573 nm, and 667 nm. (b) A bird's eye view image of our autonomous laboratory and schematic illustrations of the batch NP synthesis module (left) and UV-Vis spectroscopy module (right). The synthesis module automatically synthesizes colloidal Ag NPs, while the UV-Vis spectroscopy module extracts the optical properties of the synthesized NPs. AI models optimize the synthetic recipe, although not illustrated in this figure.

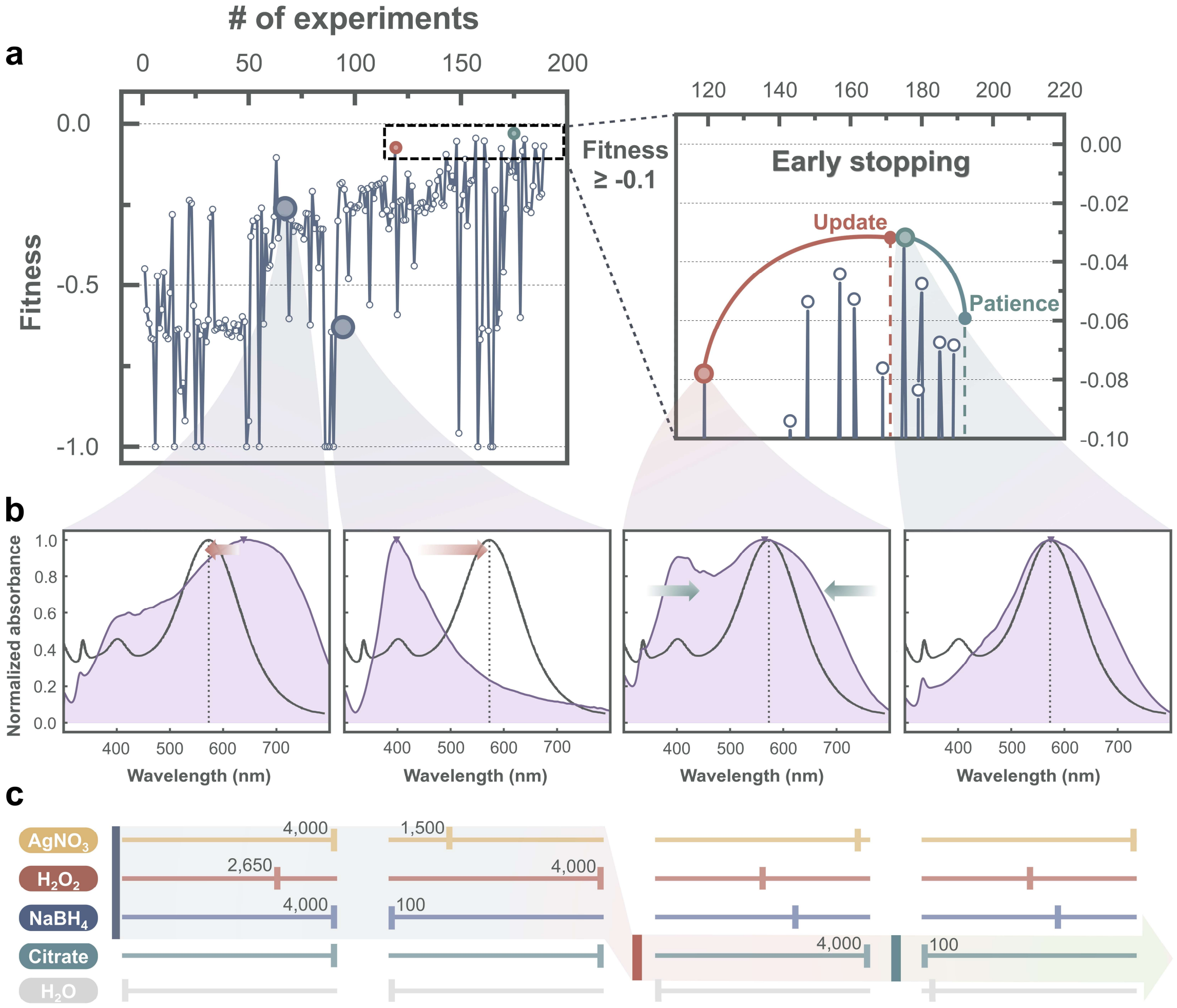

**Fig. 2 Bayesian optimizer with the early stopping criterion.** (a) The evolution of fitness values as the number of experiments progressed for the exemplary case of the 573 nm ($\lambda_{max}$) target property. The region of fitness values between -0.1 and zero is magnified on the right side. (b) Evolution of the absorption spectra at various experimental iterations. The produced absorption spectra (purple area) are compared to the target spectrum (gray line) obtained from the literature[3]. (c) The evolution of reagent volumes at various experimental iterations (same as Fig. 1b). The volume ranges were between 100 and 4,000 μL for all five reagents. The arrow in shadow follows and highlights the main changes during the optimization processes and indicates that, in the early stages, the control of $AgNO_3$, $H_2O_2$, and $NaBH_4$ is mainly observed while the control of citrate is noticeably observed in the later phase.

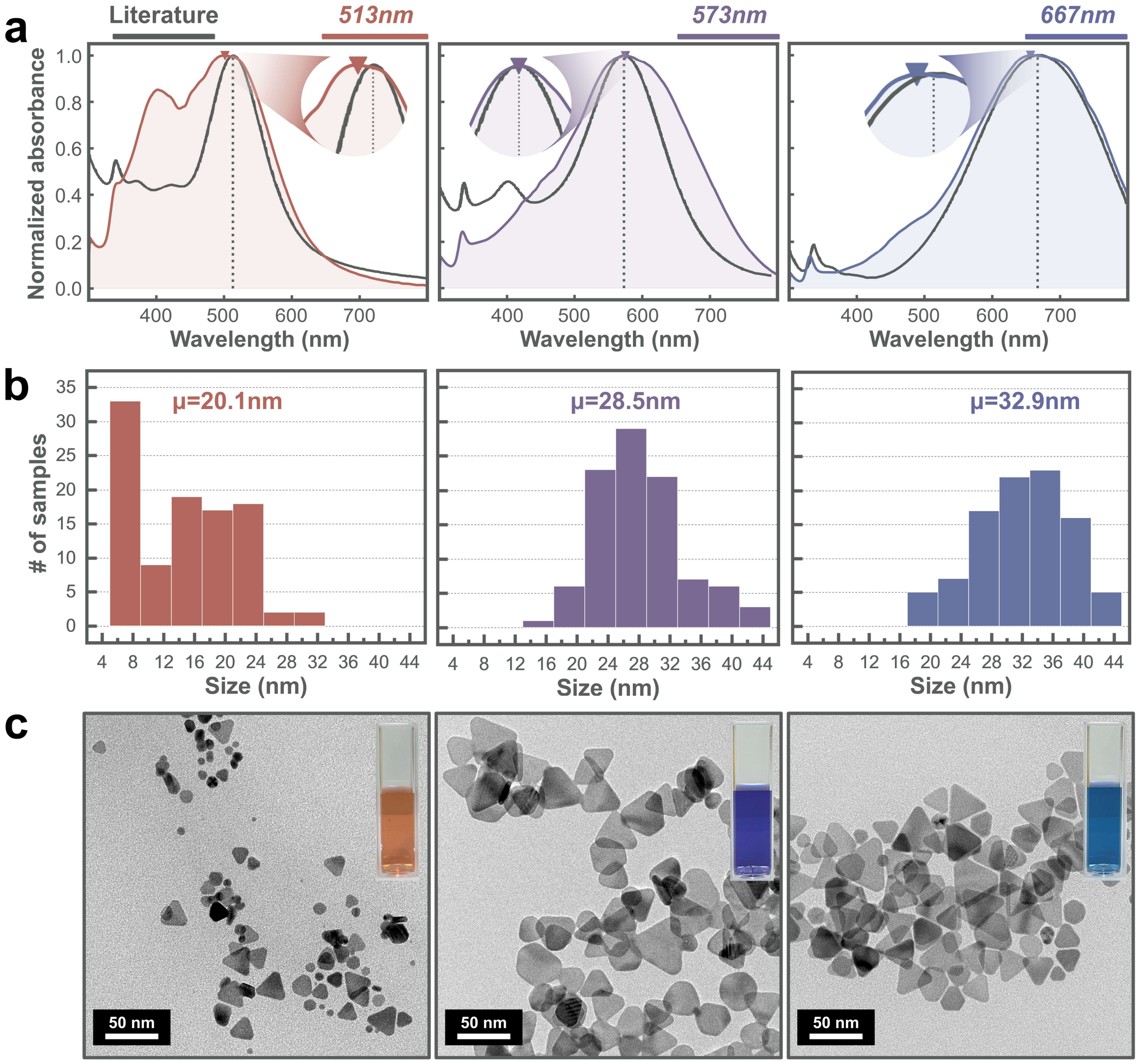

**Fig. 3 Analysis of UV-Vis absorption spectra and NP morphologies obtained by TEM.** (a) Comparison of the produced absorption spectra and target spectra obtained from the literature[3] for three cases of $\lambda_{max}$ values of 513 nm, 573 nm, and 667 nm. Spectra with the gray line represent target spectra in the literature[3]. (b) The NP size distributions for the three cases based on TEM analysis. The average NP size (μ) is included. (c) Representative TEM images for the three cases. The inset images show the real scenario of colloidal NP solutions, which are diluted in cuvettes for UV-Vis spectroscopic characterizations.

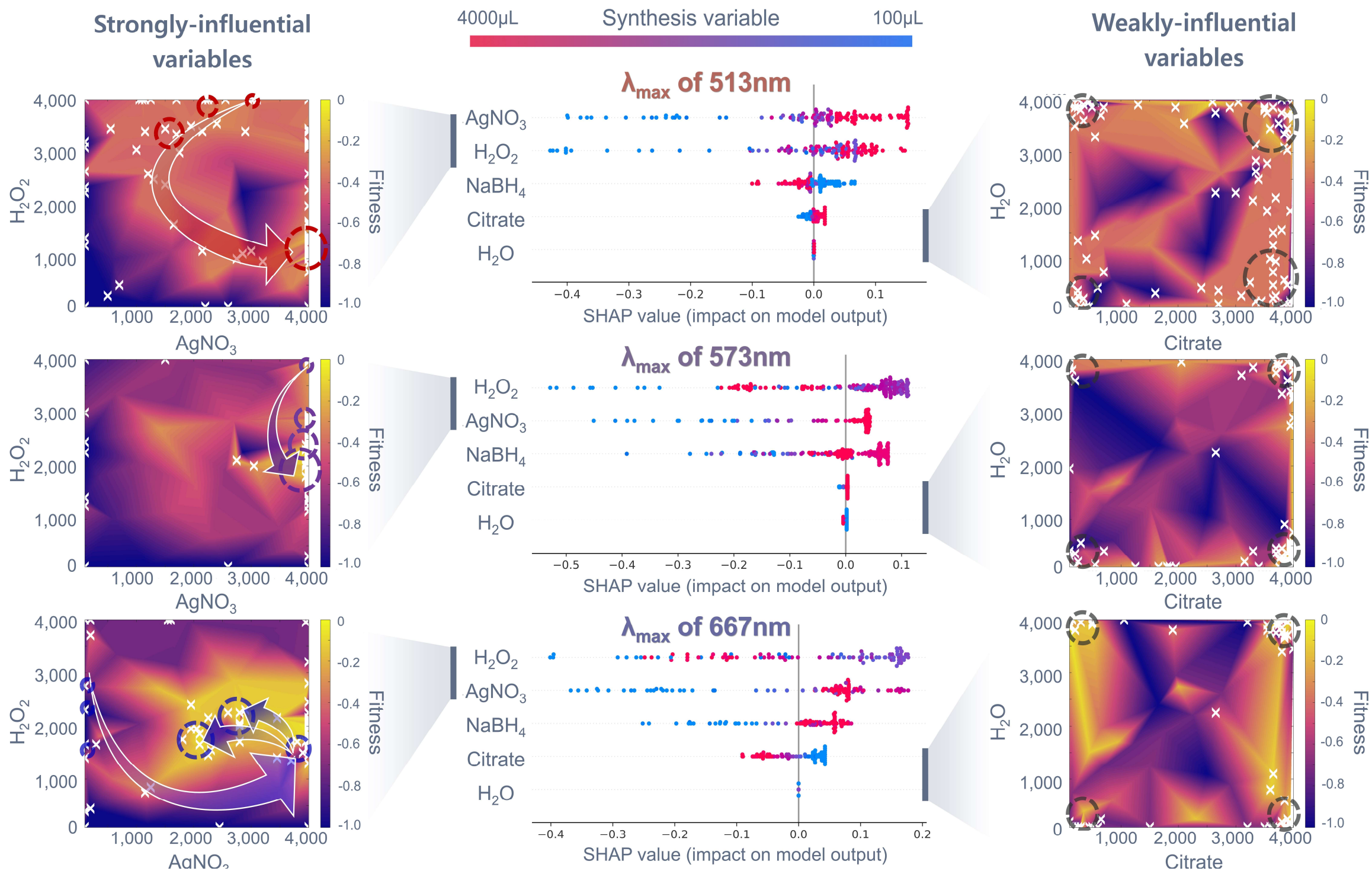

**Fig. 4 SHAP-based interpretation of synthetic variables.** The SHAP analyses are shown in the center, for three cases of $\lambda_{max}$ values of 513 nm, 573 nm, and 667 nm. In each case, the reagent with the highest impact is positioned on the top, while those with the lower impact follow. The color represents the volume of each synthetic reagent between 100 and 4,000 μL. Each left and right column shows the evolutions of two strongly-influential variables and two weakly influential variables in the fitness map, respectively. The circles and arrows in the fitness maps highlight the synthetic routes during optimizations.

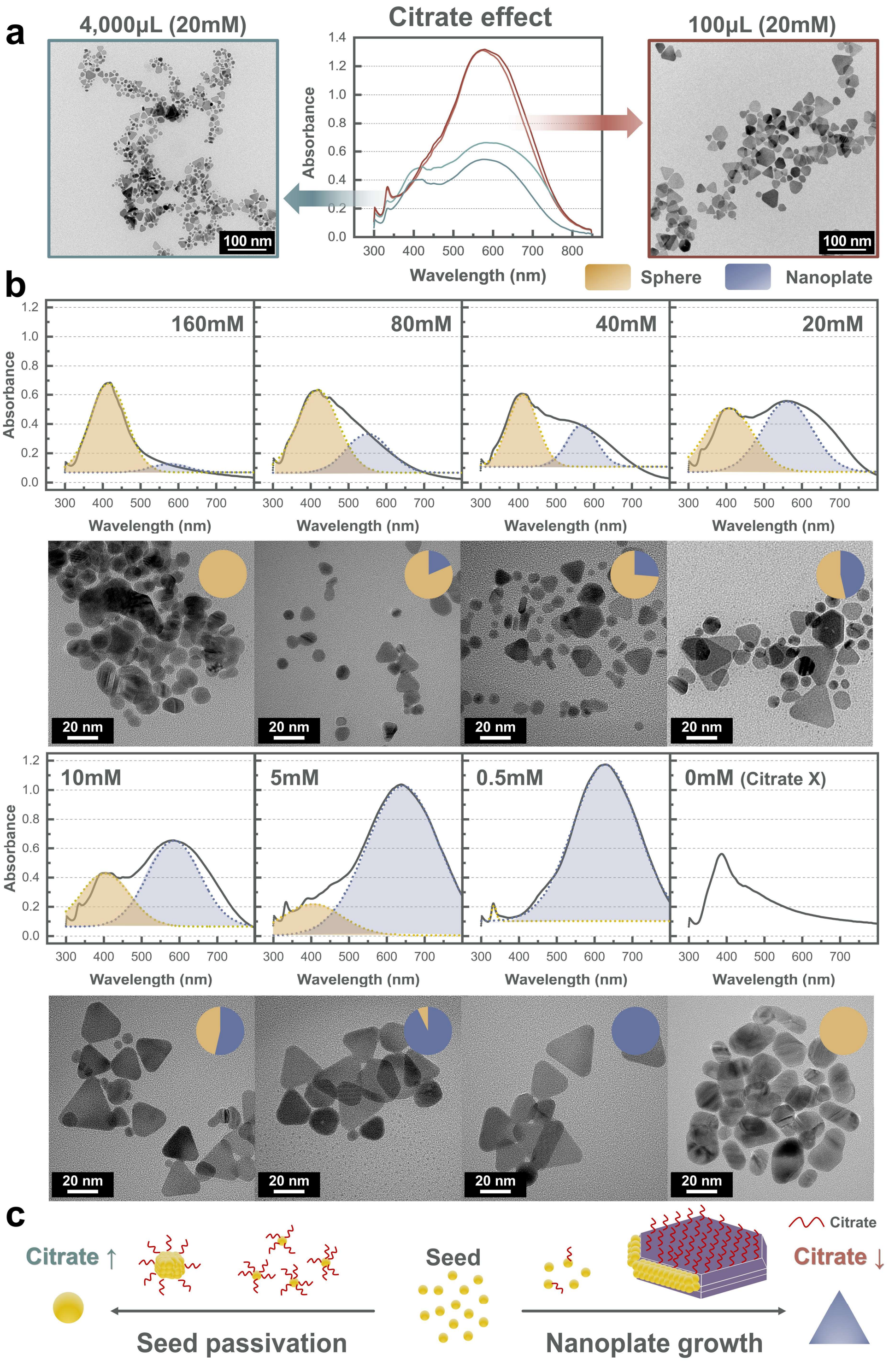

**Fig. 5 The effect of citrate on Ag NP synthesis and UV-Vis absorption spectra.** (a) Comparisons of absorption spectra and NP morphologies (TEM images) between citrate volumes of 100 μL (red lines) and 4,000 μL (green lines). (b) The changes in absorption spectra and the corresponding NP morphologies (TEM images) with variations in citrate concentrations from 0 mM (no use) to 0.5 mM (diluted) to 160 mM (extremely concentrated). The absorption spectra are deconvoluted with two peaks contributed by spherical NPs and plate-shaped NPs. The inset pie chart in the TEM images represents the ratio between two NP shapes, with blue denoting plate shapes and yellow denoting spherical shapes. (c) Schematic of citrate effects on Ag NP growth kinetics and mechanisms.

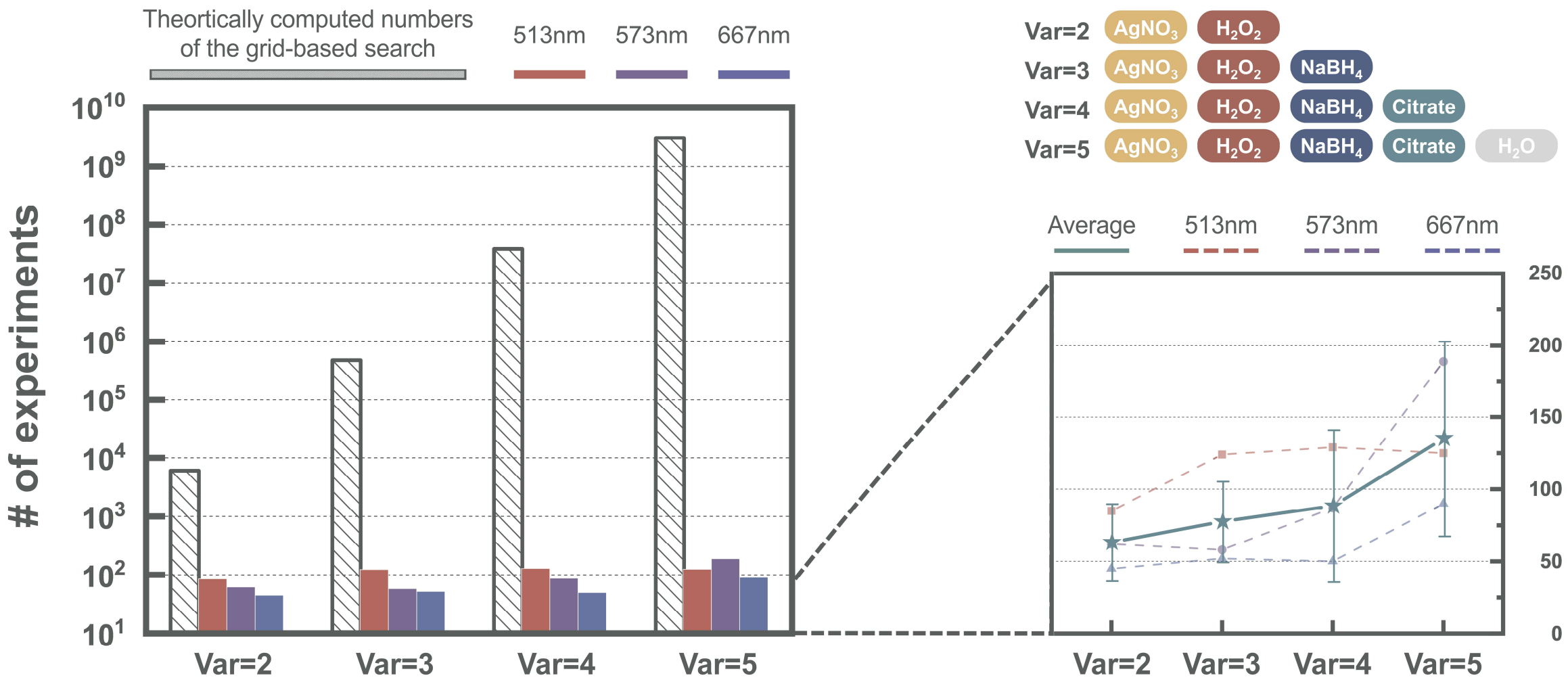

**Fig. 6 Quantitative estimations of AI-based optimization efficiency.** The number of experiments to complete the optimizations was shown with varying synthesis variables. The number of synthesis variables is progressively controlled from two to five, based on the SHAP impact order in Fig. 4. To better understand the optimization efficiency, the results were compared to the theoretically computed numbers of the grid-based search scheme. The AI modeling results are magnified on the right side.

Supplementary Information for

# Bespoke Nanoparticle Synthesis and Chemical Knowledge Discovery Via Autonomous Experimentations

*Hyuk Jun Yoo,[1,2†] Nayeon Kim,[1,3†] Heeseung Lee,[1,4] Daeho Kim,[1,2] Leslie Tiong Ching Ow,[1] Hyobin Nam,[5] Chansoo Kim,[1,6] Seung Yong Lee,[5] Kwan-Young Lee,[2*] Donghun Kim,[1*] and Sang Soo Han[1*]*

[1]Computational Science Research Center, Korea Institute of Science and Technology, Seoul 02792, Republic of Korea

[2]Department of Chemical and Biological Engineering, Korea University, Seoul 02841, Republic of Korea

[3]Department of Chemistry, Korea University, Seoul 02841, Republic of Korea

[4]Department of Materials Sciences and Engineering, Korea University, Seoul 02841, Republic of Korea

[5]Materials Architecturing Research Center, Korea Institute of Science and Technology, Seoul 02792, Republic of Korea

[6]AI-Robot Department, University of Science and Technology (UST), Seoul 02792, Republic of Korea

†These authors contributed equally.

*Correspondence to: sangsoo@kist.re.kr (S.S.H.); donghun@kist.re.kr (D.K.); kylee@korea.ac.kr (K.-Y.L.)

# Table of contents

# 1. Hardware information of autonomous laboratory

## Supplementary Figure S1. 6-axis robotic arm and gripper

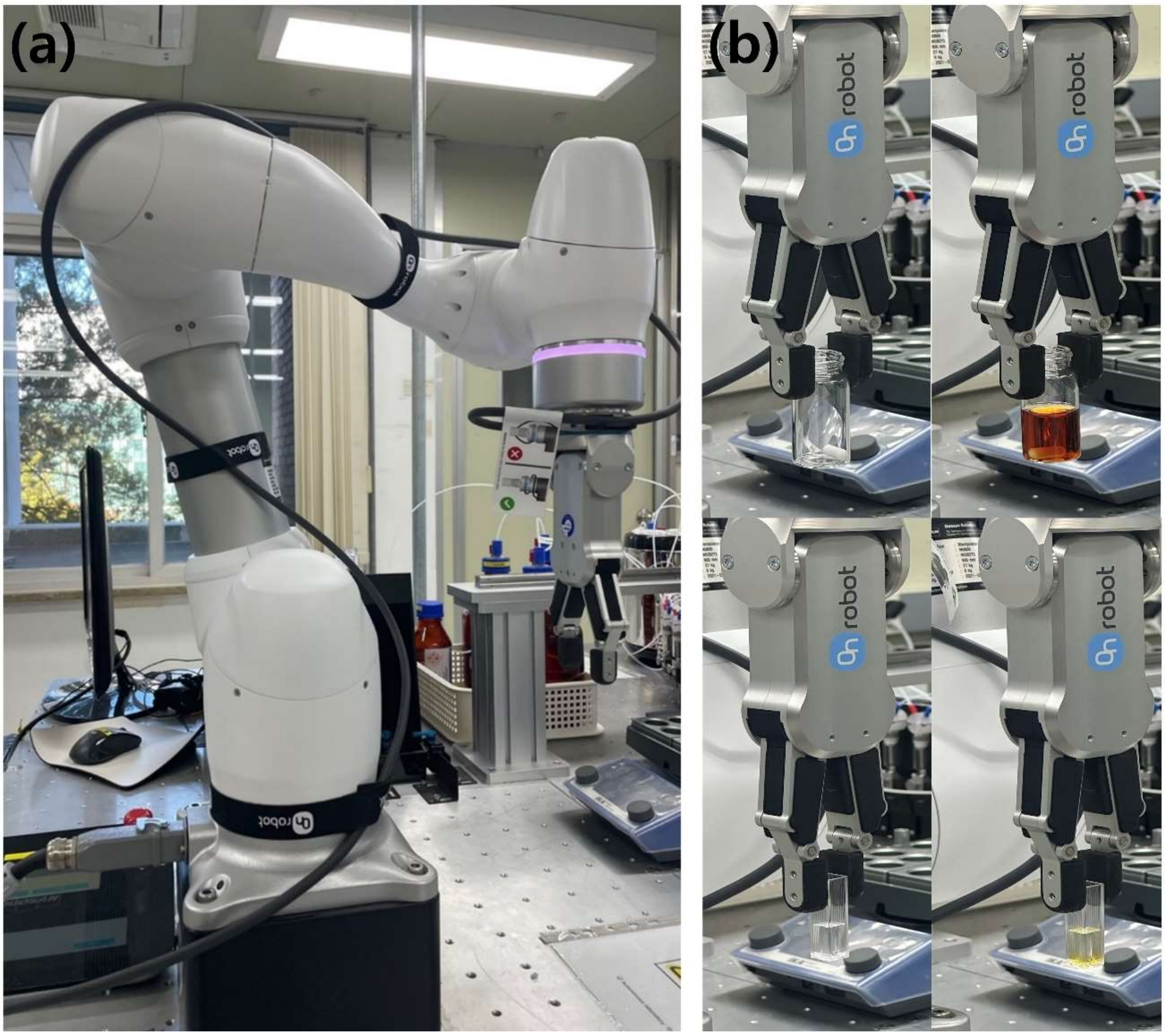

**(a)** Robotic arm (Doosan Robotics: M0609) and gripper (OnRobot: RG2).

**(b)** Pictures for gripping various chemical vessels such as vials and cuvettes to be used in our autonomous laboratory. The robotic arm has high degrees of freedom to treat various colloidal methods through the gripper, which is similar to human hands, so that it makes a more flexible system toward the scalability of an autonomous laboratory.

**Supplementary Figure S2. Batch synthesis module**

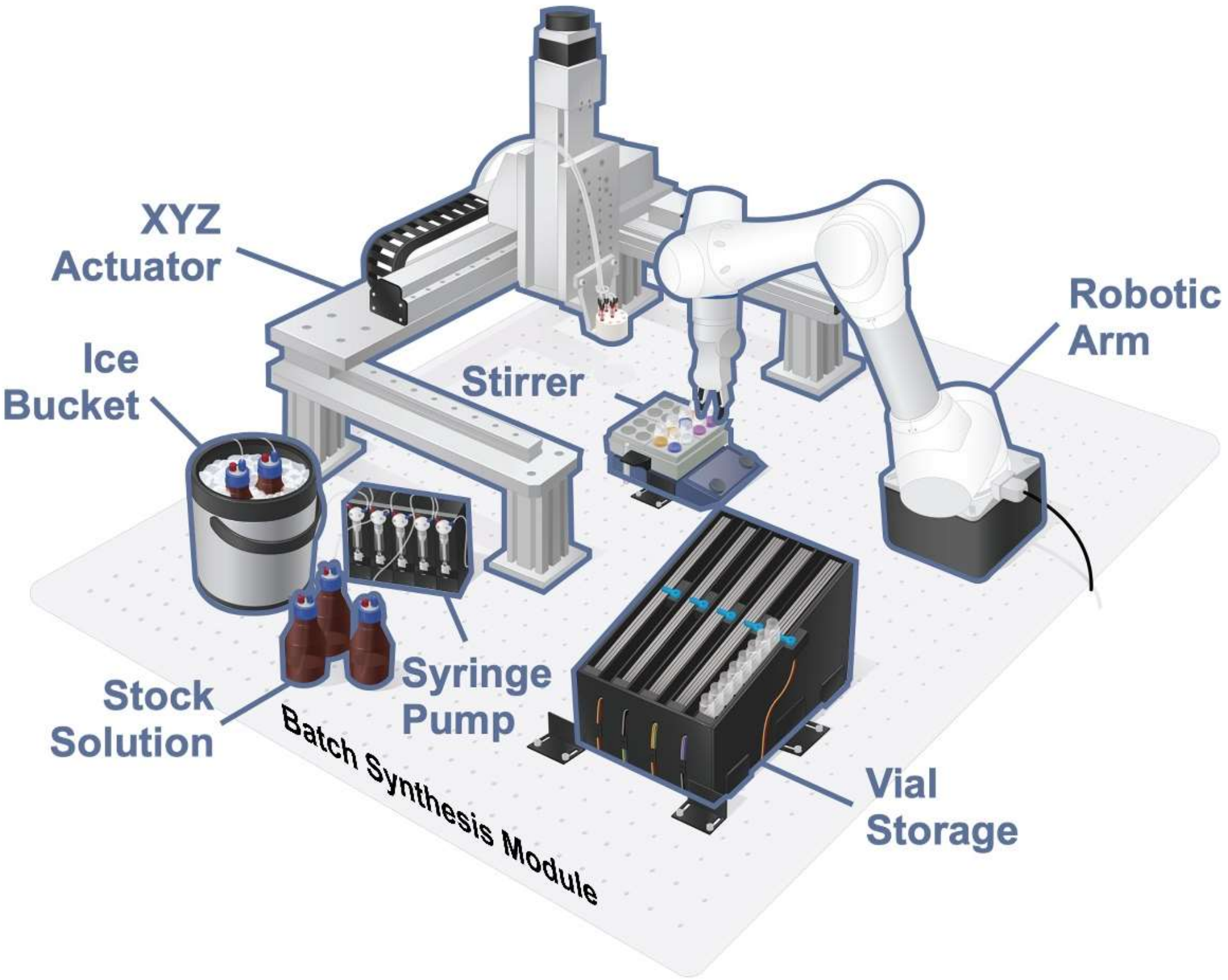

Schematic figure of the batch synthesis module. It consists of a robotic arm, a vial storage, a stirrer (hot plate), a solution dispensing system including stock solutions, a syringe pump, an ice bucket, and an XYZ actuator. This modularization platform is executed only for the synthesis process.

**Supplementary Figure S2a. Vial storage**

**Supplementary Figure S2a-1. Details on the vial storage**

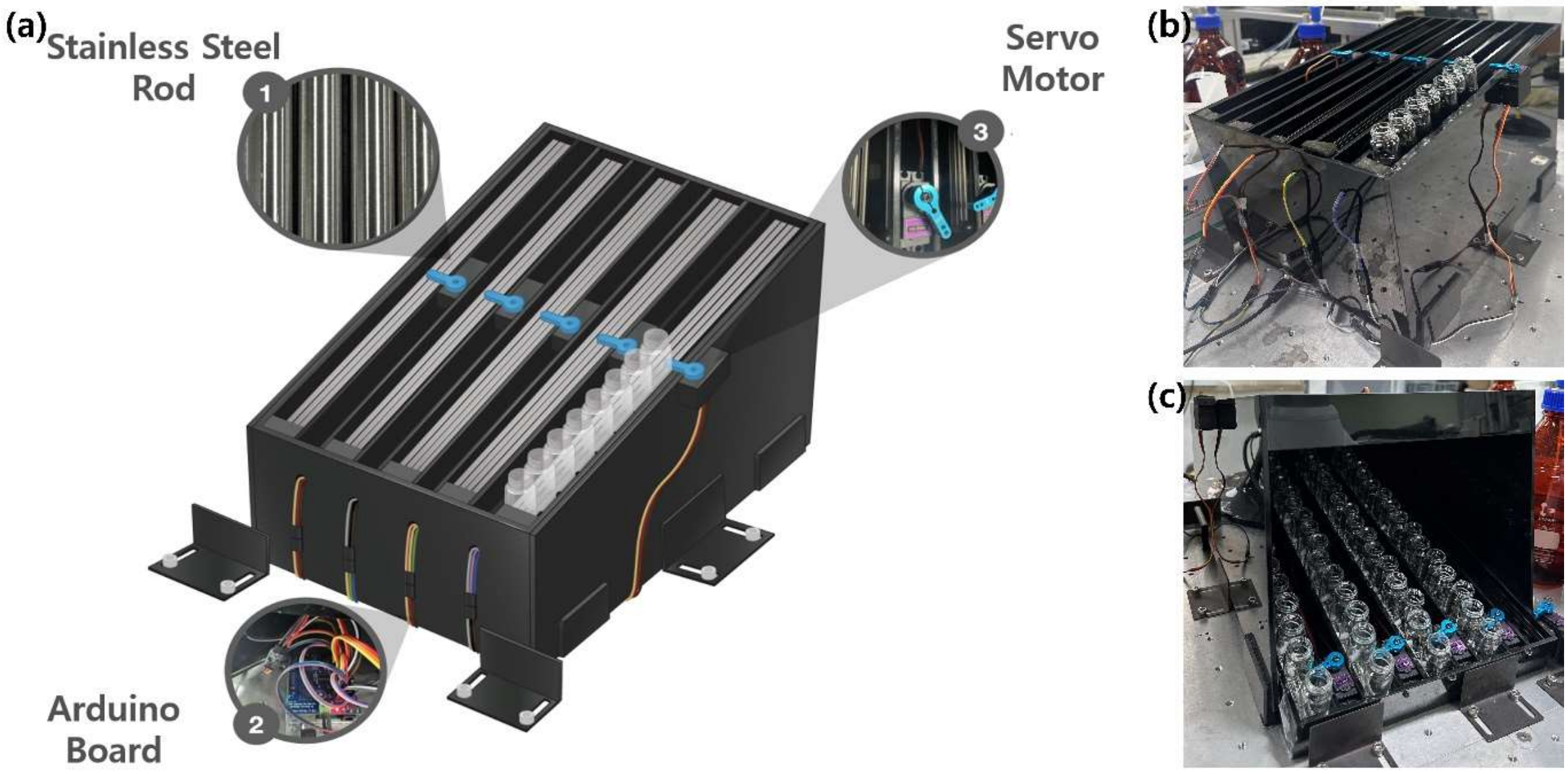

**(a)** Schematic figure of the vial storage. It consists of stainless-steel rods, Arduino boards and servo motors. The stainless-steel rod helps to smoothly move vials by reducing friction. The Arduino Uno board controls the servo motors. The servo motor (TowerPro: MG996PR) plays a role in the entrance of the vial storage. The case of vial storage is made of acrylic plates.

**(b)** The upper part is the space for safely storing vials after completion of the NP synthesis and UV-Vis characterization experiments.

**(c)** The bottom part is the space for storing empty vials before the experiments. Before a robotic arm moves vials into the vial storage, the servo motor will open and move vials behind like an entrance.

**Supplementary Figure S2a-2. Product drawing of vial storage**

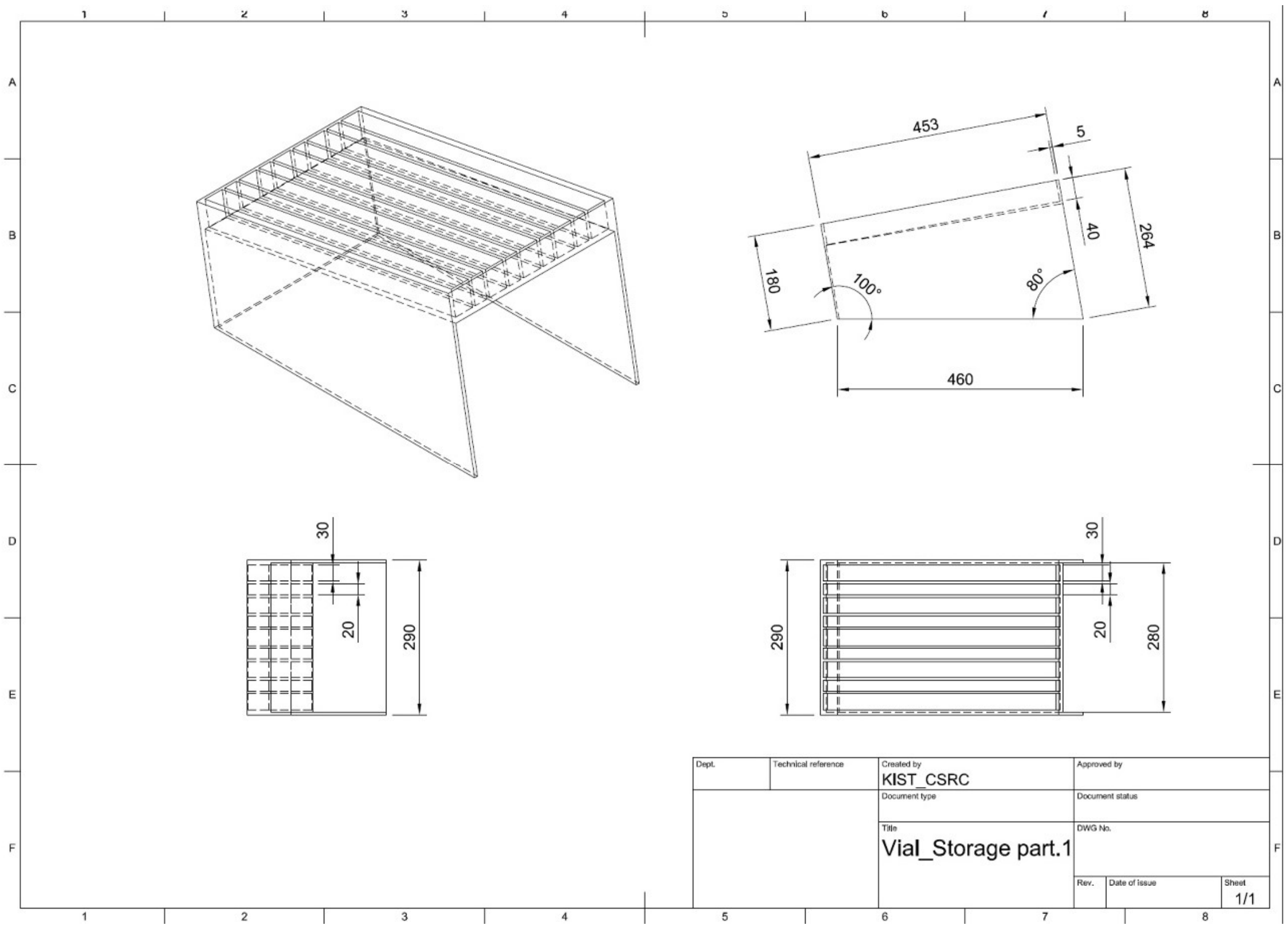

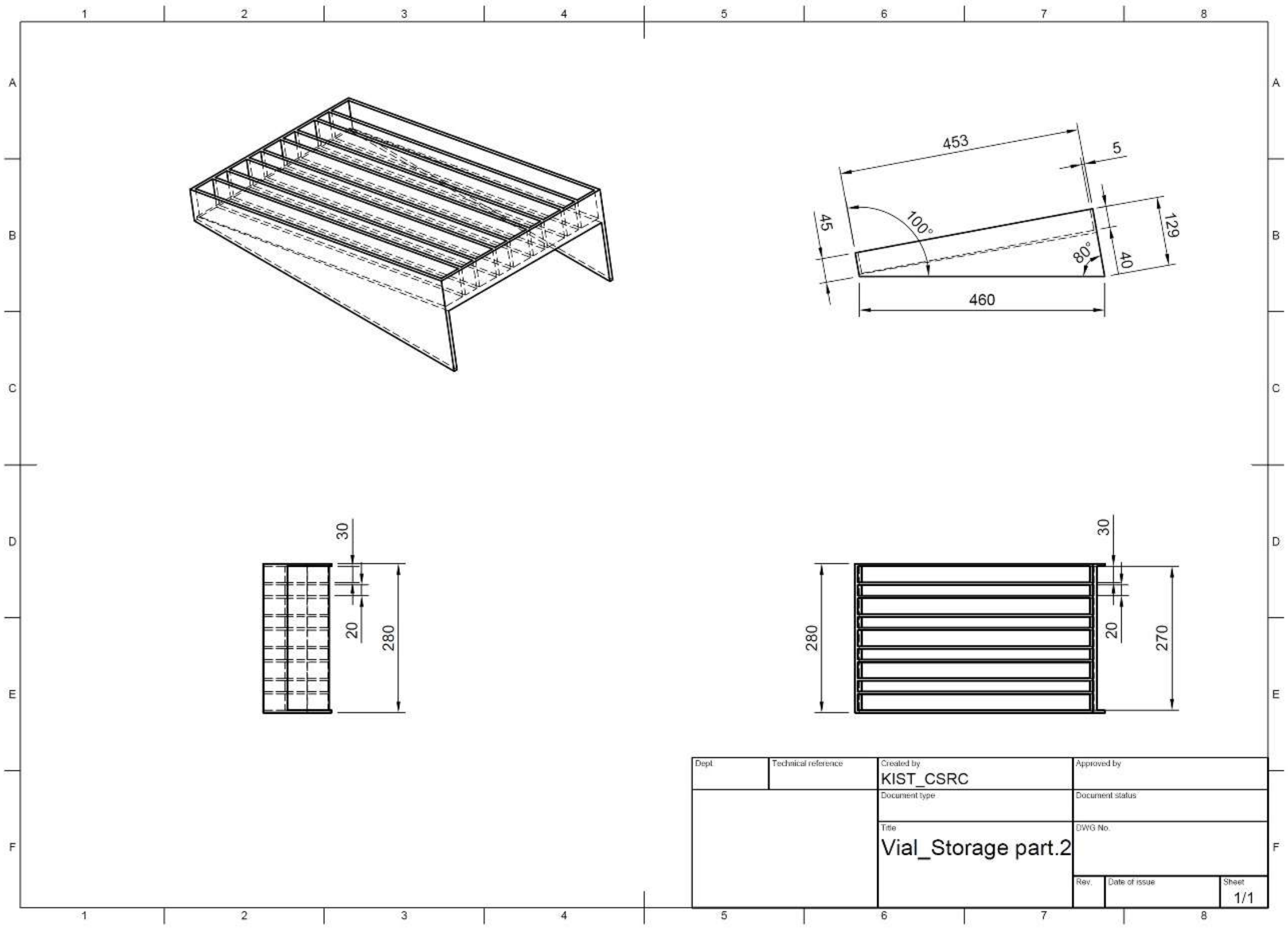

## Supplementary Figure S2b. Stirrer

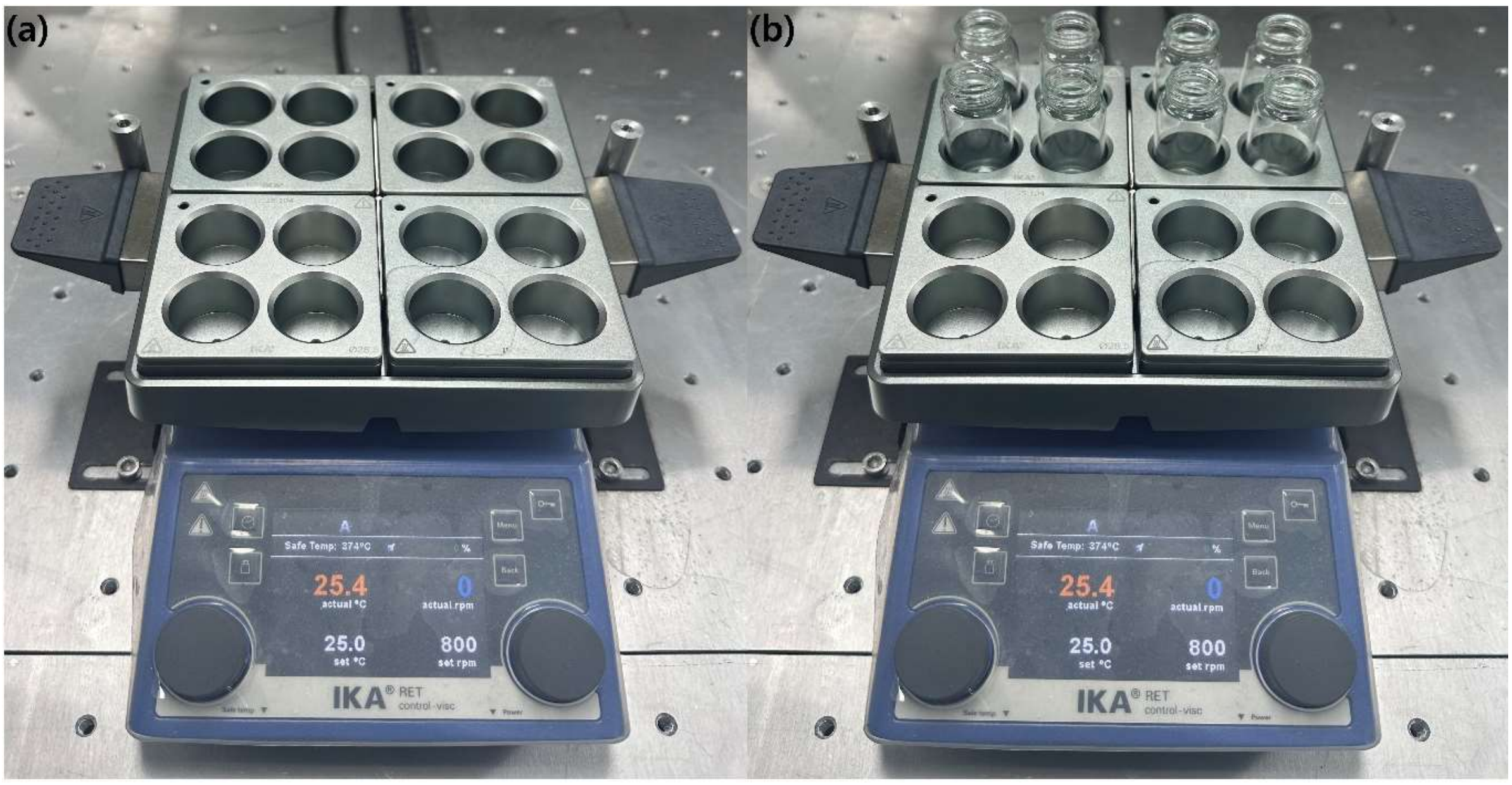

Stirrer (IKA: RET control-visc) supports RS232 communication. Operation range of temperature: 0~340 °C, Operation range of rpm: 50~1700 rpm.

**(a)** Picture of a stirrer with stirrer holders

**(b)** Picture of a stirrer with some vials on the stirrer holders

**Supplementary Figure S2c. Solution dispensing system**

**Supplementary Figure S2c-1. Details on the stock solution, syringe pump, and ice bucket**

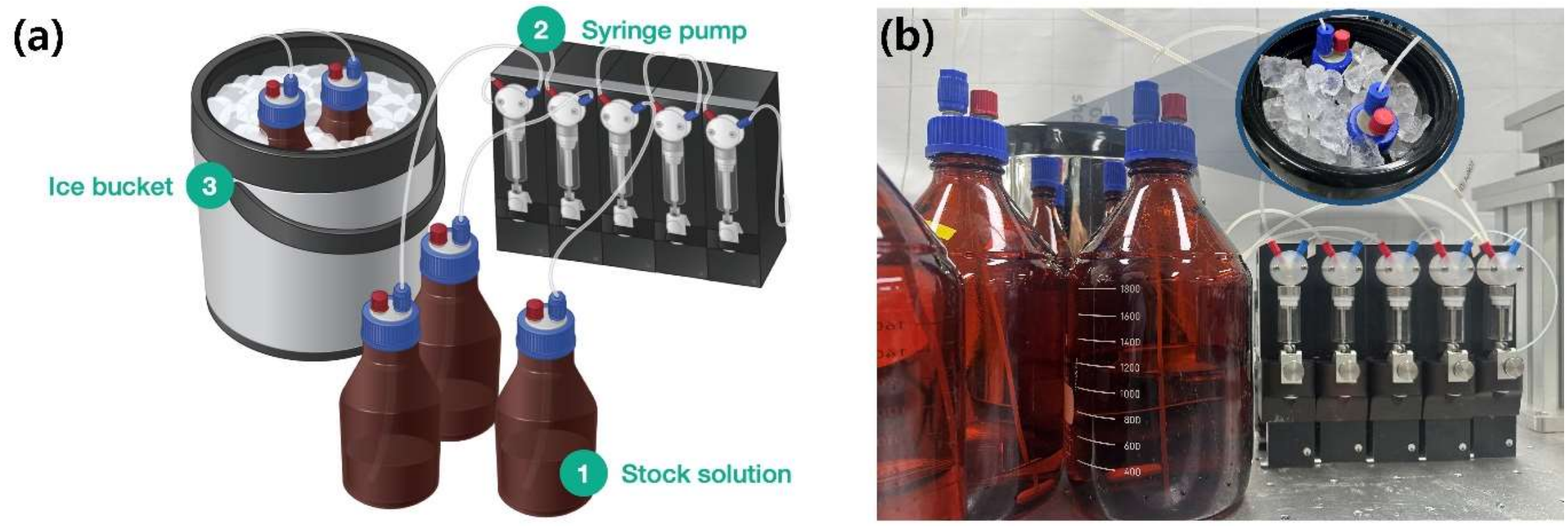

**(a)** Schematic figure of a stock solution, a syringe pump and an ice bucket. The stock solution (DaiHan Scientific) was used to prevent reagent contamination from sunlight. The syringe pump (TECAN: Cavro Centris) consisted of 5 mL syringes, 3-port valves, PTFE tubing, Omni-Lok Nuts and Cones. Each tube, nut and cone have 1/8 and 1/16 OD to use the inlet and outlet separately. The syringe pump can precisely control the volume of the solution in 181,490 increments.

**(b)** Picture of the stock solution, syringe pump and ice bucket.

## Supplementary Figure S2c-2. Details on the dispenser

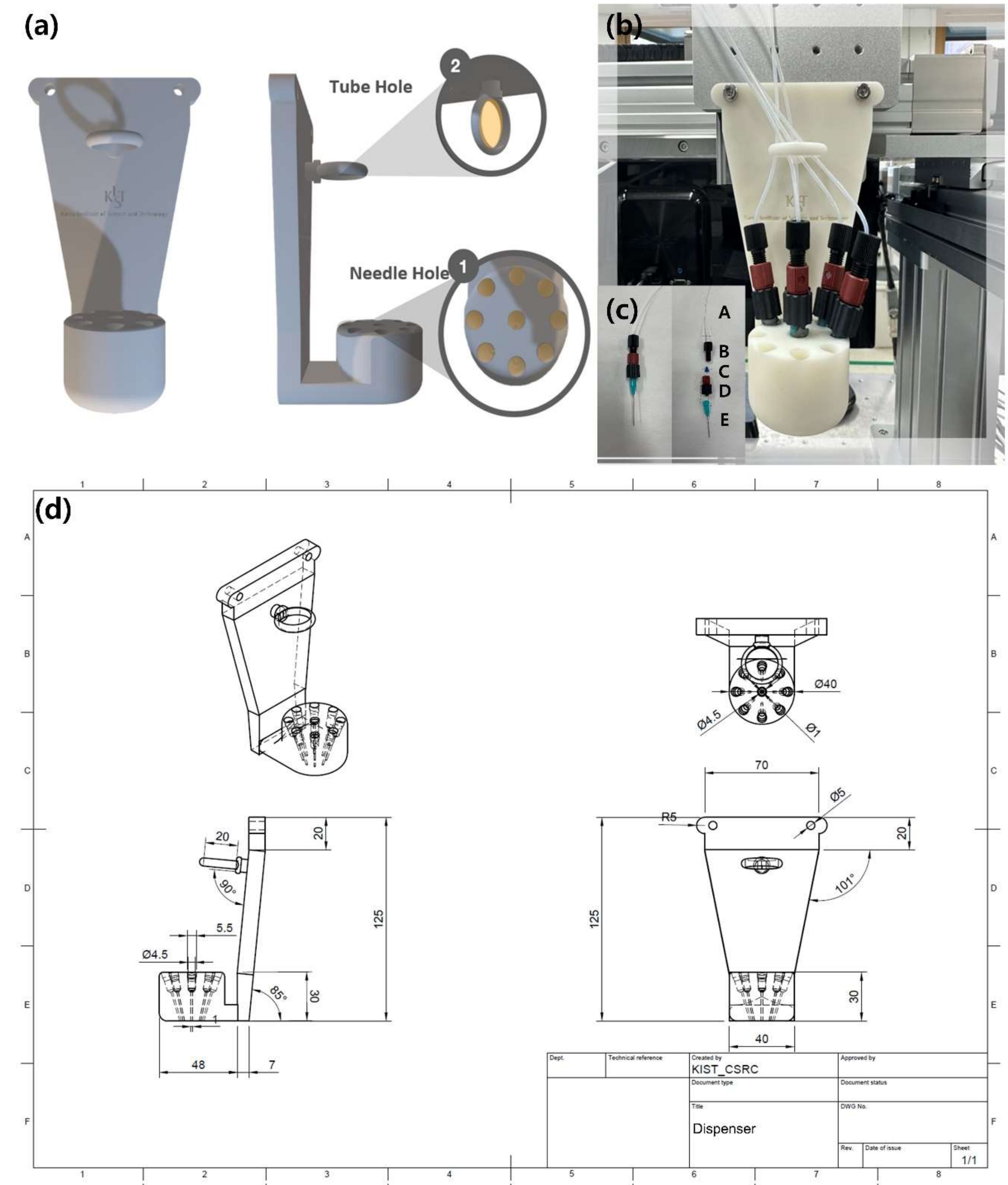

**(a)** Schematic figure for the dispenser with a tube hole and 9 needle holes. All PTFE tubes were placed together into a tube hole and all needles were placed into each needle hole.

**(b)** Pictures of the dispenser and needles. The dispenser was fabricated by a Cubicon 3D printer.

**(c)** Integration of the needle components with the dispenser.

**(d)** Product drawing of the dispenser.

## Supplementary Figure S2c-3. Details on the XYZ actuator

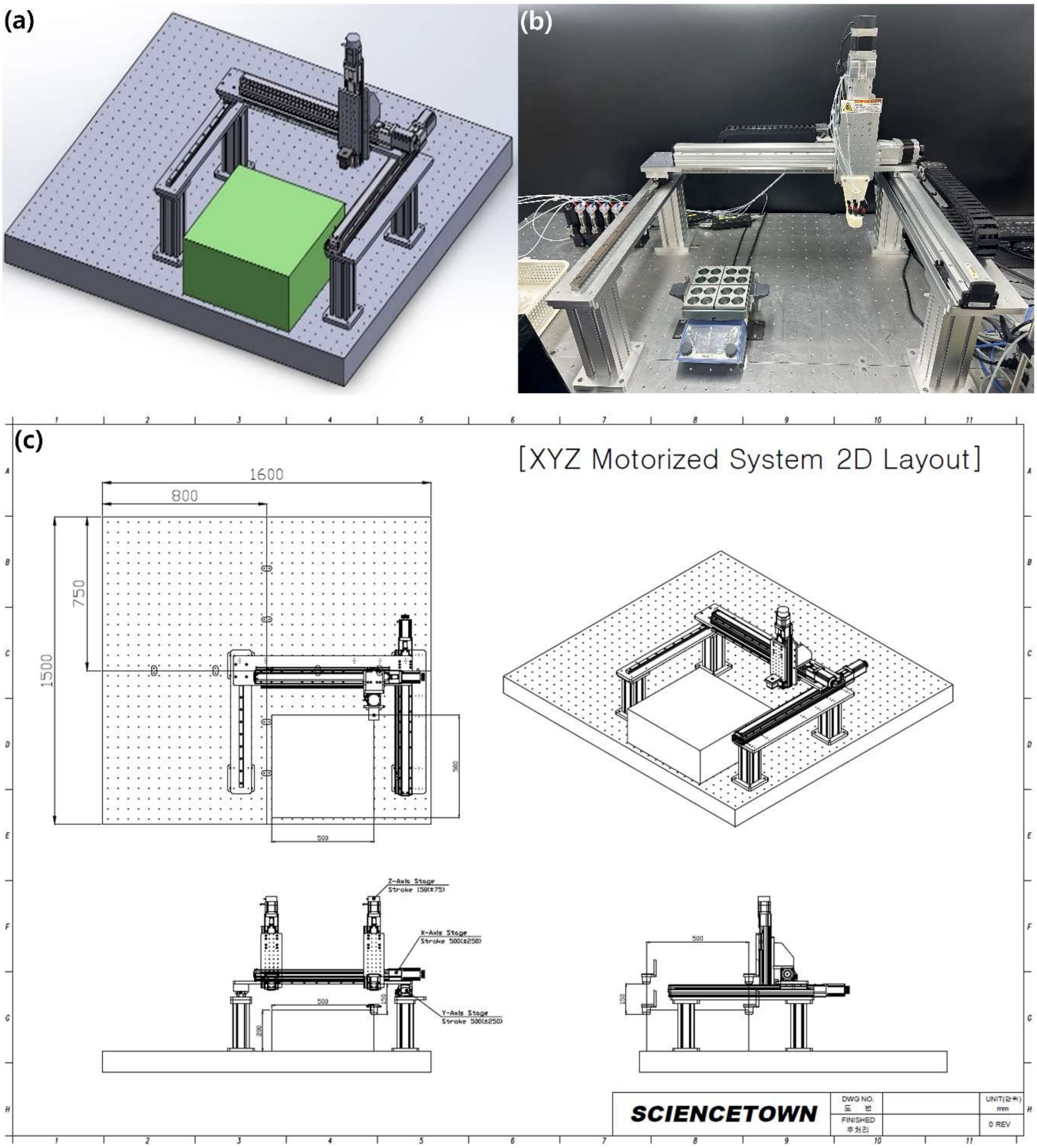

**(a)** Schematic figure of the XYZ actuator.

**(b)** Picture of the XYZ actuator combined with the dispenser.

**(c)** Product drawing of the XYZ actuator.

**Supplementary Figure S3. UV-Vis spectroscopy module**

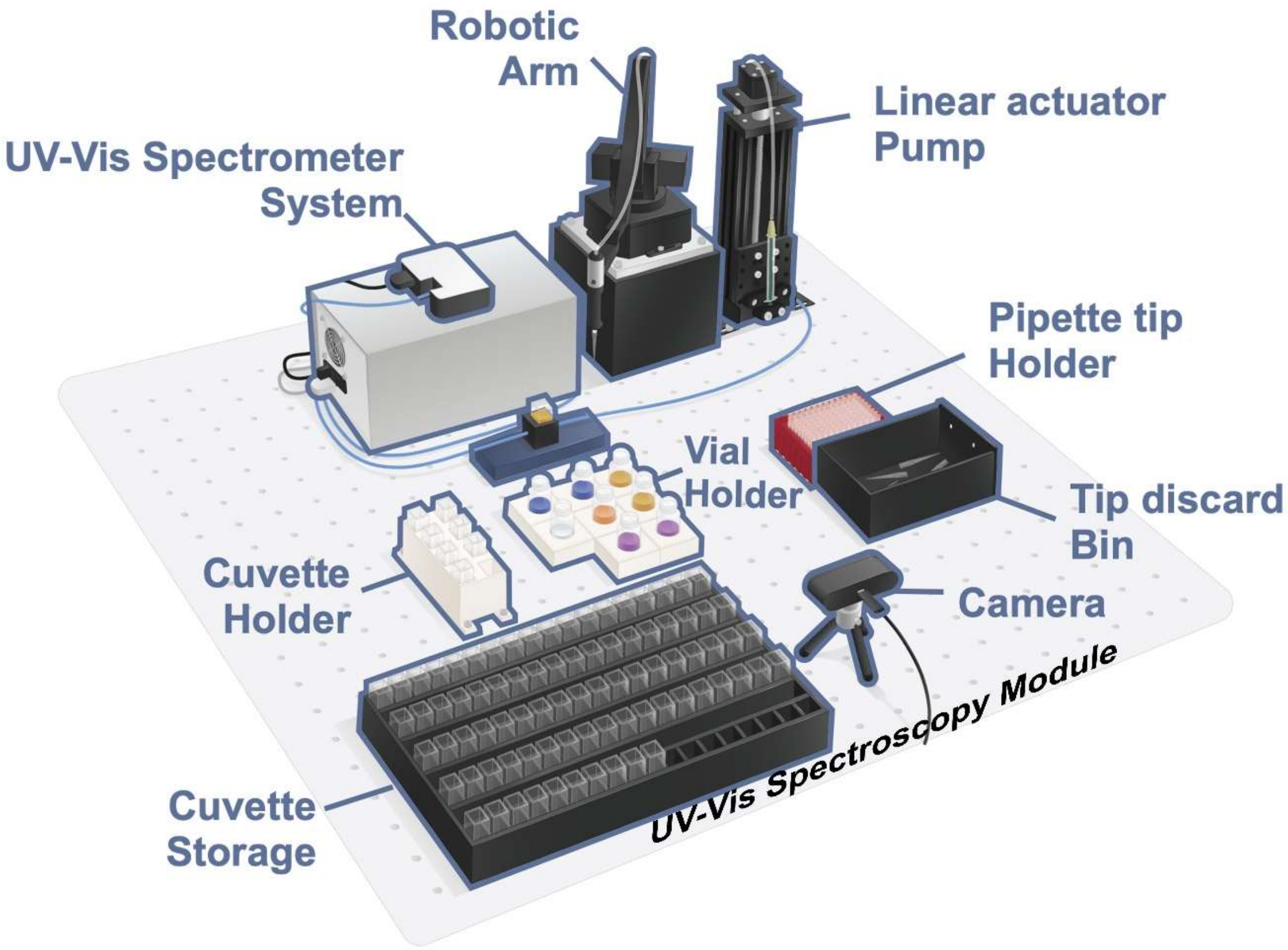

Schematic figure of the UV-Vis spectroscopy module. It consists of a cuvette storage, a cuvette holder, a vial holder, a UV-Vis spectrometer system and a pipetting machine. This modularization module is operated to evaluate the optical properties of nanoparticles.

**Supplementary Figure S3a. UV-Vis spectrometer system**

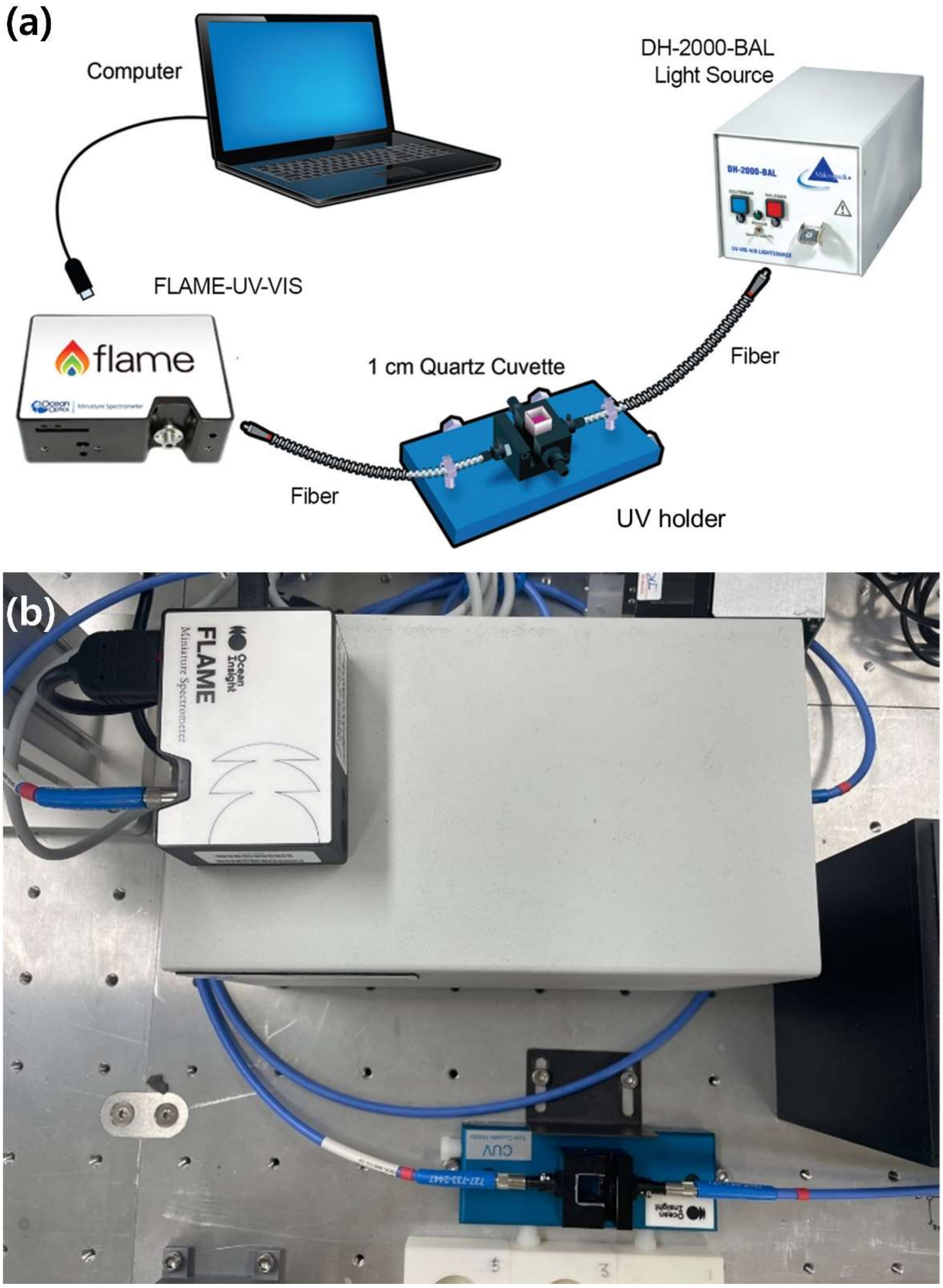

**(a)** Schematic figure of the UV-Vis spectrometer system. It consists of a light source (Ocean Optics UK, 200–2500 nm output, DH-2000-BAL included Deuterium-Tungsten Halogen lamps), a UV-Vis spectrometer (Ocean Optics UK, 200–879 nm output, Flame-UV-VIS or USB2000+), and a UV-Vis holder that contains cuvettes. They are connected by optical fibers.

**(b)** Picture of the UV-Vis spectrometer system.

**Supplementary Figure S3b. Pipetting machine with a robotic arm and a linear actuator**

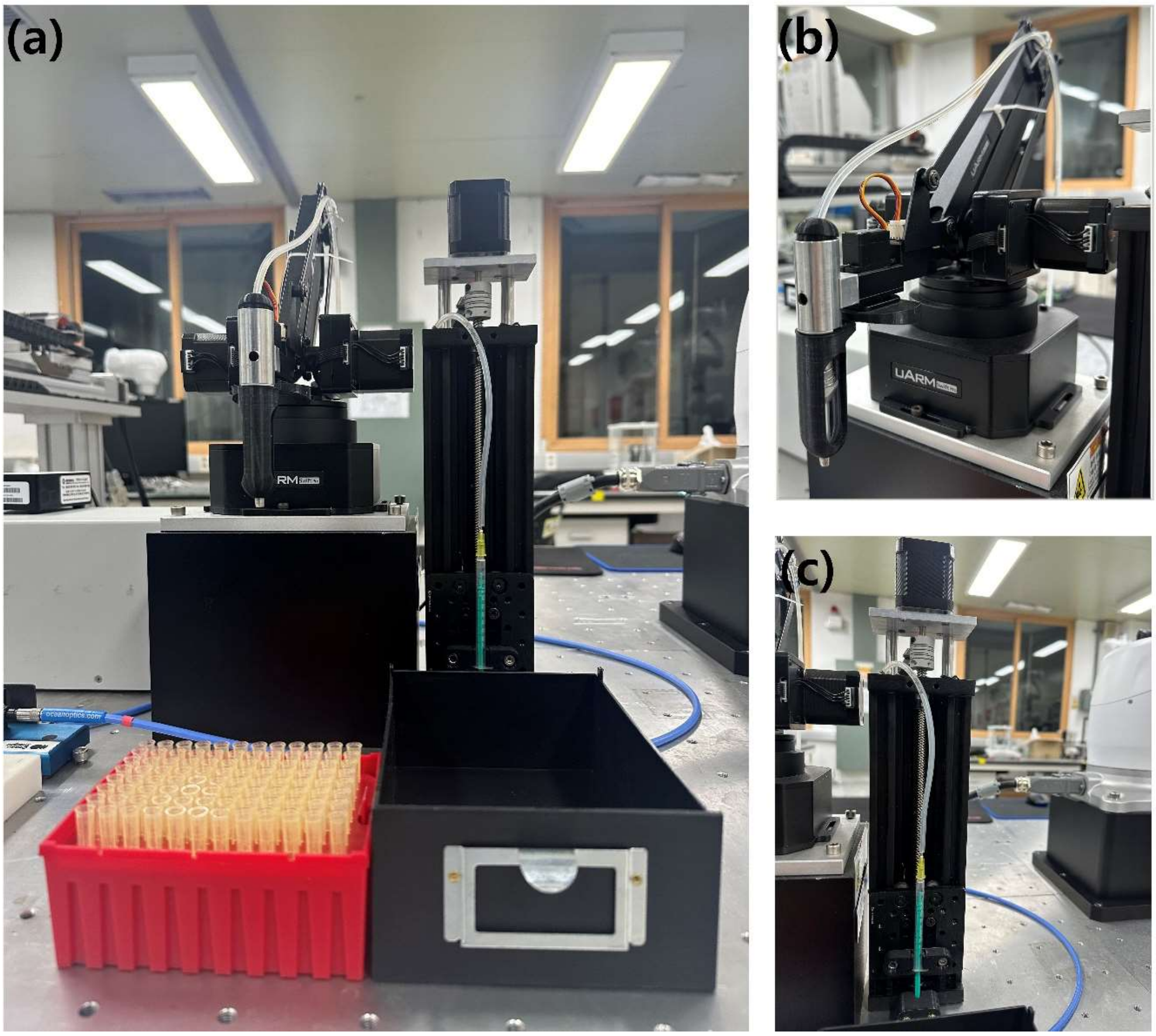

The pipetting machine was followed by OpenLH[1].

**(a)** Picture of the pipetting machine.

**(b)** A robotic arm (UFACTORY: uArm Swift Pro) is combined with the pipette dispenser. The robotic arm moves the pipette dispenser to obtain pipette tips, uses them to discharge the solution in the vials in which NPs have been synthesized, into cuvettes, and finally discards them into a bin.

**(c)** Linear actuator pump (NEMA 17 stepper motor, C-beam linear actuator bundle and shield). The linear actuator pump injects the colloidal nanoparticle solution in vials and discharges it into cuvettes by the pipette dispenser.

## Supplementary Figure S3c. Vision system for pipette tip detection via DenseSSD

**Scheme**

To ensure optical properties, we implemented object detection to prevent any failures in tip attachment in our autonomous platform. The workflow for the detection of such failures is illustrated in Supplementary Figure S3c-1. A camera provides a bird's-eye view to monitor the entire chemical synthesis process using Logitech C920. Once the vision system detects that the robotic arm has attached the tip correctly, the characterization steps proceed. However, if the system detects any issue regarding the tip attachment, it automatically repeats the step to reattach the tip correctly, ensuring that extraction of the optical properties is complete. This implementation of object detection represents safety and accuracy in our system, preventing any issues that could arise from incorrect tip attachment.

## Supplementary Figure S3c-1. Workflow of the vision system

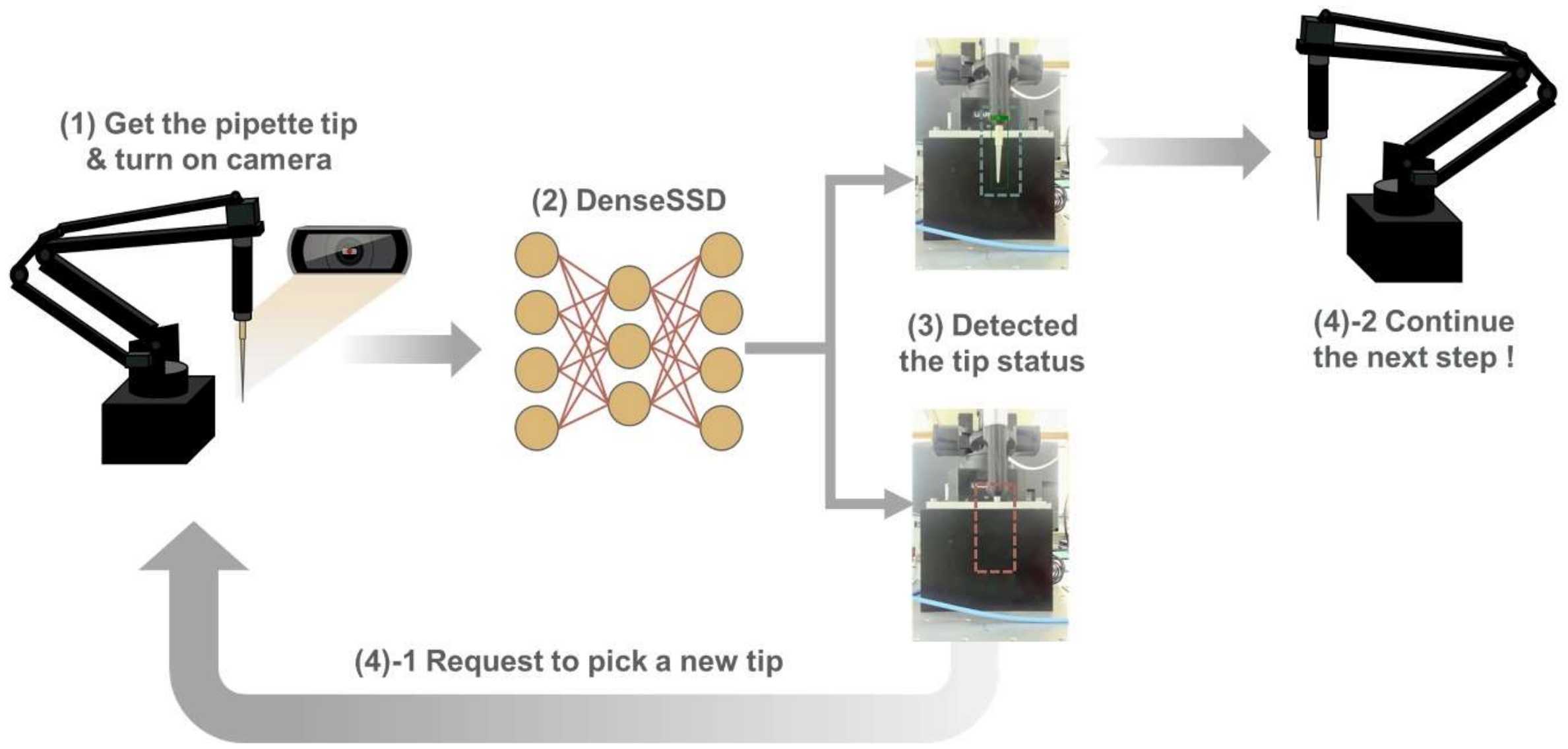

Schematic figure of pipette tip detection via DenseSSD[2].

(1) Move the robotic arm to obtain the pipette tip

(2) Input real images for tip detection

(3) Judge the status of the tip

(4)-1 If tip status is False, request to pick a new tip

(4)-2 If tip status is True, continue the next step

All processes evaluate the tip status for accurate extraction of the optical properties.

**Supplementary Figure S3c-2. Picture of the vision system**

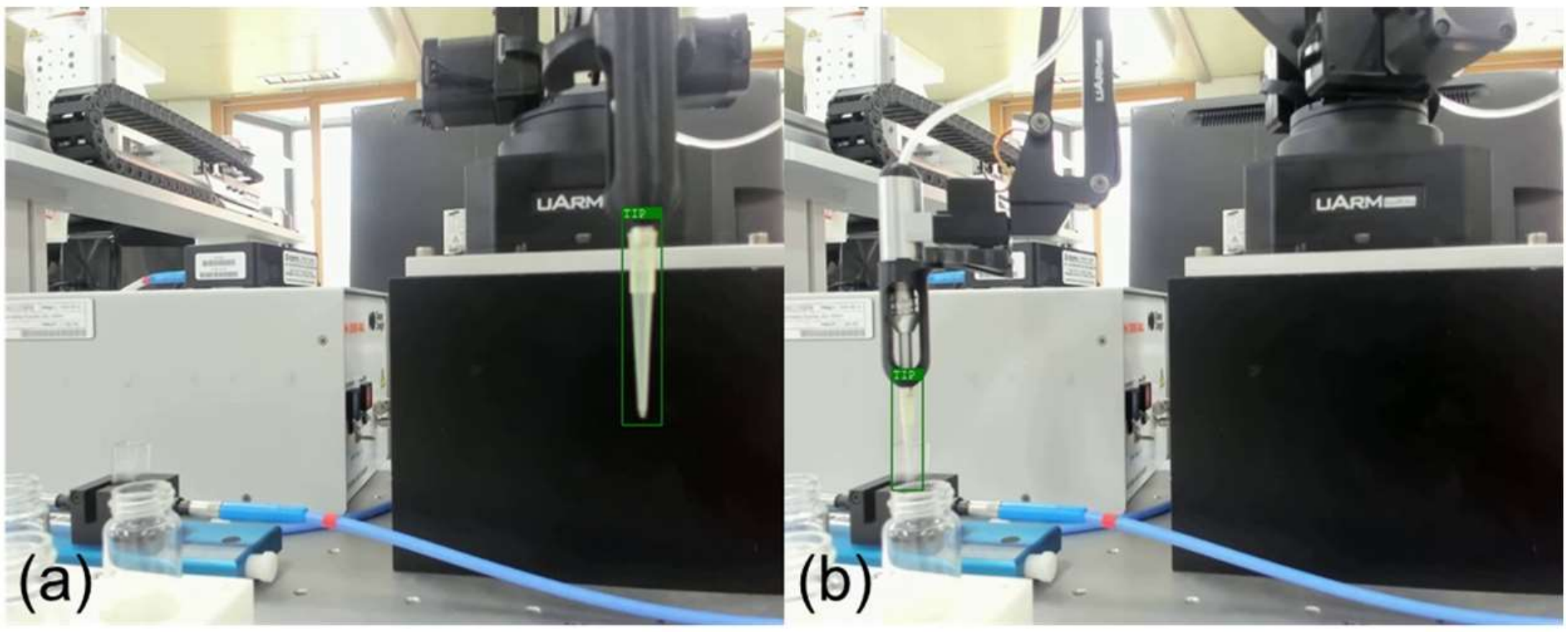

**(a)** Picture of the detection image before proceeding to the next step.

**(b)** Picture of the detection image while a pipette tip injects the solvent in a vial.

## Supplementary Figure S3d. Chemical vessel container

## Supplementary Figure S3d-1. Details on the chemical vessel container

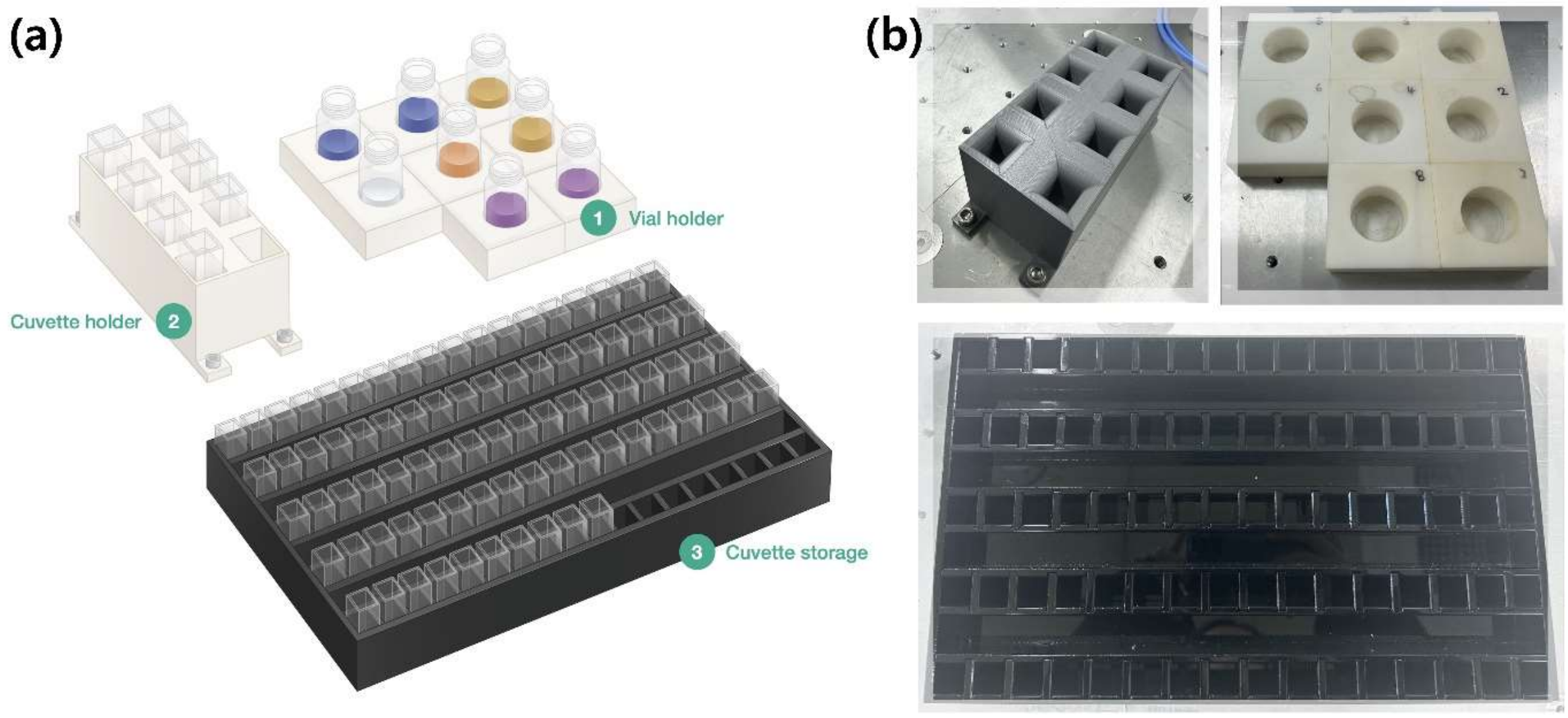

**(a)** Schematic figure of the chemical vessel container to position vials and cuvettes.

**(b)** Pictures of the chemical vessel container. All chemical vessel containers were fabricated by a Cubicon 3D printer.

**Supplementary Figure S3d-2. Product drawing of the vial holder**

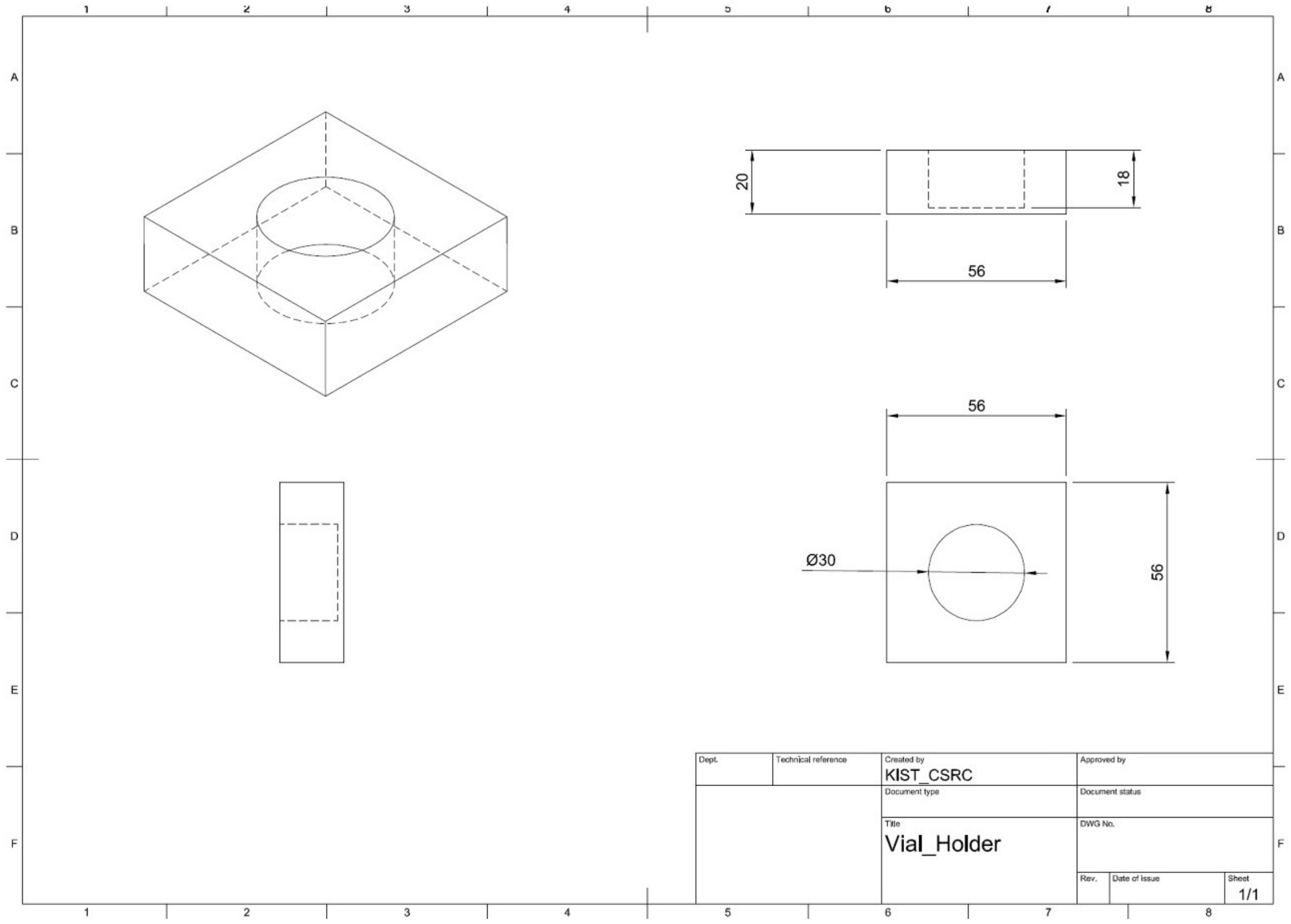

**Supplementary Figure S3d-3. Product drawing of the cuvette holder**

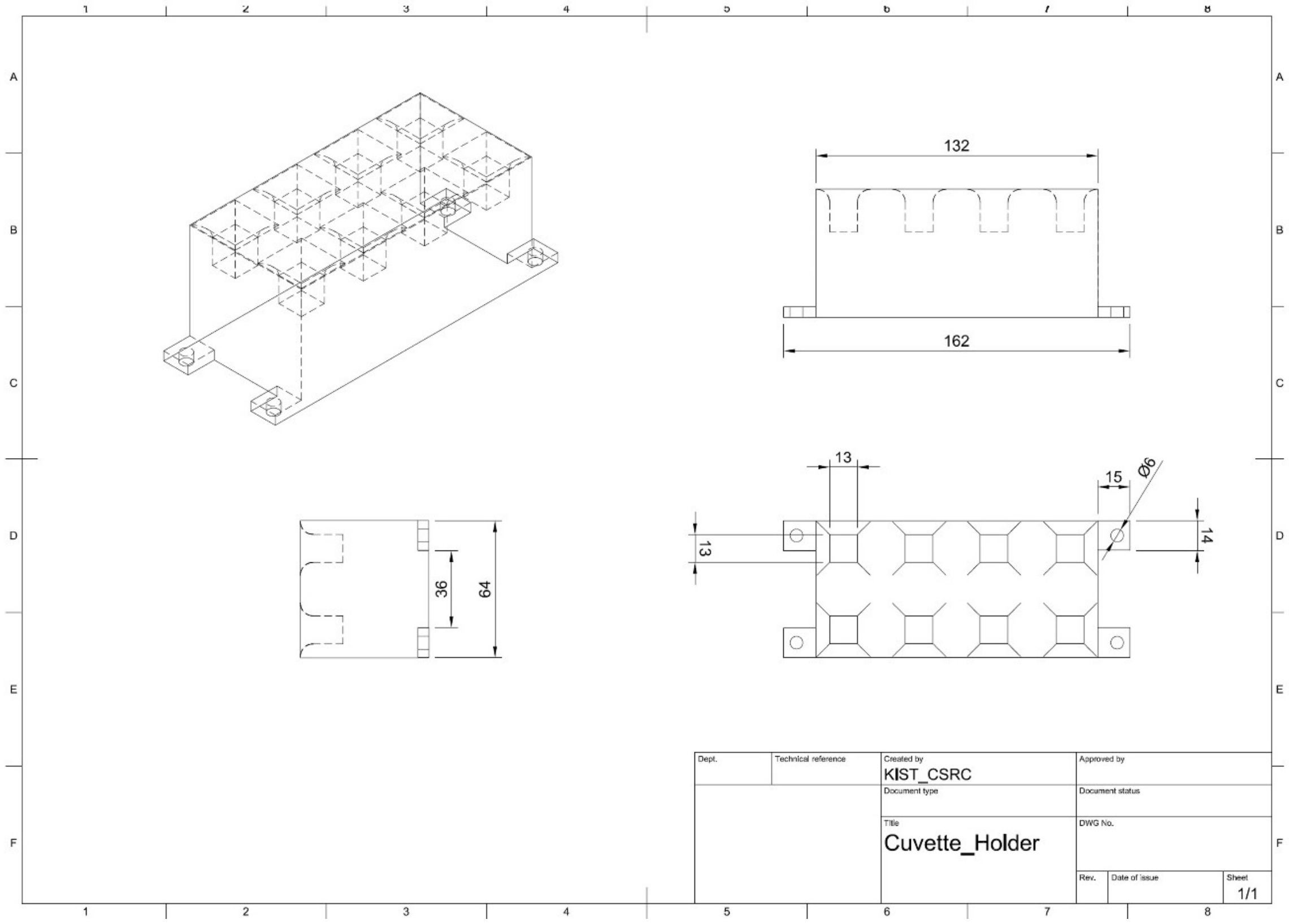

**Supplementary Figure S3d-4. Product drawing of the cuvette storage**

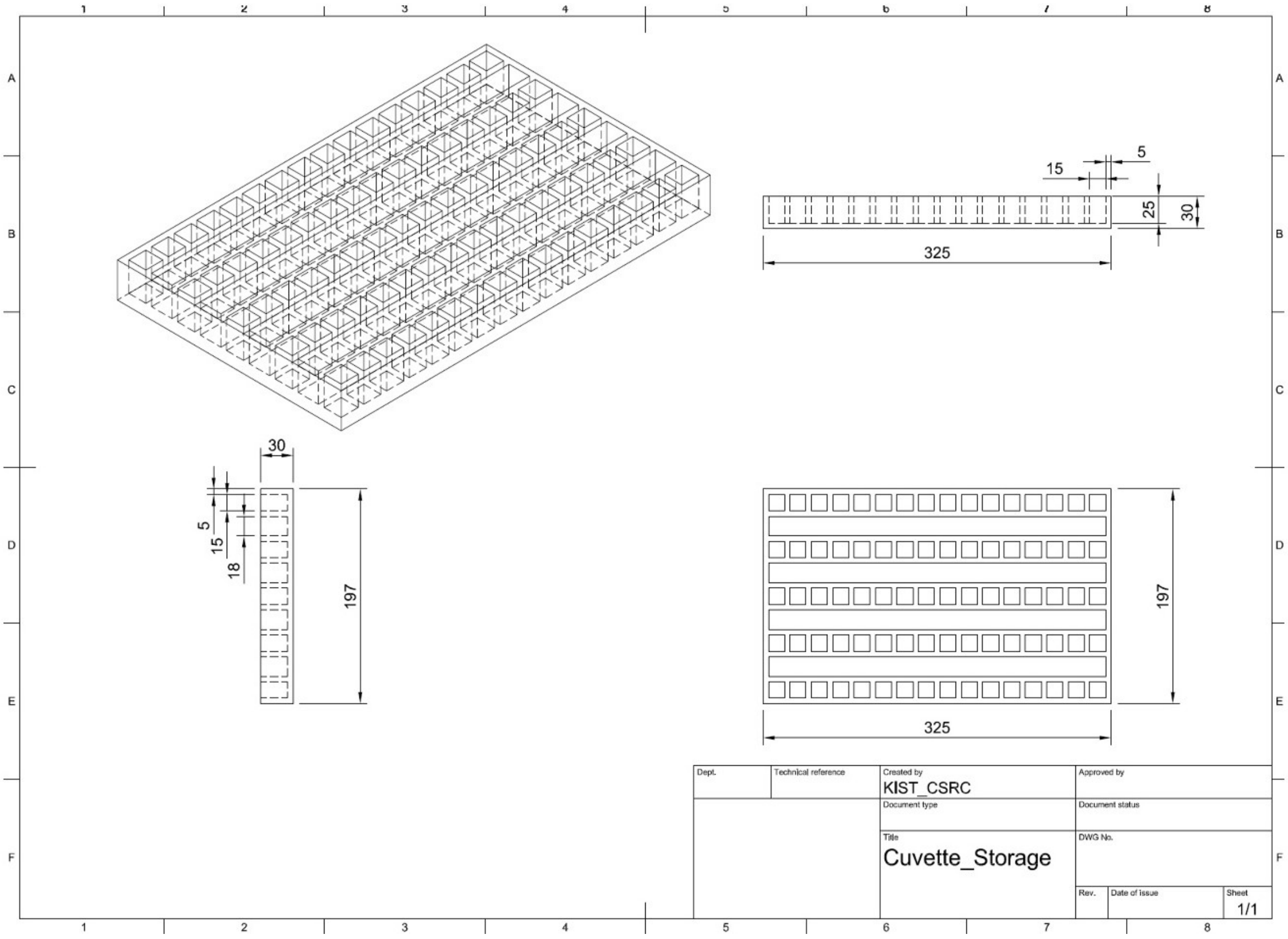

**Supplementary Figure S4. Reliability test of our autonomous laboratory**

**Supplementary Figure S4a. Precision test of syringe pumps**

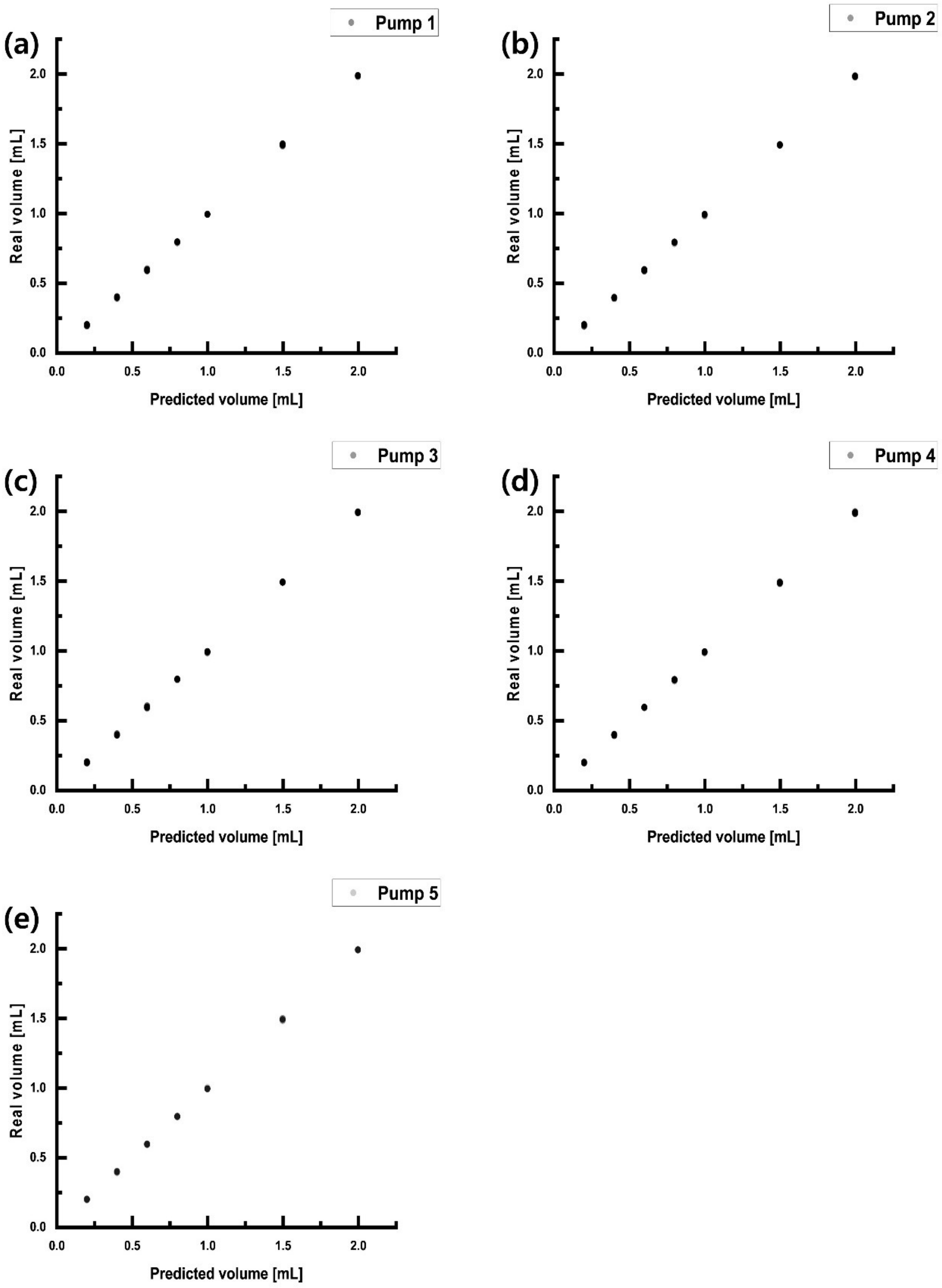

**(a)** $AgNO_3$, **(b)** citrate, **(c)** $H_2O$, **(d)** $H_2O_2$, **(e)** $NaBH_4$. All R-square values are close to 0.999.

**Supplementary Figure S4b. Precision test of the conventional pipette and solution dispensing system**

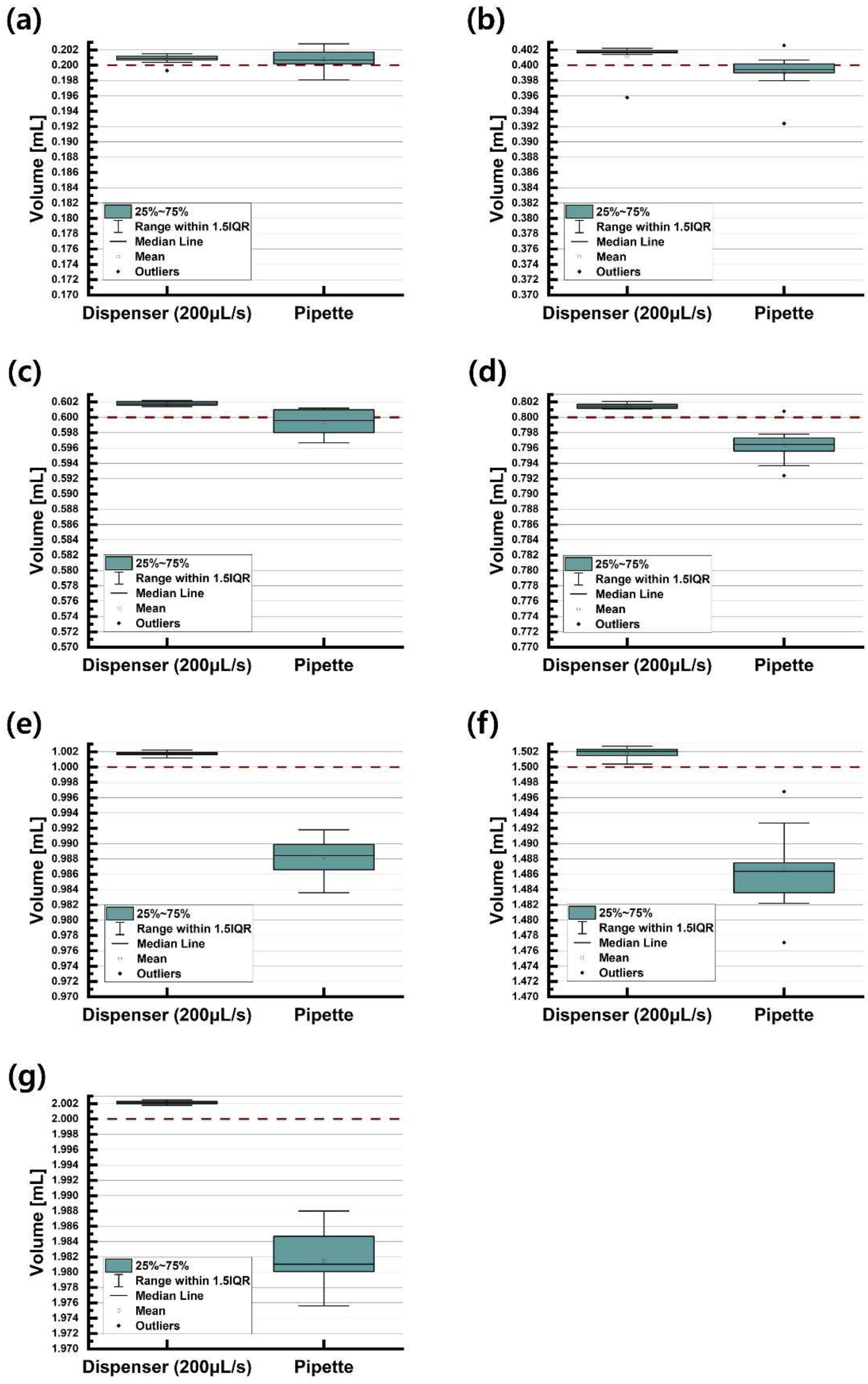

**(a)** Precision test of 0.2 mL, **(b)** 0.4 mL, **(c)** 0.6 mL, **(d)** 0.8 mL, **(e)** 1.0 mL, **(f)** 1.5 mL, and **(g)** 2.0 mL.

**Supplementary Figure S4c. Precision test of the UV-Vis spectrometer system**

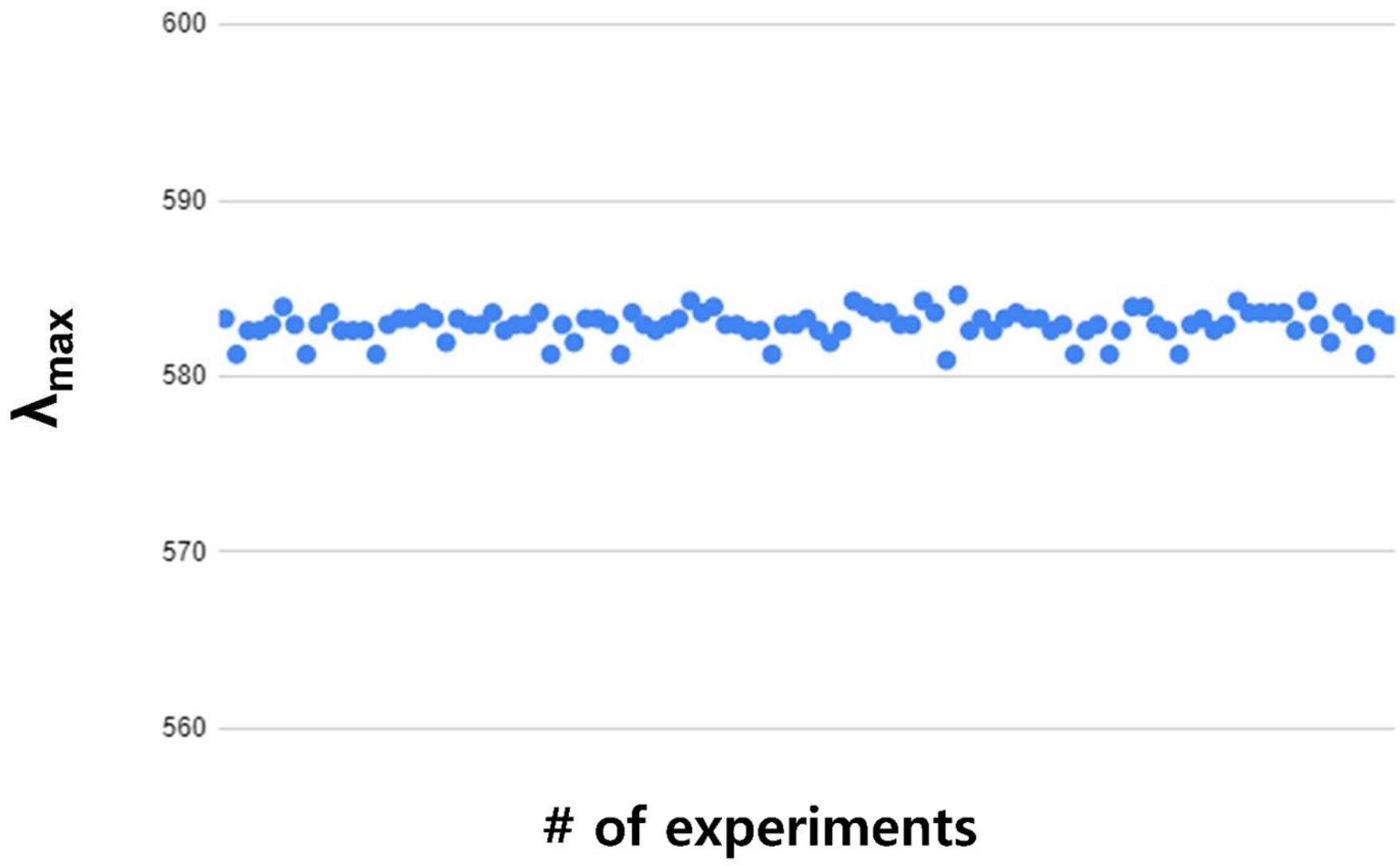

Mean of $\lambda_{max}$ = 582.9371nm, deviation of $\lambda_{max}$ = 4.35nm.

**Supplementary Figure S4d. Precision test of the linear actuator pump in the pipetting system**

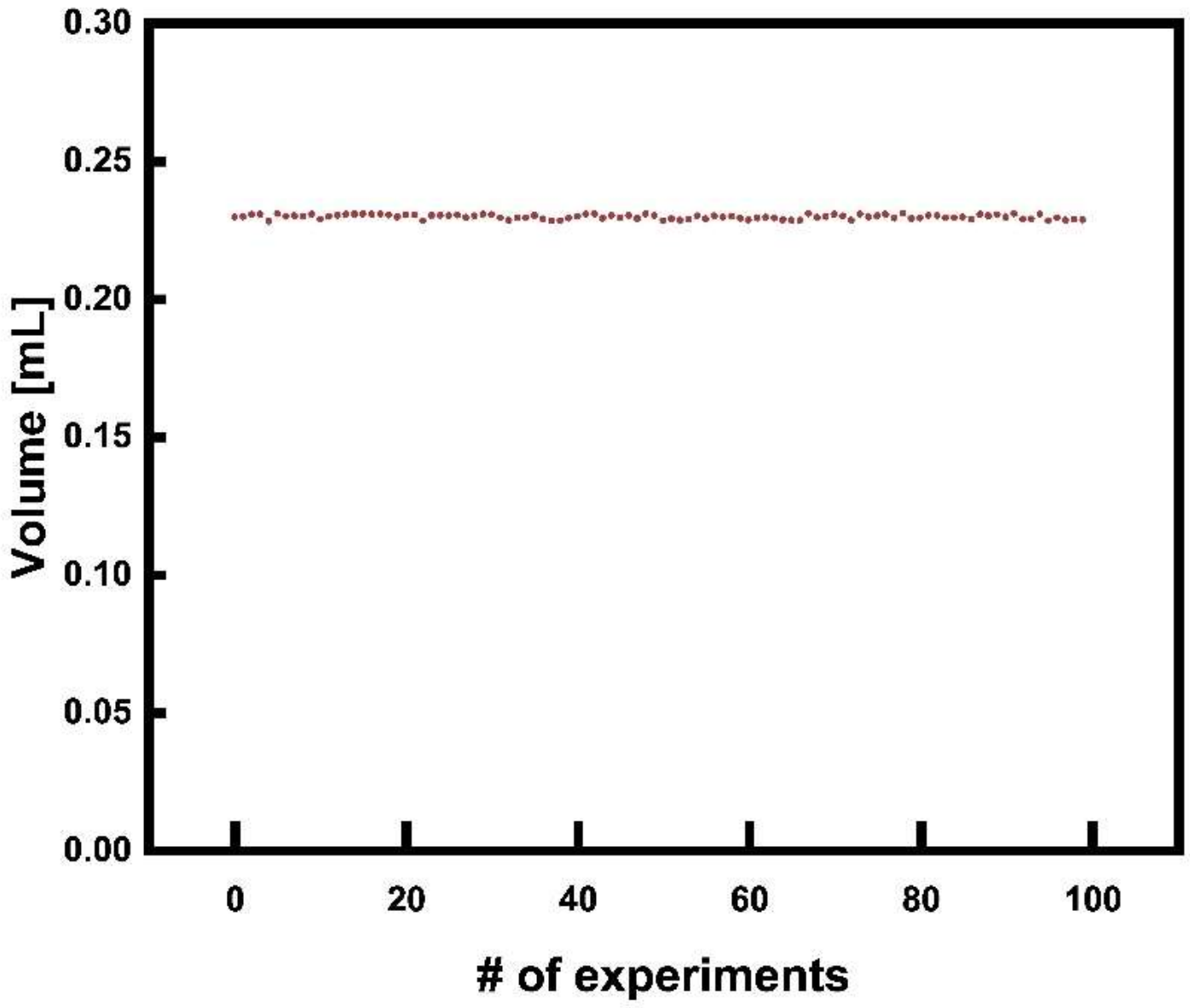

**Supplementary Figure S4e. Reproducibility test for our autonomous laboratory with the optical properties ($\lambda_{max}$, FWHM, and peak intensity)**

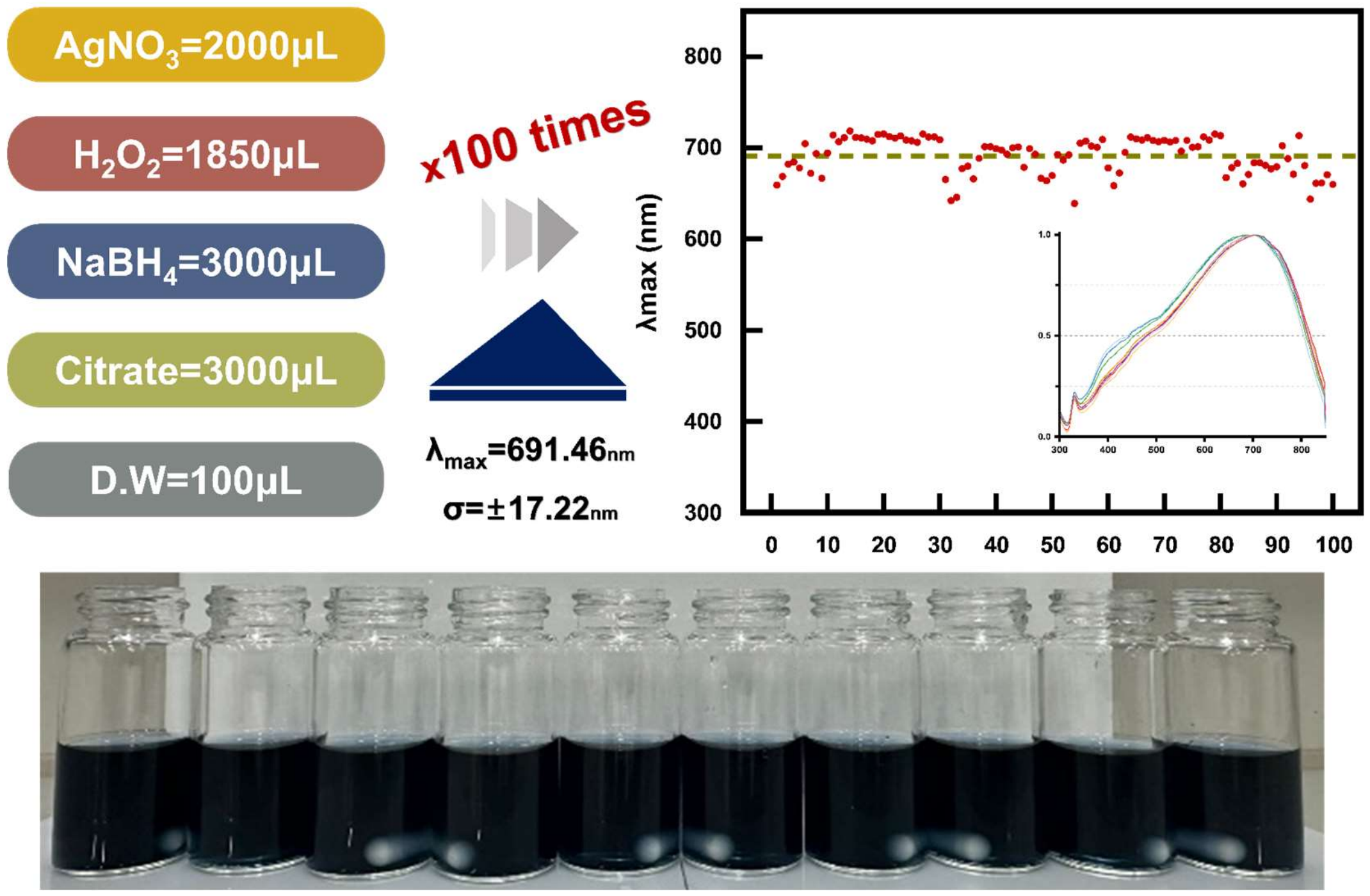

| Optical property | $\lambda_{max}$ (nm) | Intensity (a.u.) | FWHM (nm) |
|---|---|---|---|
| Average | 691.4615 | 0.38923 | 332.9369 |
| Deviation | ±17.21592 | ±0.047062 | ±39.55089 |

We repeated the experiments for Ag NP synthesis with identical synthesis conditions 100 times via our autonomous laboratory. The scatter plot readily indicates the reproducibility of our autonomous laboratory, in which a red circle corresponds to one experiment. The inset graph shows the overlapped absorption spectra for the 100 experiments. The table represents the detailed values of the reproducibility results.

## 2. Bayesian optimization with early stopping criterion

### Supplementary Figure S5. Extraction of absorption spectra in the literature

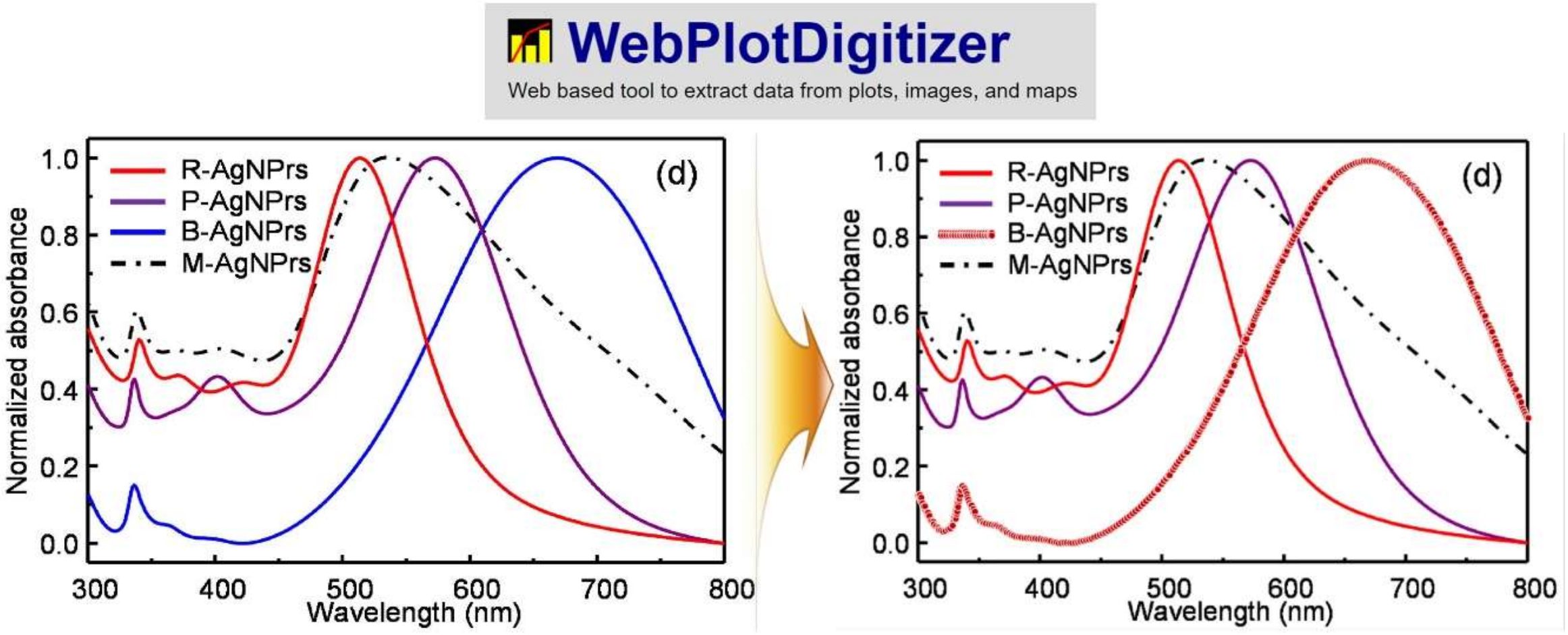

| | $\lambda_{max}$ (paper) | $\lambda_{max}$ (predicted) | FWHM (predicted) |
|---|---|---|---|
| Red | 513 | 513.5078329633614 | 96.33114981819165 |
| Purple | 573 | 573.5365991555644 | 134.41555788584895 |
| Blue | 667 | 669.9526072805022 | 191.72885826217623 |

We extracted absorption spectrum data from the literature using WebPlotDigitizer[3]. Table summarizes the comparison between values in the literature and values extracted by WebPlotDigitizer.

**Supplementary Figure S6. Scheme of the bespoke synthesis of Ag NPs**

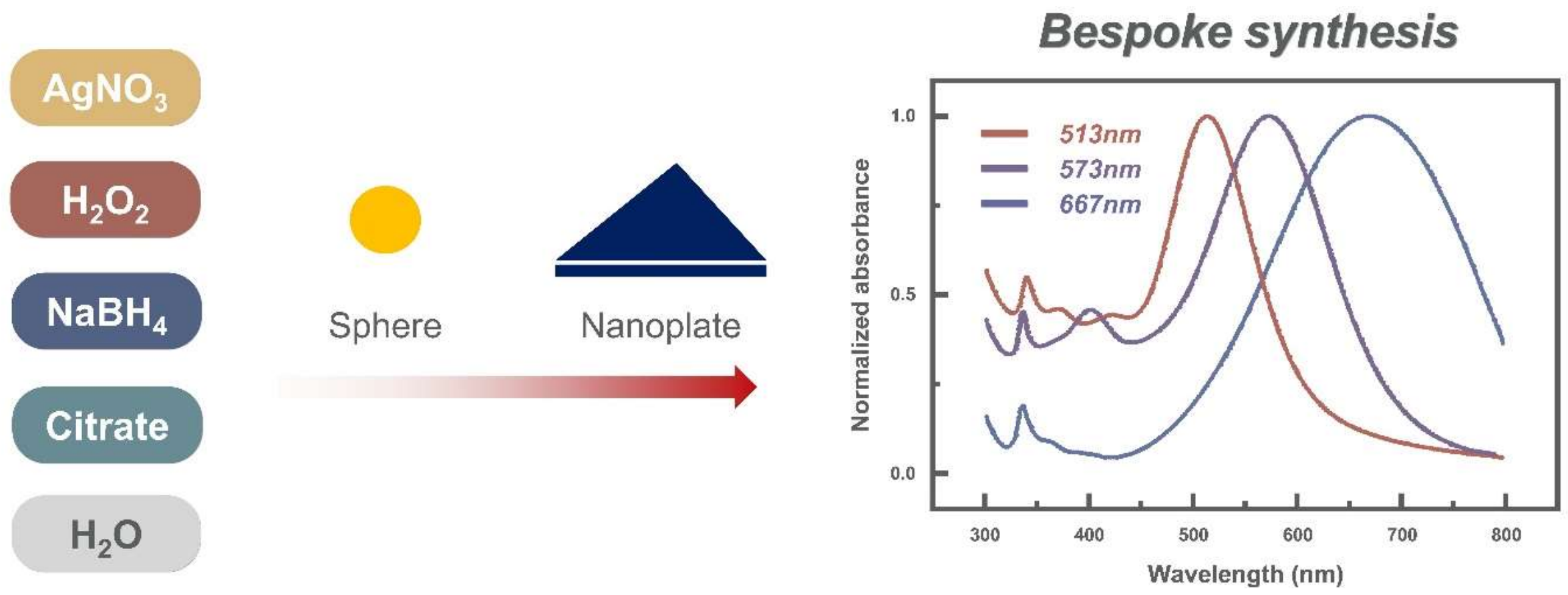

This figure shows the synthesis process of Ag NPs, in which a one-step synthesis method was considered with five experimental parameters: $AgNO_3$, $H_2O_2$, $NaBH_4$, citrate and $H_2O$. Since the one-step synthesis method does not need to consider contamination of Ag seeds, this process provides high convenience in terms of simplicity in an autonomous laboratory.

### Supplementary Figure S7. Definition of fitness function

$$\textit{Fitness function} = -0.9 * \frac{\lambda_{max} - \lambda_{max,\ target}}{A_{\lambda_{max}}}$$

$$-0.07 * (1 - intensity) - 0.03 * \frac{FWHM}{A_{FWHM}}$$

The fitness function reflects $\lambda_{max}$, FWHM and peak intensity for the absorption spectra of nanoparticles. This implies multi-objective optimization for the bespoke synthesis of NPs, although the highest weight was assigned to the $\lambda_{max}$ term because the $\lambda_{max}$ property has the highest priority for organic solar cell applications. The weights of the fitness function are regarded as hyperparameters in the experimental design. We considered the range of absorption spectra from 300 nm to 850 nm depending on the UV-Vis spectrometer. $A_{\lambda max}$ and $A_{FWHM}$ are the scaling factors used to normalize the fitness value from 0 to -1. $A_{\lambda max}$ is 337 (=|513-850|), 277 (=|573-850|), and 367 nm (=|667-300|) for the target $\lambda_{max}$. The $A_{FWHM}$ is 550 nm (=|300-850|), corresponding to the range of the absorption spectra.

### Supplementary Table S1. Criterion of filter values for early stopping

| Converted fitness | 513 nm | 573 nm | 667 nm |
|---|---|---|---|
| $\lambda_{max}$ | -0.1026 | -0.1249 | -0.09423 |
| FWHM | -0.0022 | -0.0022 | -0.0022 |
| Intensity | -0.0033 | -0.0033 | -0.0033 |
| **Filter criterion** | **-0.1081** | **-0.1304** | **-0.09973** |

This table shows the criteria of the filter values which are based on the precision test. We utilize the deviation of the precision test for the UV-Vis spectrometer system in Supplementary Figure S4c (±4.35 nm) and the deviation value of the reproducibility test in Supplementary Figure S4e (±17.21592 nm, ±0.047062 nm, ±39.55089 nm) and then substitute them into the fitness function to generate the converted fitness.

**Supplementary Figure S8. Pseudo code for implementing the early stopping criterion in the Bayesian optimizer**

**Algorithm 1** Bayesian optimization with early stopping

**Require:**
$i = 1$: ▷ $i$: iteration number
$y_i \in Y$ ▷ $y_i$: $i$th element of fitness value set, $Y$: set of fitness value
$y_{max} = -1$ ▷ $y_{max}$: the best of fitness value
$Count = 0$ ▷ $Count$: the number of proper fitness in earlystopping
$filter = -0.1$ ▷ $filter$: filter fitness value
$Patience = 5$ ▷ $Patience$: Number of epochs with no improvement

**Ensure:**
**while** $Count \geq Patience$ **do**
  $Y \leftarrow y_i$
  **if** $y_i \geq filter$ **then**
    **if** $y_i \geq y_{max}$ **then**
      $y_{max} \leftarrow y_i$
      $Count \leftarrow 0$
    **else**
      $Count \leftarrow Count + 1$
    **end if**
  **end if**
  $i \leftarrow i + 1$
**end while**
  **return** $y_{max}, i$

The figure indicates a pseudo code of the Bayesian optimizer with early stopping. Early stopping is mainly used as a regularization method to prevent an overfitting issue in machine learning. The fitness value has a range between 0 and -1. Patience means the number of epochs without improvement for which training will be stopped. The patience value and filter value are the hyperparameters for early stopping.

**Supplementary Figure S9 Optimization performance by early stopping**

**Supplementary Figure S9a. The NP case with a target $\lambda_{max}$ = 513 nm**

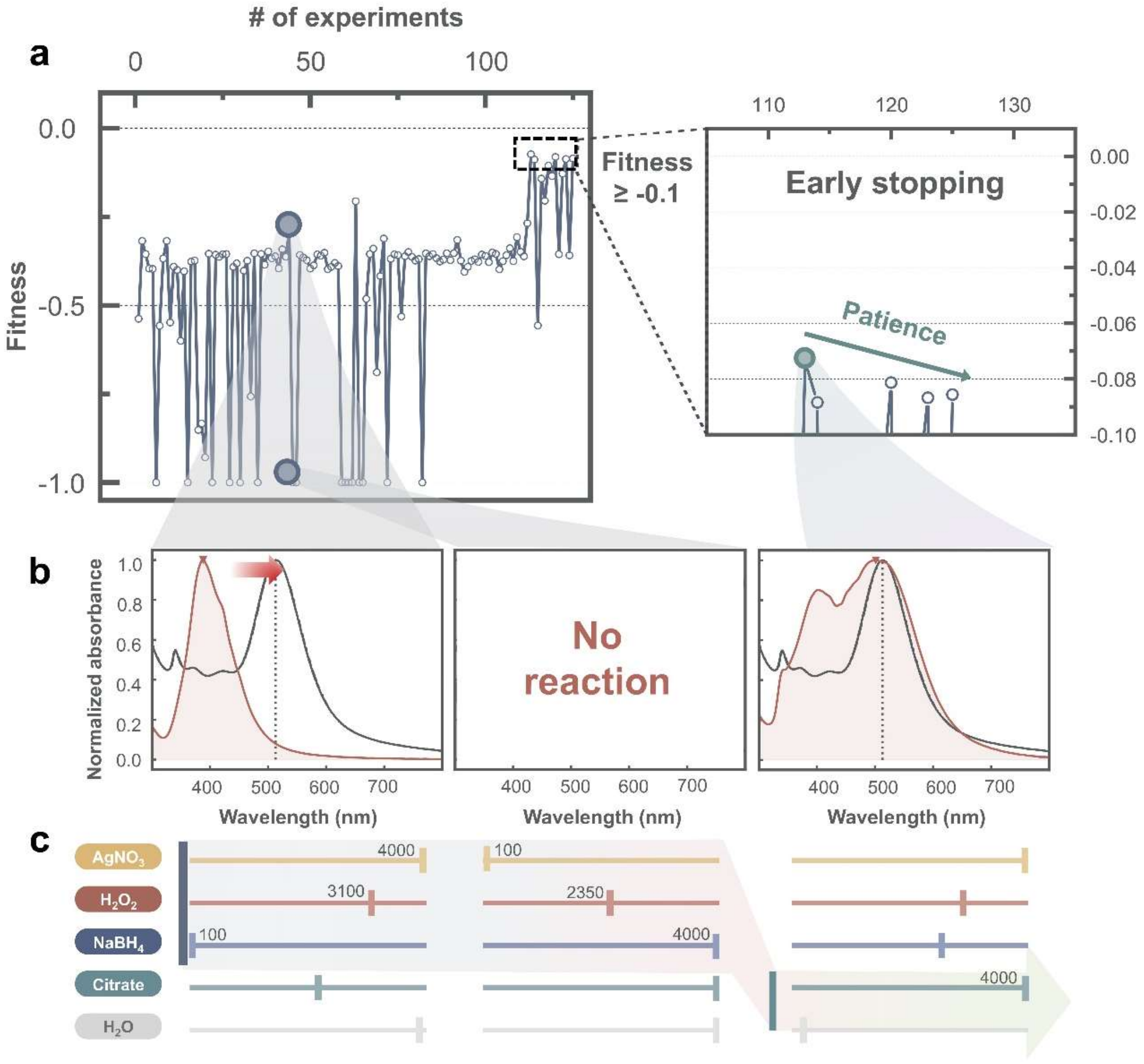

**(a)** Variation in the fitness values during autonomous experiments for bespoke synthesis of Ag NPs with $\lambda_{max}$ = 513 nm. The inset shows the zoomed-in region of fitness to clarify the effect of early stopping.

**(b)** Comparison of the absorption spectra between the literature and our experiment. The first and second spectra are obtained in the low fitness cases, while the last spectrum is obtained when the highest fitness (close to zero) is achieved. The second spectrum image means "No reaction", so that it has no absorption spectrum.

**(c)** Synthesis recipe for each absorption spectrum. Each variable can range from 100 to 4,000 μL.

**Supplementary Figure S9b. The NP case with a target $\lambda_{max}$ = 667 nm**

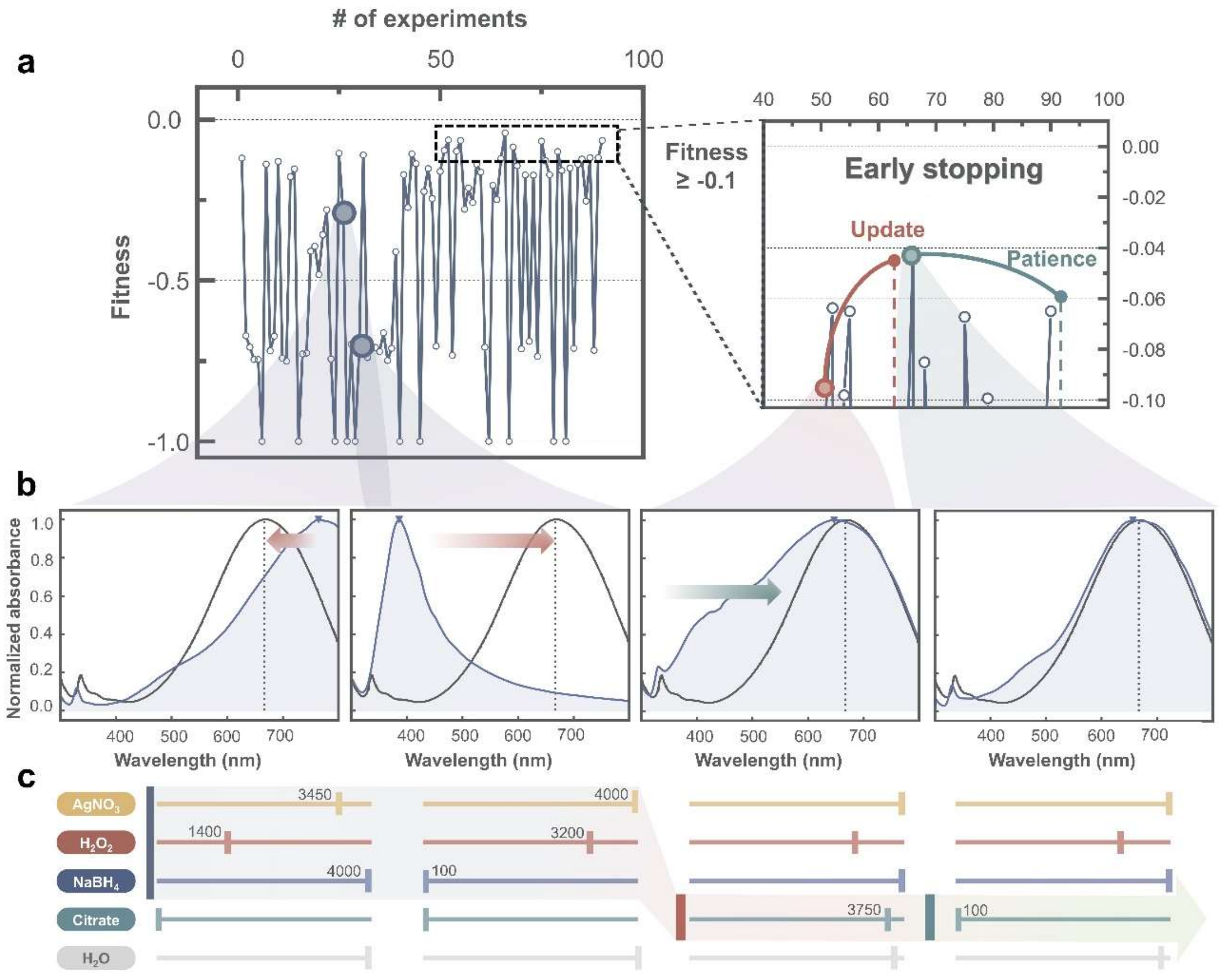

**(a)** Variation in the fitness values during autonomous experiments for bespoke synthesis of Ag NPs with $\lambda_{max}$ = 667 nm. The inset shows the zoomed-in region of fitness to clarify the effect of early stopping.

**(b)** Comparison of the absorption spectra between the literature and our experiments. The first and second spectra are obtained at low fitness, while the third and last spectra are obtained in the update and patience regions, respectively, to show the fitting process that optimizes $\lambda_{max}$ and then FWHM.

**(c)** Synthesis recipe for each absorption spectrum. Each variable can range from 100 to 4,000 μL.

**Supplementary Figure S10. Additional results on absorption spectrum and synthesis recipe**

**Supplementary Figure S10a. The NP case with a target $\lambda_{max}$ = 513 nm**

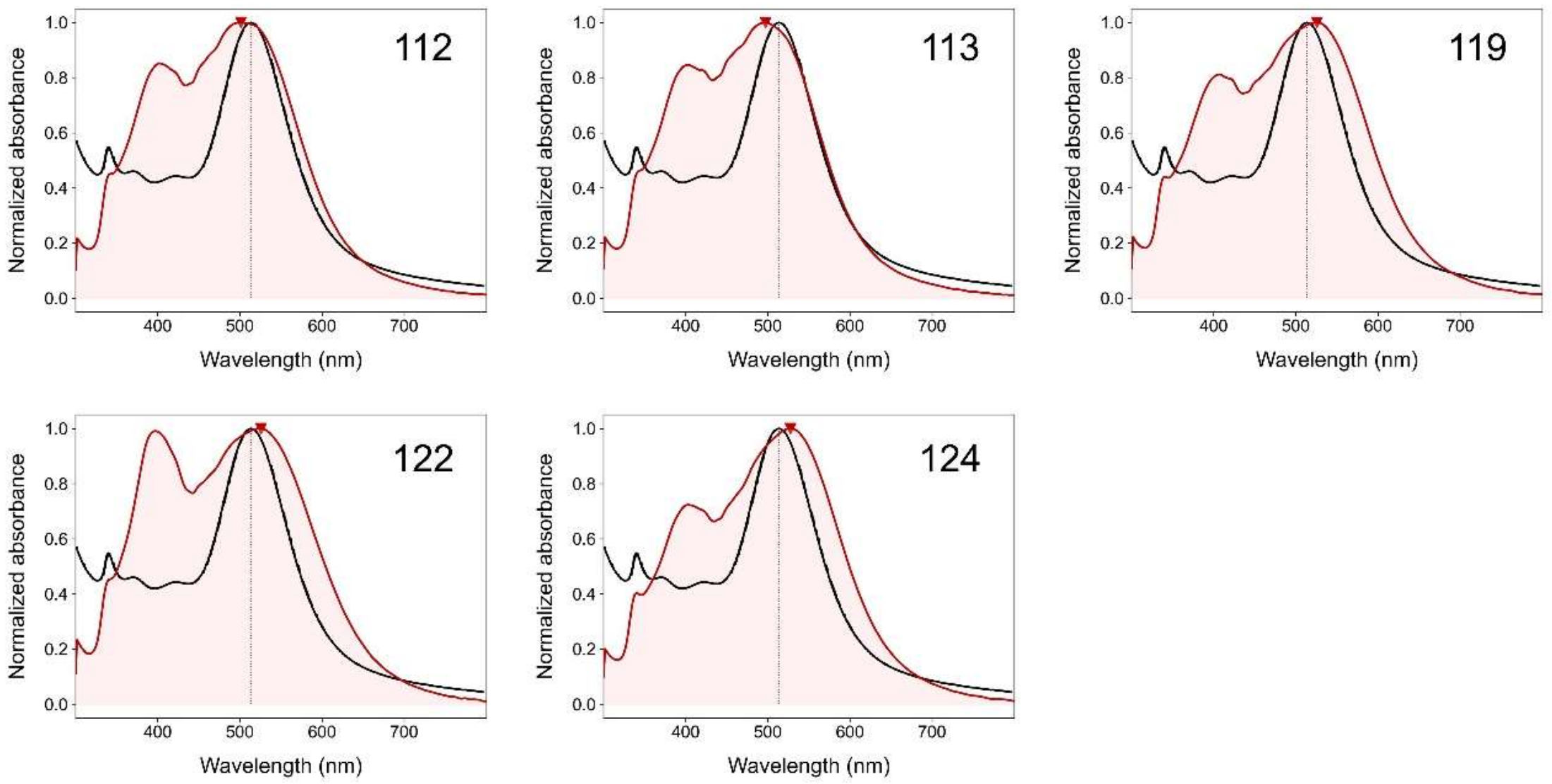

Patience

| Idx | $AgNO_3$ | Citrate | $H_2O$ | $H_2O_2$ | $NaBH_4$ | $\lambda_{max}$ | FWHM | Intensity | Fitness |
|---|---|---|---|---|---|---|---|---|---|
| 112 | 4000 | 2600 | 3950 | 1050 | 4000 | 501.3103 | 214.2446 | 0.561421 | -0.07361 |
| 113 | 4000 | 3500 | 3850 | 1050 | 4000 | 497.1225 | 205.857 | 0.503179 | -0.08841 |
| 119 | 4000 | 4000 | 3800 | 1350 | 4000 | 525.604 | 227.4289 | 0.496972 | -0.08128 |
| 122 | 4000 | 4000 | 4000 | 1450 | 4000 | 525.604 | 239.1663 | 0.42791 | -0.08675 |
| 124 | 4000 | 4000 | 4000 | 1350 | 4000 | 527.6754 | 220.4935 | 0.509183 | -0.08558 |

The figures show absorption spectra obtained in the early stopping range of our autonomous experiments with target $\lambda_{max}$ = 513 nm, where the number in each figure indicates the iteration number of the experiments. The black line corresponds to the literature and the colored line corresponds to our autonomous experiment. Table summarizes the synthesis recipes, optical properties, and fitness values at each experimental iteration. Here, "idx" means the iteration number of experiments ("idx=0" means the first experiment).

## Supplementary Figure S10b. The NP case with a target $\lambda_{max}$ = 573 nm

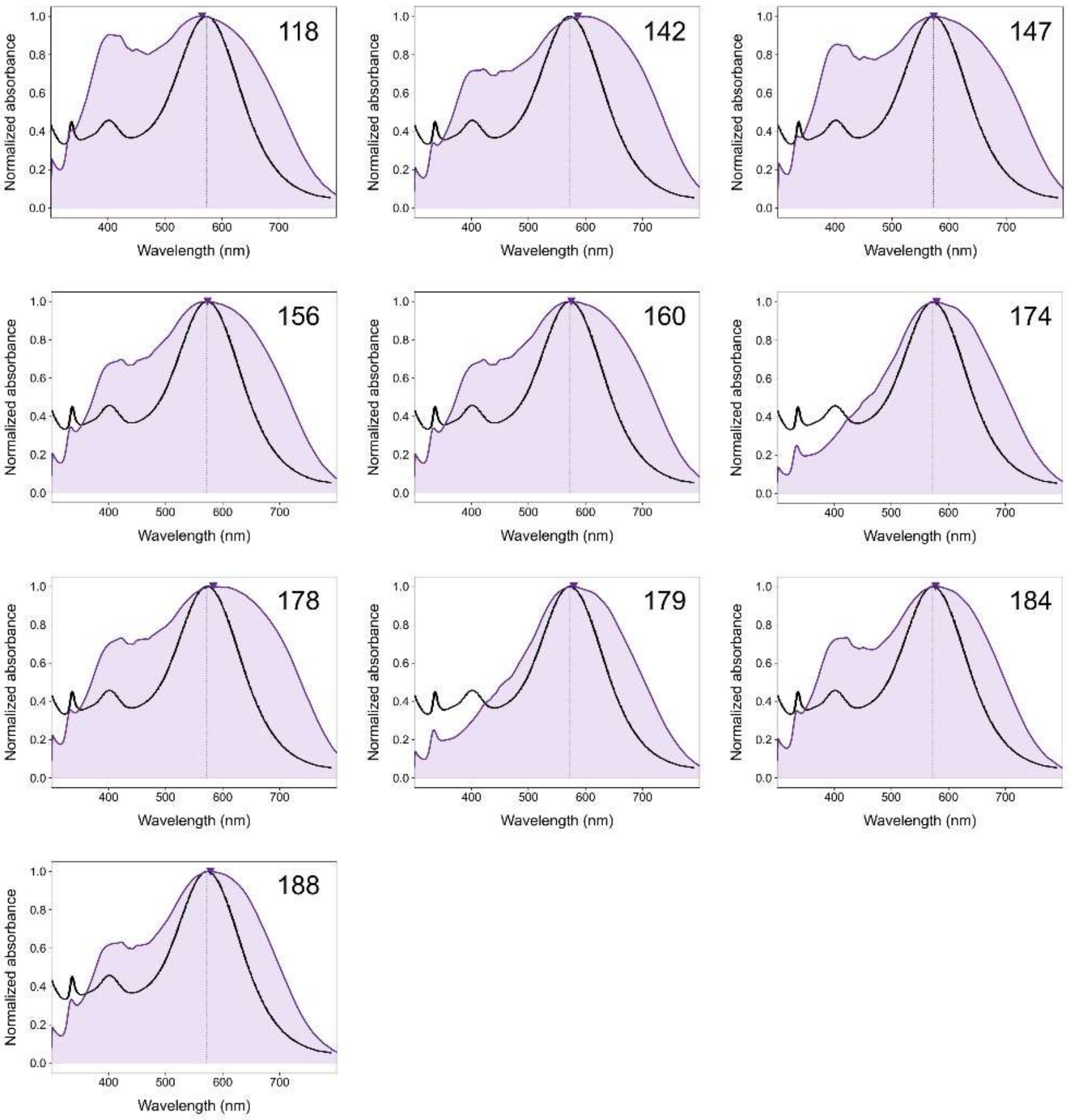

| Group | Idx | $AgNO_3$ | Citrate | $H_2O$ | $H_2O_2$ | $NaBH_4$ | $\lambda_{max}$ | FWHM | Intensity | Fitness |
|---|---|---|---|---|---|---|---|---|---|---|
| Update | 118 | 3950 | 3900 | 250 | 2050 | 2950 | 565.0043 | 335.621 | 0.513147 | -0.07837 |
| Update | 142 | 3950 | 3950 | 450 | 2300 | 3350 | 586.3132 | 351.992 | 0.548284 | -0.09408 |
| Update | 147 | 4000 | 4000 | 100 | 2250 | 3300 | 573.1453 | 355.4606 | 0.517874 | -0.05361 |
| Update | 156 | 4000 | 4000 | 100 | 2150 | 3300 | 573.4839 | 329.9111 | 0.64762 | -0.04423 |
| Update | 160 | 4000 | 4000 | 100 | 2000 | 3250 | 575.1761 | 333.346 | 0.60734 | -0.05274 |
| Update | 168 | 4000 | 4000 | 100 | 1750 | 3250 | 565.6838 | 341.9396 | 0.518408 | -0.07613 |
| Patience | 174 | 4000 | 100 | 100 | 1900 | 3250 | 574.8377 | 215.5066 | 1.207559 | -0.03226 |
| Patience | 178 | 4000 | 4000 | 100 | 1950 | 3250 | 583.6183 | 355.2353 | 0.574327 | -0.08367 |
| Patience | 179 | 4000 | 100 | 100 | 1950 | 3250 | 579.2323 | 229.2848 | 1.211081 | -0.04753 |
| Patience | 184 | 4000 | 4000 | 4000 | 1950 | 3250 | 577.2051 | 320.0108 | 0.48047 | -0.06748 |
| Patience | 188 | 4000 | 4000 | 3250 | 1950 | 3250 | 578.5567 | 311.5932 | 0.523256 | -0.06842 |

The figures show absorption spectra obtained in the early stopping range of our autonomous experiments with target $\lambda_{max}$ = 573 nm, where the number in each figure indicates the iteration number of the experiments. The black line corresponds to the literature and the colored line corresponds to our autonomous experiment. Table summarizes the synthesis recipes, optical properties, and fitness values at each experimental iteration. Here, "idx" means the iteration number of experiments ("idx=0" means the first experiment).

## Supplementary Figure S10c. The NP case with a target $\lambda_{max}$ = 667 nm

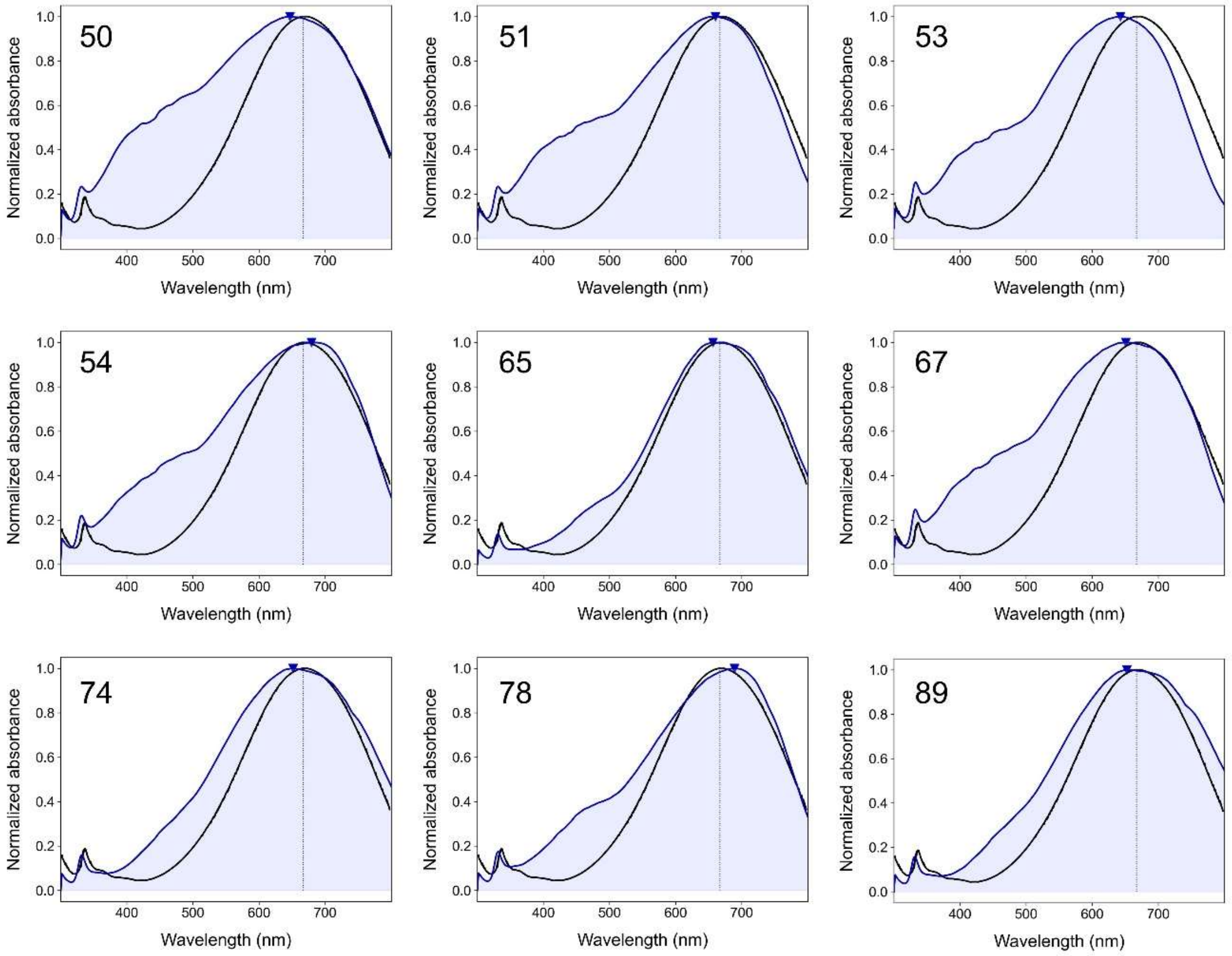

| | Idx | $AgNO_3$ | Citrate | $H_2O$ | $H_2O_2$ | $NaBH_4$ | $\lambda_{max}$ | FWHM | Intensity | Fitness |
|---|---|---|---|---|---|---|---|---|---|---|
| Update | 50 | 4000 | 3950 | 150 | 2100 | 4000 | 646.74 | 363.759 | 0.62285 | -0.09593 |
| | 51 | 4000 | 3900 | 3950 | 2150 | 4000 | 660.114 | 307.169 | 0.5704 | -0.06371 |
| | 53 | 4000 | 3850 | 3700 | 1800 | 4000 | 642.481 | 256.667 | 0.65678 | -0.09816 |
| | 54 | 4000 | 3550 | 3900 | 1800 | 4000 | 679.202 | 275.981 | 0.71364 | -0.06502 |
| Patience | 65 | 4000 | 100 | 3850 | 2000 | 4000 | 656.5345 | 230.3557 | 1.046031 | -0.04145 |
| | 67 | 4000 | 3750 | 3900 | 1950 | 4000 | 650.6634 | 298.4945 | 0.589346 | -0.08509 |
| | 74 | 2100 | 100 | 100 | 1850 | 4000 | 651.3166 | 263.4668 | 0.794329 | -0.06723 |
| | 78 | 2800 | 3950 | 4000 | 2250 | 4000 | 688.8361 | 248.1703 | 0.539129 | -0.09935 |
| | 89 | 2100 | 100 | 100 | 1650 | 4000 | 652.296 | 263.7493 | 0.79277 | -0.06495 |

The figures show absorption spectra obtained in the early stopping range of our autonomous experiments with target $\lambda_{max}$ = 667 nm, where the number in each figure indicates the iteration number of the experiments. The black line corresponds to the literature and the colored line corresponds to our autonomous experiment. Table summarizes the synthesis recipes, optical properties, and fitness values at each experimental iteration. Here, “idx” means the iteration number of experiments (“idx=0” means the first experiment).

## 3. Correlation between the absorption spectrum and NP morphology by TEM analysis

**Supplementary Figure S11. Deconvolution of an absorption spectrum for the Ag NP case with target $\lambda_{max}$ = 513 nm**

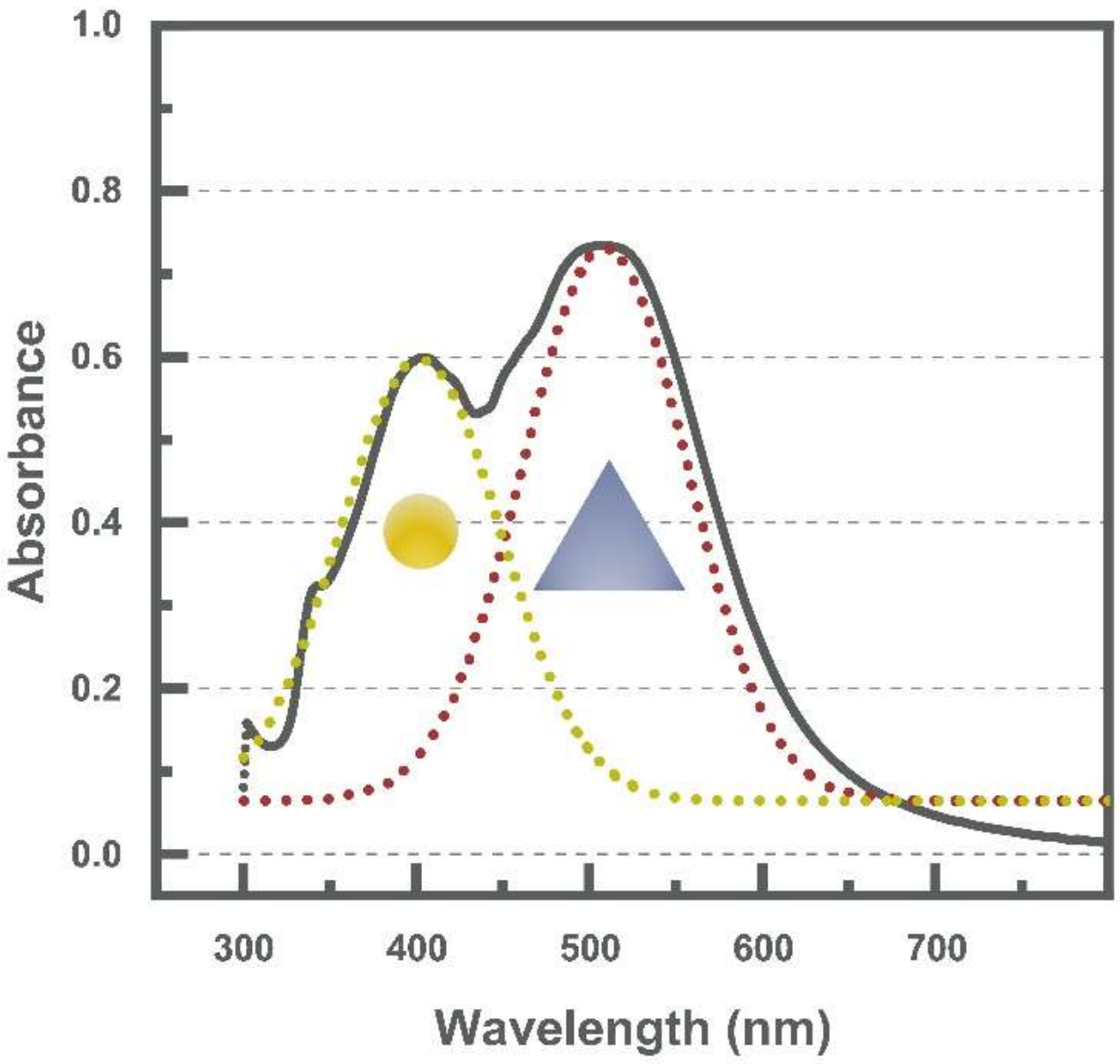

This figure shows deconvolution of the absorption spectrum for Ag NPs with target $\lambda_{max}$ = 513 nm into two peaks at ~400 nm and 500~600 nm, which correspond to a sphere shape[4] and a nanoplate shape[5], respectively.

**Supplementary Figure S12. TEM analysis to verify the shape of synthesized Ag nanoparticles**

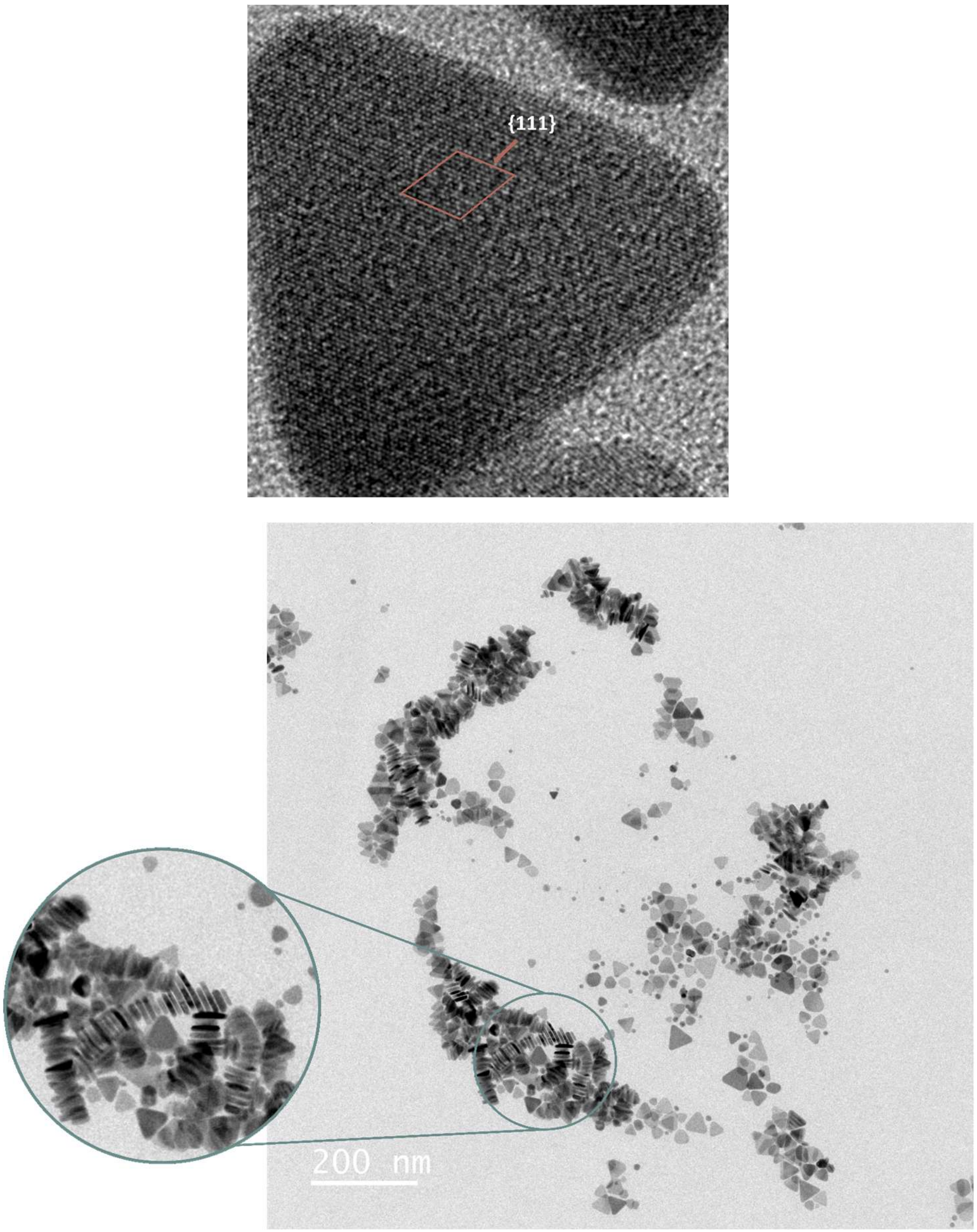

# 4. Interpretation of synthesis variables and visualization of the AI-based synthetic route

**Supplementary Figure S13. Correlation between $\lambda_{max}$ and the ratio of $H_2O_2/AgNO_3$**

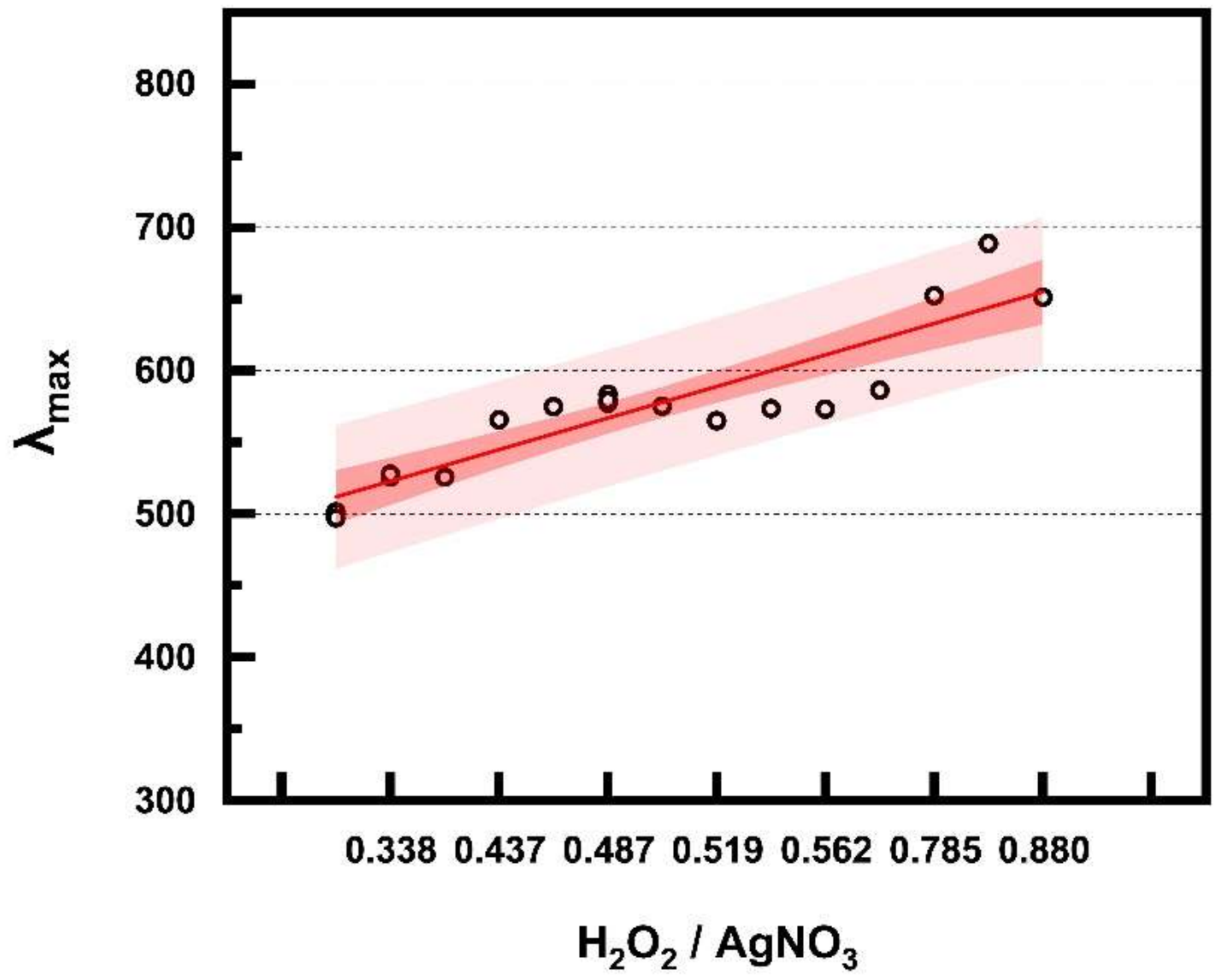

# 5. Effect of citrate on the shape evolution of Ag nanoparticles

## Supplementary Figure S14. Effect of citrate concentration on the absorption spectra

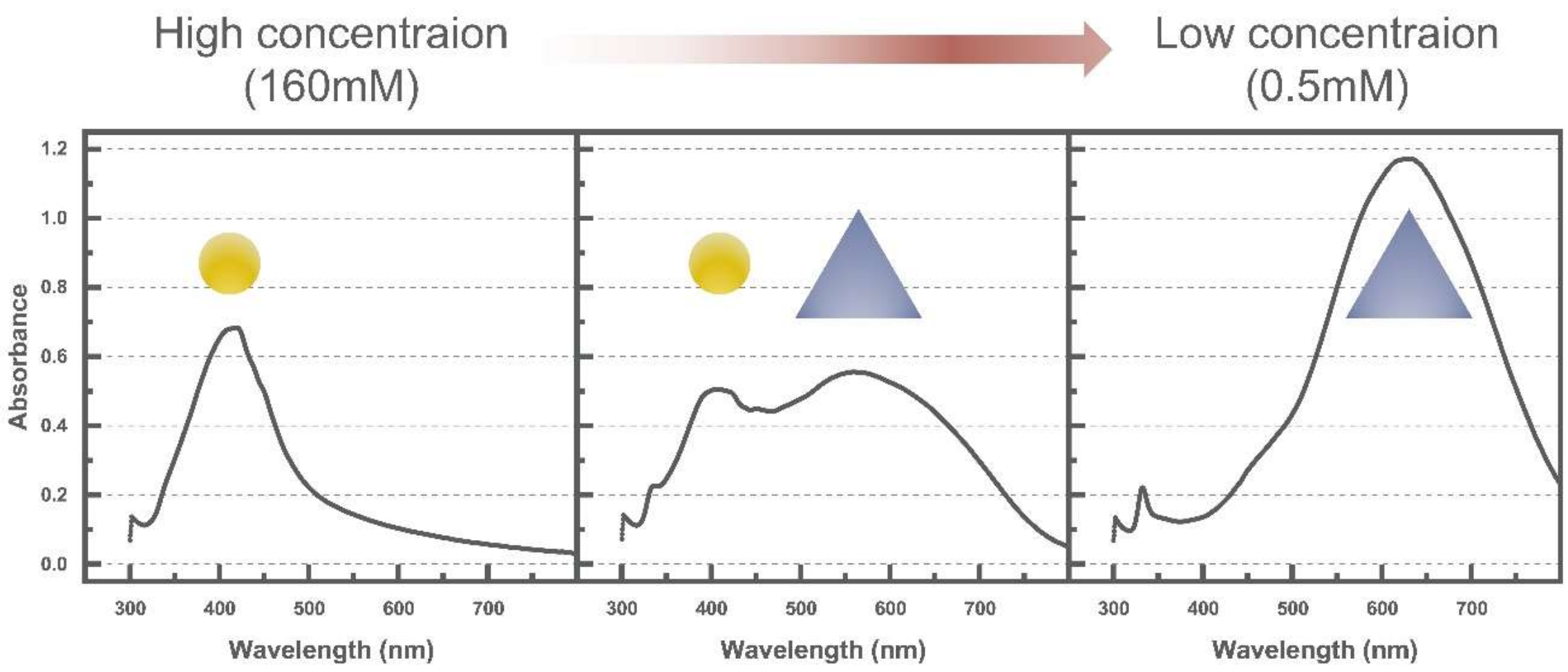

This figure shows the trend between the absorption spectrum and the concentration of citrate reagent. A yellow circle indicates the spherical morphology of the Ag nanoparticles, and a purple triangle indicates the morphology of the Ag nanoparticles.

**Supplementary Figure S15. Trend analysis of the FWHM and peak intensity of the absorption spectra as a function of citrate concentration**

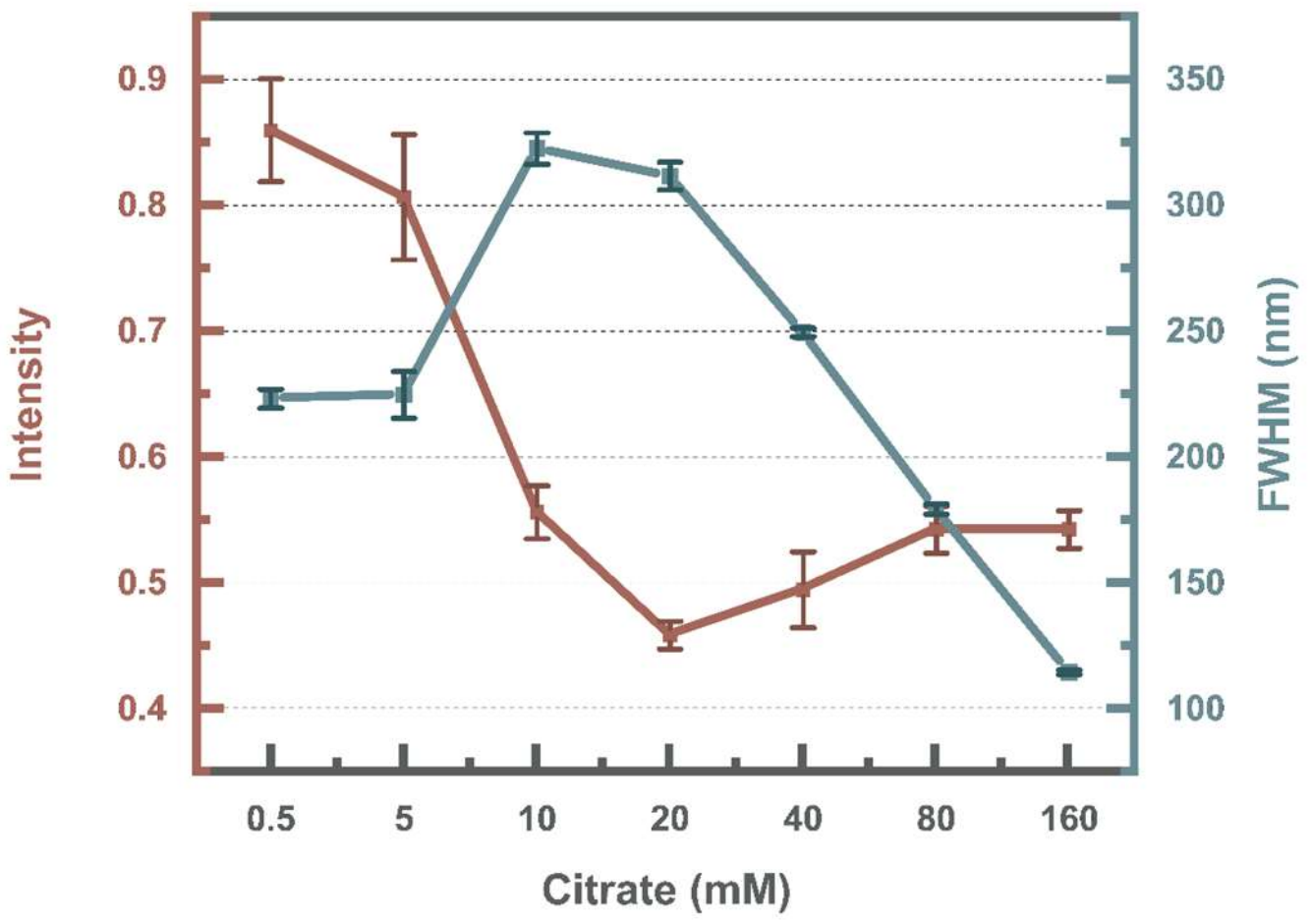

This interval plot shows the trend of the FWHM and peak intensity with citrate concentration. It shows the 95% confidence interval during ten experiments for each concentration. The red line plot represents the intensity and the blue line plot represents the FWHM.

## 6. Quantitative estimations of AI-based synthesis planning

**Supplementary Table S2. The number of experiments required to complete the optimizations with the number of variables**

| ***Target*** | ***Var=2*** | ***Var=3*** | ***Var=4*** | ***Var=5*** |
|---|---|---|---|---|
| Theoretically computed numbers of the grid-based search | $79^2$ = 6,241 | $79^3$ = 493,039 | $79^4$ = 38,950,081 | $79^5$ = 3,077,056,399 |
| 513 nm | 85 | 124 | 129 | 125 |
| 573 nm | 62 | 58 | 87 | 189 |
| 667 nm | 45 | 52 | 50 | 90 |
| **Average of experiments** | **64** | **78** | **89** | **135** |

Each synthesis variable has 79 discrete points by adjusting within the range of 100~4,000 μL by 50 μL. Therefore, theoretically computed numbers of the grid-based search for all candidates should increase exponentially. In controlling two variables, $AgNO_3$ and $H_2O_2$, we fixed volumes of $NaBH_4$, citrate, and $H_2O$ as 3,000 μL, 1,200 μL, and 2,400 μL, respectively. In controlling three synthesis variables, $AgNO_3$, $H_2O_2$, and $NaBH_4$, we fixed volumes of citrate and $H_2O$ as 1,200 μL and 2,400 μL, respectively. In controlling four synthesis variables, $AgNO_3$, $H_2O_2$, $NaBH_4$, and citrate, we fixed the volume of $H_2O$ at 2,400 μL. Here, the fixed volumes of synthesis variables are referred to in a reported literature[5].

# Reference

1. Gome, Gilad, *et al.* OpenLH: open liquid-handling system for creative experimentation with biology. in *Proceedings of the Thirteenth International Conference on Tangible, Embedded, and Embodied Interaction*, (2019).

2. Tiong, L.C.O, & Yoo, H.J, *et al.* Machine vision for vial positioning detection toward the safe automation of material synthesis. in *arXiv preprint arXiv:2206.07272*, (2022).

3. Rohatgi, A. WebPlotDigitizer user manual version 3.4. in *http://arohatgi.info/WebPlotDigitizer/app*, 1-18 (2014).

4. Kelly, K. L., Coronado, E., Zhao, L. L. & Schatz, G. C. The optical properties of metal nanoparticles: The influence of size, shape, and dielectric environment. *J. Phys. Chem. B* 107, 668–677 (2003).

5. Zhang, Q., Li, N., Goebl, J., Lu, Z. & Yin, Y. A systematic study of the synthesis of silver nanoplates: Is citrate a “magic” reagent? *J. Am. Chem. Soc.* 133, 18931–18939 (2011).